\documentstyle[12pt, epsf]{article}
\input psfig.tex
\begin{document}
\def\sm{\mbox{$SU(3)_C\times SU(2)_L \times U(1)_Y$}\,}
\def\suu{$SU(2)_L \times U(1)_Y$\,}
\def\su{$SU(2)_l \times SU(2)_h\times U(1)_Y$\,}
\def\slash{\not{}{\mskip-3.mu}}
\def\ra{\rightarrow}
\def\lra{\leftrightarrow}
\def\bea{\begin{eqnarray}}
\def\ena{\end{eqnarray}}
\def\beq{\begin{equation}}
\def\enq{\end{equation}}
\def\cs{\cos{\theta}}
\def\sn{\sin{\theta}}
\def\tw{\tan\theta}
\def\css{\cos^2{\theta}\,}
\def\sns{\sin^2{\theta}\,}
\def\tns{\tan^2{\theta}\,}
\def\eg{{\it e.g.},\,\,}
\def\ie{{\it i.e.},\,\,}
\def\etc{{\it etc}}
\def\MWs{M^2_W}
\def\MZs{M^2_Z}
\def\MHs{m^2_H}
\def\mf{m_f}
\def\mb{m_b}
\def\mt{m_t}
\def\snshel{\sin^2{\hat{\theta}}\,}
\def\csshel{\cos^2{\hat{\theta}}\,}
\def\snsms{\overline{s}^2_{\theta}\,}
\def\cssms{\overline{c}^2_{\theta}\,}
\def\kln{\kappa_{L}^{NC}}
\def\krn{\kappa_{R}^{NC}}
\def\klc{\kappa_{L}^{CC}}
\def\krc{\kappa_{R}^{CC}}
\def\ttz{{\mbox {\,$t$-${t}$-$Z$}\,}}
\def\bbz{{\mbox {\,$b$-${b}$-$Z$}\,}}
\def\tta{{\mbox {\,$t$-${t}$-$A$}\,}}
\def\bba{{\mbox {\,$b$-${b}$-$A$}\,}}
\def\tbw{{\mbox {\,$t$-${b}$-$W$}\,}}
\def\tltlz{{\mbox {\,$t_L$-$\overline{t_L}$-$Z$}\,}}
\def\blblz{{\mbox {\,$b_L$-$\overline{b_L}$-$Z$}\,}}
\def\brbrz{{\mbox {\,$b_R$-$\overline{b_R}$-$Z$}\,}}
\def\tlblw{{\mbox {\,$t_L$-$\overline{b_L}$-$W$}\,}}
\def\ppbar{ \bar{{\rm p}} {\rm p} }
\def\pp{ {\rm p} {\rm p} }
\def\ipb{ {\rm pb}^{-1} }
\def\ifb{ {\rm fb}^{-1} }
\def\stds{\strut\displaystyle}
\def\SST{\scriptscriptstyle}
\def\TT{\textstyle}
\def\D0{D\O~}
\def\sqrts{\sqrt{s}}

\def\tbW{{\mbox {\,$t$-${b}$-$W$}\,}}
\def\qgtb{q' g \ra q t \bar b}
\def\Wgtb{q' g (W^+ g) \ra q t \bar b}
\def\ubdt{q' b \ra q t}
\def\udbt{q' \bar q \ra W^* \ra t \bar b}
\def\Wbt{W^+ b \ra t}
\def\ggtt{q \bar q, \, g g \ra t \bar t}
\def\ttb{t \bar t}
\def\Wt{W t}
\def\gbtW{g b \ra W^- t}
\def\width{\Gamma(t \ra b W^+)}
\def\pbarp{ \bar{{\rm p}} {\rm p} }
\def\flong{f_{\rm Long}}
\def\rts{\sqrt{s}}

\def\ov{\overline}
\def\ra{\rightarrow}
\def\aslash{\not{\hbox{\kern -1.5pt $a$}}}
\def\bslash{\not{\hbox{\kern -1.5pt $b$}}}
\def\Dslash{\not{\hbox{\kern -4pt $D$}}}
\def\wslash{\not{\hbox{\kern -4pt $\cal W$}}}
\def\zslash{\not{\hbox{\kern -4pt $\cal Z$}}}
\setcounter{footnote}{1}
\renewcommand{\thefootnote}{\fnsymbol{footnote}}

\begin{titlepage}

hep-ph$\backslash$9609482 
 \hfill {\small MSUHEP-60922}
\begin{flushright}
{ September 1996}
\end{flushright}

\vspace*{1.2cm}
\vspace*{0.5cm} 
\centerline{\Large\bf  
Probing the Electroweak Symmetry Breaking}
\baselineskip=18pt
\centerline{\Large\bf  
Sector with the Top Quark\footnote{Lecture given by C.--P. Yuan
at the XXXVI Cracow School of Theoretical Physics, 
June 1-10, 1996, Zakopane, Poland; and at the Workshop on 
Physics at TeV Energy Scale, July 15-26, 1996, Beijing, China.}}

\vspace*{1.2cm}
\baselineskip=17pt
\centerline{\normalsize  
{\bf  F. Larios$^{(a,b)}$~~ E. Malkawi$^{(c)}$~~
and~~ C.--P. Yuan$^{(a)}$ }}
 
\vspace*{0.4cm}
\centerline{\normalsize\it
(a) Department of Physics and Astronomy, Michigan State University }
\centerline{\normalsize\it
East Lansing, Michigan 48824 , USA.}
\centerline{\normalsize\it
(b) Departmento de F\'{\i}sica, CINVESTAV, }
\centerline{\normalsize \it 
Apdo. Postal 14-700, 07000 M\'exico, D.F., M\'exico. }
\centerline{\normalsize\it
(c) Dept. of Physics, Jordan University of Science I\& Technology,}
\centerline{\normalsize\it
P.O. Box 3030-IRBID-22110, Jordan.}

\vspace{0.4cm}
\raggedbottom
\setcounter{page}{1}
\relax

\begin{abstract}
\noindent
A study on the effective anomalous interactions, up to dimension 5, of
the top quark with the electroweak gauge bosons is made in the
non-linear Chiral Lagrangian approach.  Bounds on the anomalous
dimension four terms are obtained from their
contribution to low energy data. Also,
the potential contribution to the production of top quarks at 
hadron colliders (the Tevatron and the LHC) and the electron 
Linear Collider from both dimension 4 and 5 operators is analyzed.
\end{abstract}
\vspace*{3.4cm}
\end{titlepage}
\normalsize\baselineskip=15pt
\renewcommand{\thefootnote}{\arabic{footnote}}
\setcounter{footnote}{0}

\newpage

\section{Introduction}
\indent\indent

Despite the unquestionable significance of 
its achievements, like that of
predicting the existence of the top quark \cite{topdisc},
there is no reason to believe that the Standard Model (SM) is the final
theory.  For instance, the SM contains many arbitrary parameters with no
apparent connections.  In addition,  the SM provides no
satisfactory explanation for the symmetry-breaking mechanism which takes
place and gives rise to the observed mass spectrum of the gauge bosons
and fermions.   Because the top 
quark is heavy relative to other observed
fundamental particles,\footnote{
As of the summer of 1996, 
the mass of the top quark has been measured 
at the Fermilab Tevatron to be
$m_t = 175.6 \pm 5.7 \, {\rm (stat.)} \pm 7.1 {\rm (sys.)}$\,GeV 
by the CDF group and
$m_t = 169 \pm 8 \, {\rm (stat.)} \pm 8 {\rm (sys.)}$\,GeV 
by the ~\D0 group, through the detection of $t \bar t$ events.} 
one expects that any underlying theory; 
to supersede the SM at some high energy scale $\Lambda \gg m_t$,
will easily reveal itself at lower energies through
the effective interactions of
the top quark to other light particles.
Also because the top quark mass ($\sim {v/ {\sqrt{2}}}$) 
is of the order of the Fermi scale $v={(\sqrt{2}G_F)}^{-1/2}=246$\,GeV,
which characterizes the electroweak symmetry-breaking scale, the 
top quark system could be a useful probe
for the symmetry-breaking sector. Since the 
fermion mass generation can be closely related to the electroweak
symmetry-breaking, one expects some residual effects of this breaking to
appear in accordance with the mass hierarchy \cite{pczh,sekh,ehab}. This
means that new effects should 
be more apparent in the top quark sector than
in any other light sector of the theory.  Therefore, 
it is important to study the
top quark system as a direct 
tool to probe new physics effects \cite{kane}.
 
Many attempts to offer alternative 
scenarios for the electroweak symmetry
breaking mechanism are discussed in 
literature. A general trend among all 
alternatives is that new physics appear at or below the TeV scale.
Examples include Supersymmetry models \cite{ch3_haber}, technicolor
models \cite{ch3_haber,lane} and possibly extended technicolor sectors
to account for the fermion masses \cite{sekhar2,sekhar3}. 
Other examples include top-mode condensate models \cite{tana}
and a strongly interacting Higgs sector \cite{ch3_casa}.

An attempt to study the nonuniversal interactions of the top quark has 
been carried out in Ref.~\cite{pczh,ehab} by Peccei et al. However, 
in that study only the vertex \ttz was considered based on the 
assumption that this is the only vertex which gains a significant 
modification due to a speculated dependence of
the coupling strength on the fermion mass:
$\kappa_{ij} \leq {\cal O} 
\left ( \frac{\sqrt{m_{i}m_{j}}}{v} \right )$,
where $\kappa_{ij}$ parameterizes some new 
dimensional--four interactions among gauge bosons and fermions $i$ 
and $j$. However, this is not the only possible 
pattern of interactions, 
{\it e.g.}, in some extended technicolor models 
\cite{sekhar3} one finds the nonuniversal residual interactions 
associated with the vertices \blblz, \tltlz, and \tlblw
to be of the same order.

Because of the great diversity of models proposed for possible 
new physics (beyond the SM), it 
has become necessary to be able to study 
these possible new interactions in a model 
independent approach \cite{effec}.
This approach has proved to render relevant non-trivial information 
about the possible deviations from the standard couplings of the 
heavier elementary particles (heavy scalar  bosons, the bottom and 
the top quarks, etc.) \cite{miguel}.   
Our study focuses on  the top quark, which 
because of its remarkably higher mass is the best candidate 
(among the fermion particles) for the manifestation of these
anomalous interactions at high energies \cite{toptalk}.

A common approach to study these anomalous couplings is by 
considering the most general on-shell vertices (form factors) 
 involving the bottom and the top quarks together with
the interaction bosons \cite{kane}. 
In this work we will incorporate the effective 
chiral Lagrangian approach \cite{chiral,howard},  
which is based on the principle of gauge symmetry, 
but the symmetry is realized in the most general 
(non-linear) form so as to encompass all the possible interactions 
consistent with the existing experimental data.  
The idea of using this approach 
is to exploit the linearly realized ${\rm{U(1)}_{em}}$ symmetry
and the non-linearly realized $\rm{SU(2)}_L \times \rm{U(1)}_Y$ symmetry
to make a systematic characterization of all the anomalous couplings.  
In this way, for example, different couplings which otherwise would be 
considered as independent become related through the equations of 
motion.    

We show that in general low energy data (including $Z$ 
pole physics) do not impose any stringent constraints on the anomalous
dimension four coefficients $\kappa$ of ${\cal L}^{(4)}$
( see Eq.~(\ref{eq2}) )\footnote{
For simplicity, we will only construct the complete set of dimension
4 and 5 effective operators for the 
fermions $t$ and $b$, although our results
can be trivially extended for 
the other fermion fields, e.g. flavor changing
neutral interactions $t$-$c$-$Z$, etc.}.
This means that low 
energy data do not exclude the possibility of new physics
whose effects come in through the deviations 
from the standard interactions of
the top quark, and these deviations have to be directly 
measured via production
of top quarks at the colliders.  
For instance, the couplings ${\kappa}^{CC}_{L,R}$ 
can be measured from the decay 
of the top quarks in $t\overline {t}$ pairs 
produced either at hadron colliders ( the Fermilab Tevatron and the 
CERN Large Hadron Collider (LHC) ) 
or at the electron linear collider ( LC ).
They can also be studied from the 
production of the single-top quark events via, for example,
 $W$-gluon or $W$-photon fusion process \cite{effw}.
 The coupling ${\kappa}^{NC}_{L,R}$ can only 
be sensitively probed at a future linear collider via the 
$e^{+}\;e^{-}\;\ra \gamma , Z \;\ra t\ov {t}$ process because at hadron 
colliders the $t\ov {t}$ production rate 
is dominated by QCD interactions 
( $q\ov {q}, gg\;\ra\; t\ov {t}$ ).   However, at the LHC 
${\kappa}^{NC}_{L,R}$ may also be studied 
via the associated production of 
$t\ov {t}$ with $Z$ bosons (this requires a separate study).

Also, we will include the next higher order
dimension 5 fermionic operators and then 
examine the precision with which 
the coefficients of these operators can be measured in high energy 
collisions.   Since it is the 
electroweak symmetry breaking sector that we 
are interested in, we shall concentrate 
on the interaction of the top quark 
with the longitudinal weak gauge bosons; which are equivalent to the 
would-be-Goldstone bosons in the high energy 
limit.  This equivalence is 
known as the Goldstone Equivalence Theorem \cite{et1pol}-\cite{et}.

Our strategy for probing these anomalous dimension 5 operators 
( ${\cal L}^{(5)}$ ) is to 
study the production of $t \ov t$ pairs as well as 
single-$t$ or $\ov t$ via the $W_L W_L$, $Z_L Z_L$ and $W_L Z_L$ 
(denoted in general as $V_L V_L$) processes in the 
TeV region.  As we shall show later, 
based on a power counting method \cite{poc},
the leading contribution of the 
scattering amplitudes at high energy goes as $E^3$ for the anomalous 
operators  ${\cal L}^{(5)}$, 
where $E\,=\,\sqrt{s}$ is the CM energy of the $W W$ or $Z Z$ system 
(that produces $t\ov t$), or the $W Z$ system 
(that produces $t\ov b$ or $b\ov t$).   On the other hand, when the
$\kappa$ coefficients are set equal to zero the dimension 4 operators
${\cal L}^{(4)}$ can at most contribute with the first power $E^1$
to these scattering $V_L V_L$ processes .  In other words, the high
energy $V_L V_L \ra f \ov f$ scatterings are more sensitive to 
${\cal L}^{(5)}$ than to ${\cal L}^{(4)}$ (with $\kappa$'s $=0$).
If the $\kappa$'s are not set equal to zero, then the high energy
behaviour can at most grow as $E^2$ as compared to $E^3$ for the 
dimension 5 operators \cite{top5}.  
Furthermore, the dimension 4 anomalous couplings 
$\kappa$'s are better measured at the scale of $M_W$ or $m_t$ by  
studying the decay or the production of the top quark at either the
Tevatron and the LHC as mentioned before, or the LC at the $t\ov t$
threshold (for the study of $Z$-$t$-$t$). 

We show that there are 19 independent dimension 5 
operators (with only $t$, $b$ 
and gauge boson fields) in ${\cal L}^{(5)}$ 
after imposing the equations of 
motion for the effective chiral lagrangian. 
The coefficients of these operators 
can be measured at either the LHC or the  
LC to magnitudes of order $10^{-2}$ 
or $10^{-1}$ after normalizing (the  
coefficients) with the factor\footnote{$\Lambda$ is 
the cut-off scale of the effective theory.  It could be the lowest new 
heavy mass scale, or something around $4 \pi v \simeq 3.1$ TeV if no 
new resonances exist below $\Lambda$.} $\frac {1}{\Lambda}$ based on the
naive dimensional analysis \cite{georgi,howard}.  It 
is expected that at the LHC or the 
LC there will be  about a few hundreds to a few thousands of $t\ov t$ 
pairs or single-$t$ or single-$\ov t$ events produced via the $V_L V_L$ 
fusion process.

This work is organized as follows:  
In section 2 we will introduce the basic
framework of the non-linearly realized chiral Lagrangian, in which the
$SU(2)_L\times U(1)_Y$ gauge symmetry is nonlinearly realized.  In this
approach, only the $U(1)_{EM}$ symmetry remains unbroken and thus
the realization under this subgroup is 
linear as usual.   We will set up a
Lagrangian with dimension 4 terms that will reproduce the couplings
of the Standard Model type, as well as possible deviations.  Then, in
sections 3 and 4 we discuss the constraints on the dimension 4 anomalous
couplings from low energy data and 
the strategies to directly measure these
couplings at the hadron or electron colliders.   In sections 5 and 6 we 
construct the complete set of dimension 5 couplings and discuss
their effects to the production of top quarks in high energy regime
via weak boson fusion processes.  
Finally our conclusions are given in section 7.

\section{The Non-linearly Realized Electroweak Chiral Lagrangian}
\indent\indent

We consider the electroweak theories in which the gauge symmetry 
$G \equiv {\rm{SU(2)}}_{L}\times {\rm{U(1)}}_{Y}$ is spontaneously 
broken 
down to $H={\rm{U(1)}}_{em}$\cite{malkawi,cole,chan}.  
There are three 
Goldstone bosons, $\phi^{a}$ ($a=1,2,3$), generated by this 
breakdown of $G$ into $H$, which are eventually {\it eaten} by the 
$W^{\pm}$ and $Z$ gauge bosons and become their longitudinal 
degrees of freedom.

In the non-linearly realized chiral 
Lagrangian formulation, the Goldstone 
bosons transform non-linearly under $G$ but linearly
under the subgroup $H$. A convenient 
way to implement this is to introduce
the matrix field
\begin{equation}
\Sigma ={\rm{exp}}\left( i\frac{\phi^{a}\tau^{a}}{v_{a}} \right)
\label{sigfield}\, ,
\end{equation}
where $\tau^{a},\, a=1,2,3$, are the Pauli matrices normalized as
${\rm{Tr}}(\tau^a \tau^b)=2 \delta_{ab}$.
The matrix field $\Sigma$ transforms under $G$ as
\begin{equation}
\Sigma\rightarrow {\Sigma}^{\prime}=\,g_L
\Sigma \,g^{\dagger}_R\, ,
\end{equation}
with
\begin{eqnarray}
g_L =&& {\rm {exp}}\left ( i\frac{\alpha^{a}\tau^{a}}{2}\right )\; ,\\
g_R =&& {\rm {exp}}(i\frac{y\tau^3}{2})\; ,\nonumber
\end{eqnarray}
where $\alpha^{1,2,3}$ and $y$ are the group parameters of $G$.
Because of the ${\rm{U(1)}}_{em}$ invariance, $v_1=v_2=v$ in
Eq.~(\ref{sigfield}), but they are not necessarily equal to $v_3$.
In the SM, $v$ ($=246$\,GeV) is 
the vacuum expectation value of the Higgs
boson field, and characterizes the scale of the symmetry-breaking.
Also, $v_3=v$ arises from the approximate custodial symmetry
present in the SM.
It is this symmetry that is responsible for the
tree-level relation
\begin{equation}
\rho= \frac{M_W^2}{M_Z^2\,\cos^2 \theta_W}=1\,
\label{yeq3}
\end{equation}
in the SM, where $\theta_W$ is the electroweak mixing angle,
$M_W$ and $M_Z$ are the masses of $W^\pm$ and $Z$ boson, respectively.
In this study we assume the underlying theory guarantees that
$v_1=v_2=v_3=v$.

In the context of this non-linear formulation of 
the electroweak theory, 
the massive charged and neutral weak bosons can be 
defined by means of the {\it composite} field:
\begin{equation}
{{\cal W}_{\mu}^a}=-i{\rm{Tr}}(\tau^{a}\Sigma^{\dagger}
D_{\mu}\Sigma)\,\label{compw}
\end{equation}
where\footnote{This is not the covariant derivative of $\Sigma$.   The 
covariant derivative is \\
$D_{\mu}\Sigma=\partial_{\mu}\Sigma-ig\frac{\tau^a}{2}W_{\mu}^a
\Sigma+ig^{\hspace{.5mm}\prime}\Sigma \frac{\tau^3}{2}B_{\mu}$.}
\begin{equation}
D_{\mu}\Sigma=\left (\partial_{\mu}-ig\frac{\tau^a}{2}W_{\mu}^a\right)
\Sigma\,\,\, .\label{covderw}
\end{equation}
Here, $W_\mu^a$ is the gauge boson associated with the 
${\rm{SU(2)}}_L$ group, and its transformation is the usual one 
($g$ is the gauge coupling). 
\begin{equation}
\tau^a W_\mu^a \ra \tau^a W_\mu^{'a} \;=\;
 g_L\; \tau^a W_\mu^a\; g_L^{\dagger}\;+\;
\frac{2i}{g} g_L\partial_\mu g_L^{\dagger}
\end{equation}
The $D_{\mu}\Sigma$ term transforms under $G$ as
\begin{equation}
D_{\mu}\Sigma \ra D_{\mu}\Sigma^{'} = 
g_L \left( D_{\mu}\Sigma \right) g^{\dagger}_R +
g_L \Sigma \partial_\mu g^{\dagger}_R \; .
\end{equation}
Therefore, by using the commutation rules 
for the Pauli matrices and the 
fact that $Tr(AB)=Tr(BA)$ we can prove that the composite field 
${{\cal W}_{\mu}^a}$ will transform under $G$ in the following manner:
\begin{equation}
 {{\cal W}_{\mu}^3}\rightarrow {{{\cal W}^{\prime}}_{\mu}^3}
       ={{\cal W}_{\mu}^3}-\partial_{\mu}y\, ,
\end{equation}
\begin{equation}
{{\cal W}_{\mu}^\pm}\rightarrow {{{\cal W}^{\prime}}_{\mu}^\pm}
  =e^{\pm iy}{{\cal W}_{\mu}^\pm} \label{wtransf}\; ,
\end{equation}
where
\begin{equation}
{\cal W}_{\mu}^{\pm}={\frac{{\cal W}_{\mu}^{1}\mp i{\cal W}_{\mu}^{2}}
{\sqrt{2}}}\, .\label{wdef}
\end{equation}
Also, it is convenient to define the field 
\begin{equation}
{\cal B}_{\mu}=g^{\hspace{.5mm}\prime}B_{\mu}\,\, ,
\label{b}
\end{equation}
which is really the same gauge boson field associated with the 
${\rm{U(1)}}_Y$ group. ($g^{\hspace{.5mm}\prime}$ is the gauge 
coupling.)
The field ${\cal B}_{\mu}$ transforms under $G$ as
\begin{equation}
{\cal B}_{\mu} \rightarrow {\cal B}^{\prime}_{\mu} =
 {\cal B}_{\mu}+\partial_{\mu}y\,
\label{bb}\, .
\end{equation}

We now introduce the composite fields
${\cal Z}_{\mu}$ and ${\cal A}_{\mu}$ as
\begin{equation}
{\cal Z}_\mu={\cal W}^3_\mu +{\cal B}_\mu
\label{b1}\,\, ,
\end{equation}
\begin{equation}
s_w^2{\cal A}_\mu = s_w^2{\cal W}^3_\mu 
- c_w^2 {\cal B}_\mu\, ,
\label{b2}
\end{equation}
where $s_w^2\equiv\sin^2\theta_W$, and $c_w^2=1-s_w^2$.
In the unitary gauge ($\Sigma =1$)
\begin{equation}
{\cal W}_{\mu}^a=-gW_{\mu}^a \,\, ,
\end{equation}
\begin{equation}
{{\cal Z}}_{\mu} =-\frac{g}{c_w} Z_{\mu}\,\, ,
\end{equation}
\begin{equation}
{{\cal A}}_{\mu}=-\frac{e}{s_w^2}A_{\mu} \label{afield}\,\, ,
\end{equation}
where we have used the relations\, 
$e=g s_w=g^{\hspace{.5mm}\prime} c_w$,
$W_\mu^3= c_w Z_\mu + s_w A_\mu$, and
$B_\mu= -s_w Z_\mu + c_w A_\mu$.
In general, the composite fields contain Goldstone boson fields:
\begin{eqnarray}
{{\cal Z}}_{\mu}\; =&&
-\frac{g}{c_w} Z_{\mu}+{2\over {v}} {\partial}_{\mu} {\phi}^3 + 
\cdot \cdot \cdot\; , \label{zexp} \\  
{{\cal W}}_{\mu}^{\pm}\; =&&- g W_{\mu}^{\pm}+
{2\over {v}}\partial_{\mu} {\phi}^{\pm}+\cdot \cdot \cdot\; .
\nonumber \label{wexp} 
\end{eqnarray}

The transformations of ${\cal Z}_{\mu}$ and
${\cal A}_{\mu}$ under $G$ are
\begin{equation}
{\cal Z}_{\mu}\rightarrow {\cal Z}_{\mu}^{\prime}=
{\cal Z}_{\mu}\label{ztransf}\, ,
\end{equation}
\begin{equation}
{\cal A}_{\mu} \rightarrow {\cal A}_{\mu}^{\prime} =
{\cal A}_{\mu} -\frac{1}{s_w^2}\partial_{\mu}y\label{atransf} \,\, .
\end{equation}
Hence, under $G$ the fields ${\cal W}_\mu^\pm$ and ${\cal Z}_\mu$
transform as vector fields, but 
${\cal A}_\mu$ transforms as a gauge boson
field which plays the role of the photon field $A_\mu$.

Using the fields defined as above, one may construct
the ${\rm{SU(2)}}_L \times {\rm{U(1)}}_Y$ gauge invariant interaction
terms in the chiral Lagrangian
\begin{eqnarray}
{\cal L}^B =&-&\frac{1}{4g^2} {{\cal W}_{\mu \nu}^a}
{{\cal W}^{a}}^{\mu \nu}
 -\frac{1}{4{g^\prime}^2} {\cal B}_{\mu \nu}{\cal B}^{\mu \nu}
\nonumber \\
&+&\frac{v^2}{4}{\cal W}^{+}_{\mu}{{\cal W}^{-}}^{\mu}+\frac{v^2}{8}
{\cal Z}_{\mu}{\cal Z}^{\mu}+{\dots }\,\, ,
\label{eq4}
\end{eqnarray}
where
\begin{equation}
{\cal W}^{a}_{\mu \nu}=\partial_{\mu}{\cal W}^{a}_{\nu}
-\partial_{\nu}{\cal W}^{a}_{\mu}+\epsilon^{abc}{\cal W}^{b}_{\mu}
{\cal W}^{c}_{\nu}\label{wstrength} \,\, ,
\end{equation}
\begin{equation}
{\cal B}_{\mu \nu}=\partial_{\mu}{\cal B}_{\nu}-
\partial_{\nu}{\cal B}_{\mu}
\,\, ,\end{equation}
and where ${\dots}$ denotes other possible four-
or higher-dimension operators \cite{fer,app}.

It is easy to show that\footnote{
Use ${\cal W}_{\mu}^a \tau^a = -2 i \Sigma^{\dagger} D_\mu \Sigma ~$,
and $[\tau^a,\tau^b]=2 i \epsilon^{abc} \tau^c $.}
\begin{equation}
{\cal W}_{\mu \nu}^a \tau^a=-g\Sigma^{\dagger}W^a_{\mu \nu}\tau^a
\Sigma\,\,
 \end{equation}
and
\begin{equation}
{{\cal W}_{\mu \nu}^a} {{\cal W}^{a}}^{\mu \nu}=g^2 W_{\mu \nu}^{a}
{{W^a}^{\mu \nu}}\,\, .
\label{eq05}
\end{equation}
This simply reflects the fact that the kinetic term is not related to
the Goldstone bosons sector, i.e., it does not originate from the
symmetry-breaking sector.

The mass terms in Eq.~(\ref{eq4}) can be expanded as
\begin{eqnarray}
\frac{v^2}{4}{\cal W}_{\mu}^{+}{{\cal W}^{-}}^{\mu}
+\frac{v^2}{8}{\cal Z}_{\mu}{{\cal Z}}^{\mu}
&=&{\partial}_{\mu}\phi^{+}\partial^{\mu}\phi^{-}
+\frac{1}{2}\partial_{\mu}\phi^{3}\partial^{\mu}\phi^{3} \nonumber \\
&&+\frac{g^2v^2}{4}W_{\mu}^{+}{W^{\mu}}^{-}
+\frac{g^2v^2}{8c_w^2}Z_{\mu}Z^{\mu}+{\dots}\,\, .
\end{eqnarray}
At the tree-level, the mass of $W^\pm$ boson is $M_W=gv/2$ and
the mass of $Z$ boson is $M_Z=gv/2c_w$.

Fermions can be included in this context by assuming that each flavor
transforms under $G={\rm{SU(2)}}_L\times {\rm{U(1)}}_{Y}$ as \cite{pecc}
\begin{equation}
f\rightarrow {f}^{\prime}=e^{iyQ_f}f \label{eq1} \, ,
\end{equation}
where $Q_{f}$ is the electric charge of $f$\footnote{
For instance, $Q_{f}=2/3$ for the top quark.}.

 Out of the fermion fields $f_1$, $f_2$ (two different flavors),
and the Goldstone bosons matrix field $\Sigma$,
the usual linearly realized fields
$\Psi$ can be constructed. For example, the left-handed
fermions [${\rm{SU(2)}}_L$ doublet] are
\begin{equation}
\Psi_{L} \equiv {\psi_1\choose {\psi_2}}_{L} \; = \Sigma F_{L}\; = 
\Sigma{f_1\choose {f_2}}_{L} \label{psi} \,
\end{equation}
with $Q_{f_1}-Q_{{f_2}}=1$.
One can easily show that $\Psi_{L}$\,transforms linearly under $G$ as
\begin{equation}
\Psi_{L}\rightarrow {\Psi}^{\prime}_{L}={\rm g} \Psi_{L}\, ,
\end{equation}
where ${\rm{g}}=
{{\rm {exp}}}(i\frac{\alpha^{a}\tau^{a}}{2})
{{\rm {exp}}}(i y\frac {Y}{2})\in G $, and 
$Y\;=\;\frac{1}{3}$ is the hypercharge of the left handed quark doublet.

In contrast, linearly realized right-handed fermions
$\Psi_{R}$  [${\rm{SU(2)}}_L$ singlet] simply coincide 
with $F_{R}$, i.e.,
\begin{equation}
\Psi_{R}\equiv {\psi_1\choose {\psi_2}}_{R} = 
F_{R}={f_1\choose {f_2}}_{R}\, .\label{psr}
\end{equation}
With these fields we can now construct the most general gauge 
invariant chiral Lagrangian that includes the electroweak 
couplings of the top quark up to 
dimension four \cite{malkawi}\footnote{
In this study we do not include possible 
flavor changing neutral current couplings, e.g. $t$-$c$-$Z$.}.

\begin{eqnarray}
{\cal L}^{(4)}&=&i\overline{t}\gamma^{\mu}\left ( \partial_{\mu}
 +i\frac{2s_w^2}{3}{\cal A}_{\mu}\right) t
+i\overline{b}\gamma^{\mu}\left (\partial_{\mu}-i\frac{s_w^2}{3}
{\cal A}_{\mu}\right ) b\nonumber \\
&&-\left (\frac{1}{2}-\frac{2s_w^2}{3}+
\frac{1}{2}\kappa_{L}^{\rm {NC}}\right)
\overline{t_{L}}\gamma^{\mu} t_{L}{{\cal Z}_{\mu}}
 -\left ( \frac{-2s_w^2}{3}+
\frac{1}{2}\kappa_{R}^{\rm {NC}}\right ) \ov {{t}_{R}}
\gamma^{\mu} t_{R}{{\cal Z}_{\mu}} \nonumber \\
&&-\left( \frac{-1}{2}+\frac{s_w^2}{3}\right)
\overline{b_{L}}\gamma^{\mu} b_{L}{{\cal Z}_{\mu}}
-\frac{s_w^2}{3}\overline{b_{R}}\gamma^{\mu} b_{R}
{{\cal Z}_{\mu}}\nonumber \\
&&-\frac{1}{\sqrt{2}}
\left (1+\kappa_{L}^{\rm {CC}}\right ) \ov {{t}_{L}}
\gamma^{\mu} b_{L}
{{\cal W}_{\mu}^+}-\frac{1}{\sqrt{2}}
\left( 1+{\kappa_{L}^{\rm {CC}}}^{\dagger}\right)
{{b}_{L}}\gamma^{\mu}t_{L}{{\cal W}_{\mu}^-} \nonumber \\
&&-\frac{1}{\sqrt{2}}\kappa_{R}^{\rm {CC}}
\overline{{t}_{R}}\gamma^{\mu} b_{R}
{{\cal W}_{\mu}^+}-\frac{1}{\sqrt{2}}{\kappa_{R}^{\rm {CC}}}^{\dagger}
 \overline{{b}_{R}}\gamma^{\mu} t_{R}{{\cal W}_{\mu}^-} \nonumber \\
&&-m_t \overline{t} t 
 \label{eq2} \,\, .
\end{eqnarray}
In the above equation $\kappa_{L}^{\rm {NC}}$, $\kappa_{R}^{\rm {NC}}$,
$\kappa_{L}^{\rm {CC}}$, and $\kappa_{R}^{\rm {CC}}$
parameterize possible deviations from 
the SM predictions~\cite{malkawi,ehab}.
In general, the charged current coefficients can be complex with the
imaginary part introducing a CP odd interaction, and the neutral current
coefficients are real so that the effective Lagrangian is hermitian.

\section{ Constraints on dimension four anomalous couplings from
the low energy data}
\indent\indent

In the chiral Lagrangian ${\cal{L}}^{(4)} $ given in Eq.~(\ref{eq2}),
there are two complex parameters ($\klc$ and $\krc$) and two
real ($\kln$ and $\krn$) all independent from each other,
which need to be constrained using precision data. 
Naturally, these parameters are not expected to be large, we assume
that their absolute values are at 
most of order one.   The imaginary parts
of the charged current couplings, which give rise to CP violation at
this level, do not contribute to the LEP observables of interest at the
one-loop level.  Hence, we will ignore imaginary parts of the
{\mbox {$\kappa$'s}}.  Also, at this level any contributions from the
right-handed charged current coupling $\krc$ are proportional to the
bottom quark's mass $m_b$ (which is much smaller than $m_t$), and are
negligible compared to the contributions from the other three couplings.
Therefore, we can only obtain bounds for $\kln$, $\krn$ and $\klc$
from LEP data at the one loop level.
However, the coupling $\krc$ can be studied independently by
using the CLEO measurement of  $b\rightarrow s\gamma$. 
For this process $\krc$ becomes the significant anomalous
coupling.  In Ref.~\cite{fuj}  the contribution of this parameter to
the branching ratio of $b\ra s\gamma$ was calculated.  From the
result given there, and the recent CLEO measurement 
$1\times 10^{-4}<Br(b\ra s\gamma)<4.2\times 10^{-4}$
\cite{recent}, we can obtain the following bounds for $\krc$ at
the $95\%$ confidence level (C.L.):
\begin{eqnarray}
-0.037 \; < \; \krc \; < 0.0015\; .\label{krcbound}
\end{eqnarray}
With these observations we will study how $\kln$, $\krn$ and
$\klc$ can be constrained by LEP data.

All contributions to low energy observables, under a few general
assumptions, can be parameterized by 4-independent parameters:
$\epsilon_1$, $\epsilon_2$, $\epsilon_3$, and $\epsilon_b$
\cite{bar,bar1,bar2}. 
In our case, the general assumptions are satisfied, namely 
all the contributions of the non-standard couplings $\kappa$'s to low 
energy observables are contained in the oblique corrections, \ie the
vacuum polarization functions of the gauge bosons, and the non-oblique
corrections to the vertex \bbz. Therefore, 
it is enough to calculate the new 
physics contribution to the $\epsilon$ 
parameters in order to isolate all effects to low energy observables.

The experimental values of the $\epsilon$ parameters are
derived from four basic \mbox{observables},
$\Gamma_{\ell}$ (the partial width of $Z$ to a charged lepton pair),
$A_{FB}^{\ell}$ (the forward--backward asymmetry at the $Z$ peak for
the charged lepton $\ell$), $M_{W}/M_{Z}$,
and $\Gamma_{b}$\,(the partial width
of $Z$ to a $b\overline{b}$ pair) \cite{alta}.

To constrain these nonstandard couplings ({\mbox {$\kappa$'s}})
one needs to have the theoretical predictions for 
the {\mbox {$\epsilon$'s}}.
The SM contribution to the $\epsilon$'s have been calculated in, for
example Ref. \cite{bar3}.   Naturally,  since 
we are considering the case
of a spontaneous symmetry breaking scenario in which there is no 
Higgs boson, we have to subtract the Higgs boson contribution
from these SM calculations.

Since the top quark will only contribute to the vacuum polarization
functions and the vertex $\bbz$, we only need to consider:
\beq
\epsilon_1 = e_1-e_5 \,,
\enq
\beq
\epsilon_2 = e_2 - c^2_w e_5\, ,
\enq
\beq
\epsilon_3 = e_3 - c^2_w e_5 \, ,
\enq
\beq
\epsilon_b = e_b \, ,
\enq
where $e_1$, $e_2$, $e_3$, $e_5$, and $e_b$ are defined as:
\beq
e_1 = \frac{A^{ZZ}(0)}{\MZs} - \frac{A^{WW}(0)}{M_W^2}\, ,
\enq
\beq
e_2 = F^{WW}(M_W^2) - F^{33}(M_Z^2)\, ,
\enq
\beq
e_3 = \frac{c_w}{s_w}F^{30}(M_Z^2)\, ,
\enq
\beq
e_5 = {M_Z^2}\frac{dF^{ZZ}}{dq^2}(M_Z^2)\, .
\enq
The vacuum polarization functions of the gauge bosons are 
written in the following form
\beq
\Pi_{\mu\nu}^{ij}(q^2)=-ig_{\mu\nu}\left( A^{ij}(q^2) +
q^2 F^{ij}(q^2) \right)
+\,\, q_\mu q_\nu\,\, {\rm {terms}}\, ,
\label{selfenergy_V}
\enq
where $i,j=W$, $Z$, $\gamma$(photon). Alternatively, instead of using 
$Z$ and $\gamma$ one can use $i,j=3,0$ for $W^3$ and $B$, respectively.
The relation between the two cases is as follows
\beq
A^{33}= c^2_w A^{ZZ} + 
2 s_w c_w A^{\gamma Z} +s^2_w A^{\gamma\gamma}\, ,
\enq
\beq
A^{30}=-c_w s_w A^{ZZ} +  (c^2_w-s^2_w) A^{\gamma Z} + c_w s_w
        A^{\gamma\gamma} \, ,
\enq
\beq
A^{00}=s^2_w A^{ZZ} - 2 s_w c_w A^{\gamma Z} +c^2_w A^{\gamma\gamma}\, ,
\enq
and similarly for $F^{ij}$.

The quantity $e_b$ is defined through the proper vertex correction 
\beq
{V_{\mu}}\left ( Z \rightarrow b\bar{b}\right ) = -\frac{g}{2c_w}e_b
\gamma_{\mu}\frac{1-\gamma_5}{2}\, .
\enq

\subsection{Radiative Corrections in Effective Lagrangians}
\indent\indent

Before presenting our results for the contributions of the non-standard
couplings to the LEP data, we will discuss a key aspect
of effective theories in general.

Non--renormalizability of the effective Lagrangian presents
a major issue of how to consistently
handle both the divergent and the finite pieces in 
loop calculations \cite{burg,mart}. Such a problem arises because one 
does not know the 
underlying theory; hence, no matching can be performed 
to extract the correct scheme to be used in the effective Lagrangian 
\cite{geor}.   One approach is to associate the divergent piece in
loop calculations with a physical cutoff scale $\Lambda$, the upper 
scale at which the effective Lagrangian is 
valid \cite{pecc}. In the chiral Lagrangian approach this cutoff 
$\Lambda$ is taken to be $4\pi v \sim 3$\,TeV \cite{geor}\footnote{
The scale $4\pi v \sim 3$\,TeV is only meant to indicate the typical
cutoff scale. It is equally probable to have, say, $\Lambda=1$ TeV.}.
For the finite piece no completely satisfactory approach is available
\cite{burg}.   We assume that there exists an underlying renormalizable
"full" theory that is valid at all scales
(or at least at scales much higher than $\Lambda$).
In this case,  $\Lambda$ serves as an infrared
cutoff scale under which the heavy degrees of freedom can be
integrated out to give rise to the effective operators in the chiral
Lagrangian. Due to the renormalizability of the full theory, and from
the renormalization group invariance, one concludes that the same cutoff
$\Lambda$ should also serve as the associated ultraviolet cutoff of the
effective Lagrangian in the calculation of the Wilson coefficients. 
Hence, in the dimensional regularization scheme, the ultraviolet
divergent piece $1/\epsilon$ is replaced by $\ln(\Lambda^2/{\mu^2})$,
where $\epsilon=(4-n)/2$ and $n$ is the space-time dimension.
Furthermore, the renormalization scale $\mu$ is set to be $m_t$, 
the heaviest mass scale in the low energy effective Lagrangian.
To study the effects to low energy observables due to a heavy top
quark, in addition to the SM contributions, 
we shall only include those non-standard contributions 
(from the $\kappa$'s) of the
order
\beq
{{m^2_t}\over {16 \pi^2 v^2}} \; \ln {{\Lambda^2}\over {m^2_t}} \, .
\nonumber
\enq

\subsection{Contributions on the low energy observables}
\indent\indent

To perform calculations using the chiral Lagrangian, 
one should arrange 
the contributions in powers of ${1\over {4\pi v}}$ and include all 
diagrams up to the desired power. In
a general $R_{\xi}$ gauge ($\Sigma \neq 1$), the couplings of 
the Goldstone bosons to the fermions should also be included
in Feynman diagram calculations.
These couplings can be easily found 
by expanding the operators in ${\cal{L}}^{(4)} $.
 
The relevant Feynman diagrams are shown in Figure~\ref{fdiags}.
Calculations can be done for a general $R_\xi$ gauge.
As it turns out, the dependence on $m_t$ for $\epsilon_1$ (which is
the deviation from $\rho=1$) and for $\epsilon_b$
is quadratic, whereas for $\epsilon_2$ and $\epsilon_3$ is only
logarithmic.  Hence, in our effective 
model, the significant constraints on
the parameters $\kln$, $\krn$, and $\klc$ 
are only coming from $\epsilon_1$
and $\epsilon_b$. 

\begin{figure}
\centerline{\hbox{
\psfig{figure=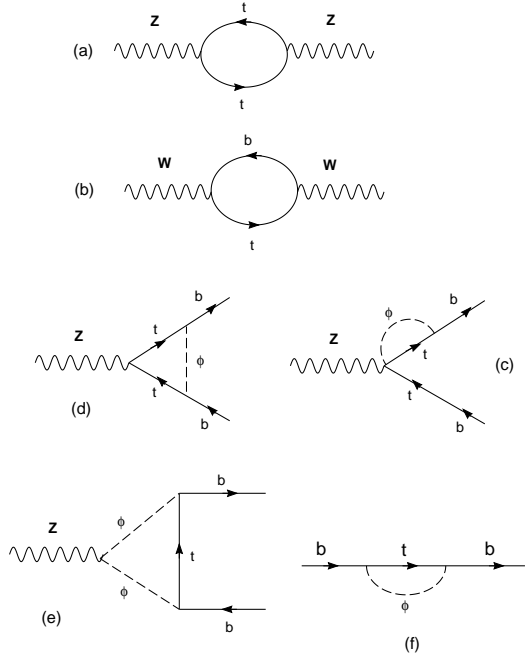,height=3.5in}}}
\caption{The relevant Feynman diagrams, for the nonstandard top quark 
couplings case and in
the 't Hooft--Feynman gauge, which contribute to the order
${\cal O}({m_t^2}\ln {\Lambda}^2)$.}
\label{fdiags}
\end{figure}

The leading contributions (of  order  $m_t^2\ln {\Lambda}^{2}$) are the
following: 

\begin{itemize}
\item
For the vacuum polarization function of the $Z$ boson
( Figure ~\ref{fdiags}(a) ),
\beq
A^{ZZ}(0)=\frac{\MZs}{4\pi^2}\frac{3m_t^2}{v^2}
\left (-\kln +\krn \right)
\frac{1}{\epsilon}
\enq
\item
For the vacuum polarization function of the $W$ boson
( Figure ~\ref{fdiags}(b) ),
\beq
A^{WW}(0)=\frac{\MWs}{4\pi^2}\frac{3m_t^2}{v^2}\left (-\klc \right)
\frac{1}{\epsilon}
\enq
\item 
The vertex corrections are depicted in Figures ~\ref{fdiags}(c),
~\ref{fdiags}(d) and ~\ref{fdiags}(e),
\beq
(c) \ra \frac{ig}{4 c_w}\frac{m_t^2}{4\pi^2 v^2}\left (-2\klc \right)
\gamma_\mu \left( 1-\gamma_5 \right)\frac{1}{\epsilon}
\enq

\item
\beq
(d) \ra \frac{ig}{4 c_w}\frac{m_t^2}{4\pi^2 v^2}\left ( -2 c^2_w \klc 
+ \frac{1}{4}\krn - \kln\right)
\gamma_\mu \left( 1-\gamma_5 \right)\frac{1}{\epsilon}
\enq 

\item
\beq
(e) \ra \frac{ig}{4 c_w}\frac{m_t^2}{4\pi^2 v^2}
\left ( -c^2_w +\frac{1}{2}\right)\klc
\gamma_\mu \left( 1-\gamma_5 \right) \frac{1}{\epsilon}
\enq 
\item
Finally, the $b$-quark self 
energy ( Figure ~\ref{fdiags}(f) ) contribution is
\beq
-\frac{3m_t^2}{16\pi^2 v^2} \gamma_\mu p^\mu 
\left (\klc \right)
\left( 1-\gamma_5 \right) \frac{1}{\epsilon}
\enq
\end{itemize}

Therefore, the net non-standard 
contributions to the $\epsilon$ parameters
are
\beq
\delta\epsilon_1=\frac{G_F}{2\sqrt{2}{\pi}^2}3{m_t^2}
 (-\kln+\krn+\klc)\ln{\frac{{\Lambda}^2}{m_t^2}}\,\, , \label{cal1}
\enq
\beq
\delta\epsilon_b=\frac{G_F}{2\sqrt{2}{\pi}^2}{m_t^2}
\left ( -\frac{1}{4}\krn+\kln \right ) \ln{\frac{{\Lambda}^2}{m_t^2}}
\,\, , \label{cal2}
\enq

It is interesting to note that $\klc$ does not contribute to 
$\epsilon_b$ up to this order ($m_t^2\ln {\Lambda}^{2}$)
which can be understood from Eq.~(\ref{eq2}). If $\klc=-1$
then there is no net \tbw coupling in the chiral Lagrangian 
after including both the standard and nonstandard contributions. Hence, 
no dependence on the top quark mass can be generated, {\it i.e.},
the nonstandard $\klc$ contribution to $\epsilon_b$ 
must cancel the SM contribution when 
$\klc=-1$, independently of the couplings of 
the neutral current.
From this observation and because the SM contribution to 
$\epsilon_b$ is finite, we conclude that $\klc$ cannot contribute
to $\epsilon_b$ at the order of interest.

Given the above results we can then compare the experimental values
of the $\epsilon$'s with the theoretical 
predictions \cite{alta3,ehabthesis}.  
For this comparison, we have included all $\epsilon^{{\rm SM}}$ and
$\delta \epsilon$, where $\epsilon^{{\rm SM}}$ is the SM
prediction\footnote{$\epsilon^{{\rm SM}}$ includes also contributions
from vertex and box diagrams.} after subtracting the contributions
due to a light Higgs boson (with mass $\sim M_Z$).  
The other term $\delta \epsilon$, is the
contribution from the dimension 4 anomalous couplings given in
Eqs.~\ref{cal1} and \ref{cal2}.
As we can see, precision data allows for all three non-standard
couplings to be different from zero.  There is a three
dimensional boundary region for these $\kappa$'s,
which we can visualize through the three
projections; on the  $\klc = 0$,  $\krn = 0$ and $\kln = 0$ planes,
presented in Figures ~\ref{fklcc}, ~\ref{fkrnc} and ~\ref{fklnc}
respectively.   As we can
see from the three projections, the only coefficient that is
constrained is $\kln$ which can only vary between $-0.5$ and
$0.5$ roughly speaking.  The other two can vary through the
whole range ($-1.0$ to $1.0$) although in a correlated manner; from
Figure~\ref{fklnc}  we can say 
that LEP data imposes $\klc \, \sim \, - \krn$
if $\kln$ is close to zero.  This 
conclusion holds for $m_t$ ranging from
$160$ GeV to $180$ GeV.

In Ref. \cite{pczh} a similar analysis was done, but in there  
the anomalous charged current contribution $\klc$ was not included,
and only the non-standard $\ttz$ couplings were considered. 
The allowed region they found in Ref. \cite{pczh} simply corresponds,
in our analysis, to the region defined by 
the intersection of the allowed
volume ( Eq.~(\ref{cal1}) ) and the plane $\klc = 0$, which 
gives a small area confined in the vicinity of the line 
$\kln = \krn$ (since $\epsilon_1 \propto 
\left (\krn - \kln \right )$ ). 
If we add the restriction given 
by $\epsilon_b$ ( Eq.~(\ref{cal2}) ),
we will realize that this sets the length of the allowed narrow area
(c.f. Figure~\ref{fcomp}).

It is also interesting to 
consider a special case in which the underlying
theory respects the global $SU(2)_L \times SU(2)_R$ custodial
symmetry that is then broken 
in such a way as to account for a negligible
deviation of the $\bbz$ vertex from 
its standard form.  This scenario will
relate the non-standard terms in our effective Lagrangian
${\cal{L}}^{(4)} $ ( Eq.~(\ref{eq2}) ).

\begin{figure}
\centerline{\hbox{
\psfig{figure=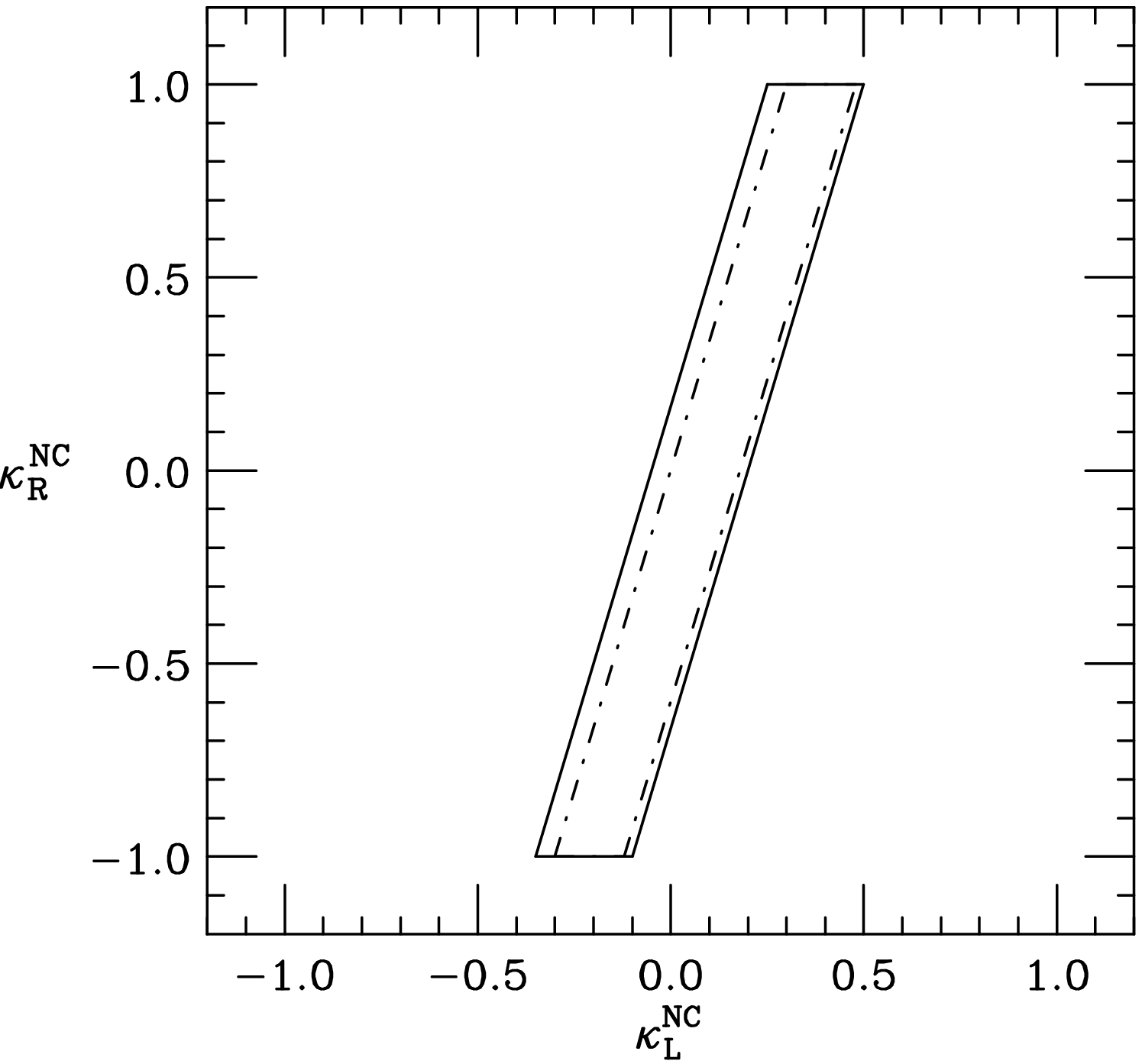,height=3.5in}}}
\caption{
A two--dimensional projection in the plane of
$\kln$ and $\krn$, for $m_t=160$ GeV (solid contour) and 180 GeV 
(dashed contour).}
\label{fklcc}
\end{figure}

\begin{figure}
\centerline{\hbox{
\psfig{figure=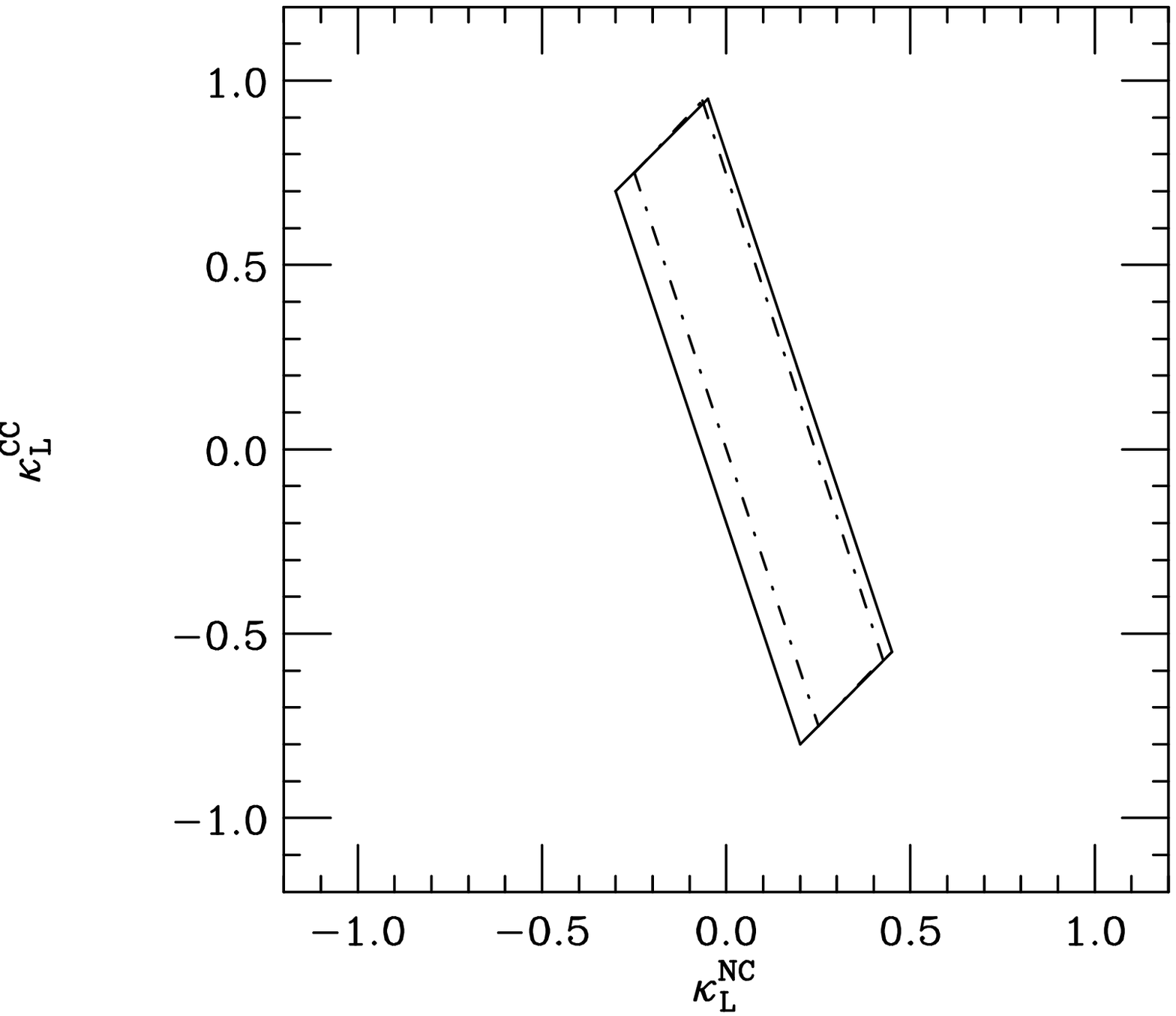,height=3.5in}}}
\caption{
A two--dimensional projection in the plane of
$\kln$ and $\klc$, for $m_t=160$ GeV (solid contour) and 180 GeV 
(dashed contour).}
\label{fkrnc}
\end{figure}

\begin{figure}
\centerline{\hbox{
\psfig{figure=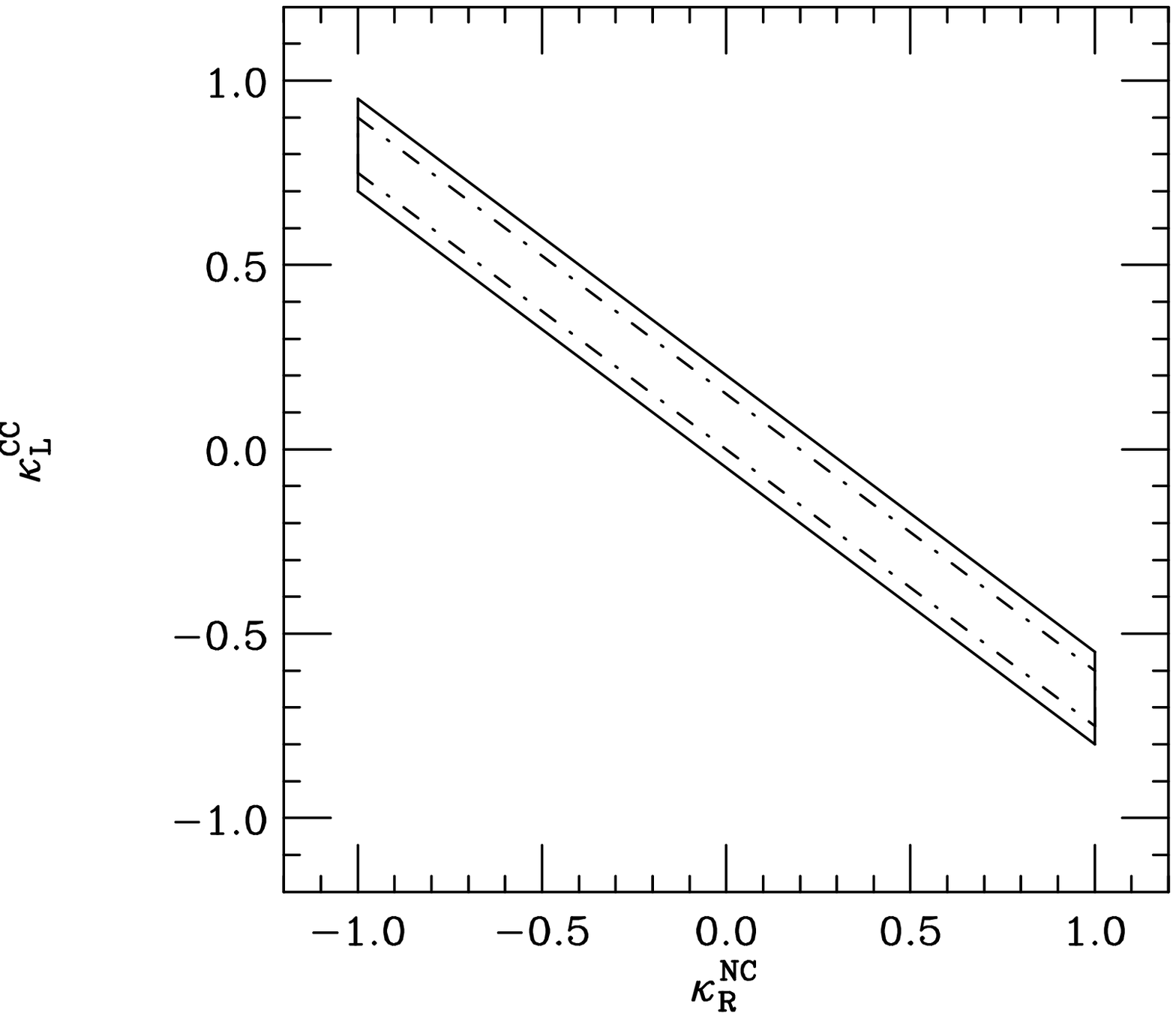,height=3.5in}}}
\caption{
A two--dimensional projection in the plane of
$\krn$ and $\klc$, for $m_t=160$ GeV (solid contour) and 180 GeV 
(dashed contour).}
\label{fklnc}
\end{figure}

\subsection{Underlying custodial symmetry case}
\indent\indent

The SM has an additional (accidental) symmetry called the custodial
symmetry which is responsible for 
the tree-level relation \cite{ehab,dono}
\beq 
\rho=\frac{M_W^2}{M_Z^2\, c^2_w}=1\,\, , \label{rho}
\enq

This symmetry is slightly broken 
at the quantum level by the $SU(2)$ doublet 
fermion mass splitting and the hypercharge
coupling $g^{\prime}$ \cite{velt}. Writing $\rho=1 +\delta \rho$, 
$\delta\rho$ would vanish to all orders if this symmetry were exact.
Low energy data indicate that
$\delta\rho$ is very close to zero,  within about 0.1\% 
accuracy \cite{langacker}.  

In the chiral Lagrangian this assumption of a custodial
symmetry sets  $v_3=v_2=v_1$ ( see Eq.~(\ref{sigfield}) ),
and forces the couplings of the top quark to the
gauge bosons ${W_{\mu}^a}$ 
to be equal after turning off the hypercharge.

Let us consider the case of 
an underlying global $SU(2)_L \times SU(2)_R$
symmetry that is broken 
in such a way as to account for a negligible deviation
of the $\bbz$ vertex from its standard 
form.   Since the top quark acquires a
mass much heavier than the other quarks' masses, we expect the new
physics effects associated with the electroweak symmetry breaking (EWSB)
sector to be substantially greater for the 
couplings (to the gauge bosons)
of this quark than for the couplings of 
all the others, including the bottom quark.
Therefore, it is natural to think 
of the initial presence of an underlying theory
that respects the custodial symmetry, and then to think of the EWSB
mechanism introducing an effective interaction that will explicitly
break this symmetry in such a way as 
to favor the deviation of the couplings
of the top quark more than the deviation 
of the other light quarks' couplings.

In the context of the chiral Lagrangian, let us think of the effective
Lagrangian  ${\cal{L}}^{(4)} $  ( Eq.~(\ref{eq2}) 
) originating from two parts:
one that reflects the underlying theory that 
respects the custodial symmetry
( denoted by ${\cal{L}}^{(custodial)} $ ), 
and another part that explicitly
breaks this symmetry but that keeps the coupling $\bbz$ essentially
unmodified ( denoted by ${\cal{L}}^{(EWSB)} $ ).  

Let us find the most general form 
for ${\cal{L}}^{(custodial)} $.   Notice that
if we set $s_w = 0$ (turn off the 
hypercharge), then the standard $SU(2)_L$
invariant term
\beq
\overline{F_L} \gamma^{\mu}\left ( \; i\partial_{\mu}
-  \frac{1}{2} \left(
\begin{array}{r}
{\cal W}_{\mu}^3 \;\;\;\; \sqrt{2}{\cal W}_{\mu}^{+} \\
 \sqrt{2} {\cal W}_{\mu}^{-} \;\;\;\; -{\cal W}_{\mu}^3
\end{array}
\right)  \; \right) F_L \; , \nonumber
\enq
with the left handed doublets
\beq
F_{L}={f_1\choose {f_2}}_{L} \; 
\enq
defined in Eq.~(\ref{psi}), respects the global
$SU(2)_L \times SU(2)_R$ symmetry\footnote{To verify this, we
just need to use the transformation rules
$\Sigma \ra \Sigma^{'} \, =\, L \Sigma R^{\dagger}$ and
$F_L \ra F_L^{'} \, =\,R F_L$ with $R$ and $L$ members
of global $SU(2)_L$ and $SU(2)_R$ respectively, as well as the identity
$i \Sigma^{\dagger} D_\mu \Sigma \, 
=\, -{\cal W}_{\mu}^{a} \frac{\tau^a}{2}$
.}
and is the only structure that 
does so (the derivative term is trivial). 
Therefore the only way in which ${\cal{L}}^{(custodial)} $ can contain
non-standard couplings is through a term proportional to the same
$W^a \tau^a$ structure:  
\beq
{\cal L}^{(custodial)}= \overline{F_L} \gamma^{\mu}
\left ( \; i\partial_{\mu}
\; - \frac{1}{2} {\cal W}_{\mu}^{a} \tau^a \; \right) 
F_L \; + \; \kappa_1
\overline{F_L} \gamma^{\mu}  {\cal W}_{\mu}^{a} \tau^a 
F_L \, . \label{custi}
\enq
where $\kappa_1$ is a real number
(so that ${\cal L}^{(custodial)}$ is hermitian).

Now, for ${\cal{L}}^{(EWSB)}$ we notice that (in the context of the
non-linearly realized $SU(2)_L\times U(1)_Y$ chiral Lagrangian) one
can break the custodial symmetry by introducing interaction terms
that involve the $\tau^3$ matrix such as
\beq
{\cal{L}}^{(EWSB)}\;=\; \kappa_2  \overline{F_L} \gamma^{\mu}
{\cal W}_{\mu}^{a} \tau^a \tau^3 F_L \;+\; \kappa_2^{\dagger}  
\overline{F_L} \gamma^{\mu}  \tau^3 {\cal W}_{\mu}^{a} \tau^a F_L \,,
\label{4ewsb}
\enq
where $\kappa_2$ is in general a complex number\footnote{
Another term could be $\overline{F_L} \gamma^{\mu}\tau^3
{\cal W}_{\mu}^{a} \tau^a \tau^3 F_L$, which contains two
symmetry breaking factors $\tau^3$. We will not consider
this term in our work.}.

When we add ${\cal{L}}^{(EWSB)}$ to the non-standard part of
${\cal L}^{(custodial)}$ we will obtain the term:
\beq
\overline{F_L} \gamma^{\mu}\left (
\begin{array}{r}
(\kappa_1+\kappa_2+\kappa_2^{\dagger}){\cal W}_{\mu}^3 \;\;\;\;
(\kappa_1-\kappa_2+\kappa_2^{\dagger}) \sqrt{2}{\cal W}_{\mu}^{+} \\ 
(\kappa_1+\kappa_2-\kappa_2^{\dagger}) \sqrt{2} {\cal W}_{\mu}^{-} \;\;\;\; 
(-\kappa_1+\kappa_2+\kappa_2^{\dagger}){\cal W}_{\mu}^3
\end{array}
\right)  \;  F_L \, . \nonumber
\enq
Therefore, by requiring $\kappa_2$ to be a real number, and by setting
$\kappa_1 = 2 \kappa_2$, the above 
result indeed describes the scenario in
which an underlying custodial symmetric theory is being broken without
modifying the coupling $\bbz$ from its standard value.  By turning the
hypercharge back on we will then see that the ${\cal{L}}^{(4)}$
Lagrangian will look like:
\begin{eqnarray}
{\cal L}^{(4')}=\overline{F_L} \gamma^{\mu} \left(\,
i\partial_{\mu}-\frac{1}{2} {\cal W}_{\mu}^a \tau^a \,\right)F_L\;
+\;\overline{F_L} \gamma^{\mu} \, \kappa_1 \left(
\begin{array}{r}
2 {\cal Z}_{\mu} \;\;\;\; \sqrt{2}{\cal W}_{\mu}^{+} \\
 \sqrt{2} {\cal W}_{\mu}^{-} \;\;\;\;\;\;\;\;\;\; 0
\end{array}
\right) \; F_L \, .\label{custol4}
\end{eqnarray}

The superscipt $(4')$ in ${\cal L}^{(4')}$ 
is just to differentiate it from
the original most general 
Lagrangian ${\cal L}^{(4)}$ of Eq.~(\ref{eq2}).  
In conclusion, if we want to consider a special case in which an
underlying custodial symmetric theory is being broken by interactions
that in the end do not modify the $\bbz$ vertex from its standard
form, we have to reproduce the matrix structure presented in
Eq.~(\ref{custol4}).  This is equivalent to just requiring the
relation\footnote{A relation like this appears in the SM 
after integrating out an ultra-heavy Higgs boson \cite{ehab}.}
\beq
\kln \; =\; 2\klc \; =\; 4\kappa_1 \;\equiv \; \kappa_L
\enq
to be satisfied in the original Lagrangian ${\cal L}^{(4)}$.
Since for the right-handed couplings
only the neutral $\kappa_R^{NC}$ participates in the radiative
corrections,  we can simplify our notation and set
$\kappa_R^{NC} \equiv \kappa_R$.

From the correlations between the effective 
couplings {\mbox {($\kappa$'s)}} 
of the top quark to the gauge bosons, one can infer if 
the symmetry-breaking sector is due to a model with
an approximate custodial symmetry or not, {\it i.e.}, 
we may be able  
to probe the symmetry-breaking mechanism in the top quark system.
To illustrate this point, we can compare our results with those in 
Ref.~\cite{pczh}. Figure~\ref{fcomp} shows the most 
general allowed region for 
the couplings $\kln$ and $\krn$, {\it i.e.}, without imposing any 
"custodial symmetry" relation 
between $\kln$ and $\klc$. This region is for a top quark mass of 
170 GeV and covers 
the parameter space $-1.0 \leq \kln \, , \krn \leq 1.0 $. One finds 
\begin{eqnarray}
-0.15 &\leq& \kln \leq 0.35 \, ,\nonumber \\
-1.0 &\leq& \krn \leq 1.0 \nonumber \, .
\end{eqnarray}
We also show on Figure~\ref{fcomp} 
the allowed regions for our special case 
$\left(\klc = {1 \over 2} \kln\right)$ 
and the model in Ref.~\cite{pczh} 
$\left(\klc=0\right)$.
The two regions overlap in the vicinity of the origin (0, 0) which 
corresponds to the SM case.
Note that for $m_{t}\leq 200$\,GeV the allowed region of 
{\mbox {$\kappa$'s}} 
in all models of symmetry-breaking should overlap 
near the origin because the SM is consistent with low energy data at  
the 95\% C.L.
For $\kln \geq 0.1$, these two regions diverge and become separable.
One notices that the allowed range predicted in Ref.~\cite{pczh} lies 
along the line $\kln=\krn$ whereas in our case the slope is 
given by the line $\kln =2\krn$. 

\begin{figure}
\centerline{\hbox{
\psfig{figure=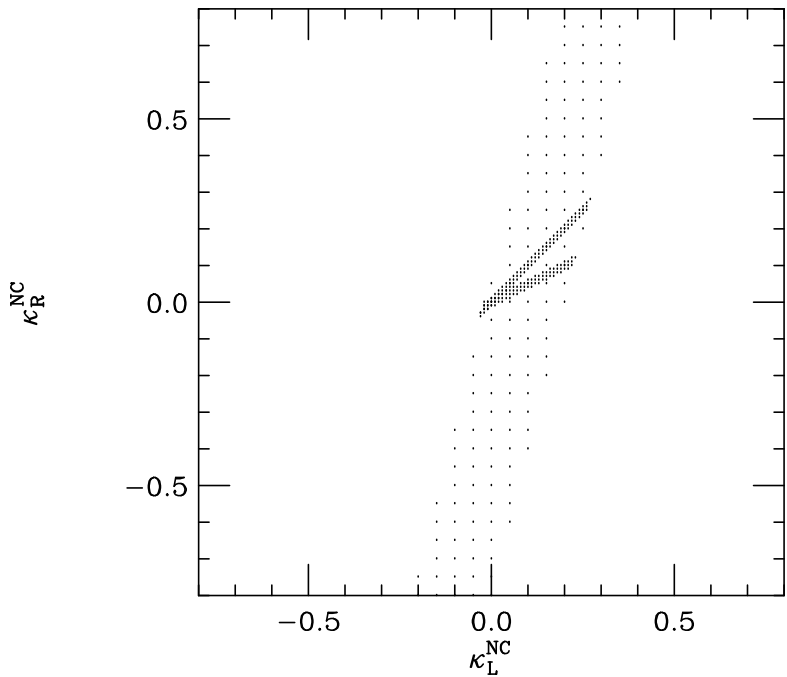,height=3.5in}}}
\caption{ A comparison between our model and the model in Ref. 
\protect{\cite{pczh}}.
The allowed regions in both models are shown on the plane of
$\kln$ and $\krn$, for $m_t=170$ GeV.}
\label{fcomp}
\end{figure}

If we imagine that any prescribed dependence between the 
couplings corresponds to a symmetry-breaking scenario, then, given the 
present status of low energy data, it is 
possible to distinguish between different 
scenarios if $\kln$, $\krn$ and 
$\klc$ are larger than 10\%. Better future measurements of 
{\mbox {$\epsilon$'s}} can further discriminate between different 
symmetry-breaking scenarios with smaller difference in the $\kappa$'s.
Next, we will discuss how the SLC precision data
can contribute to the study of the nonstandard couplings.

\subsection{At the SLC}
\indent\indent

The measurement of the left--right cross section asymmetry $A_{LR}$
in $Z$ production with a longitudinally polarized electron beam 
at the SLC provides a further test of the SM and is sensitive to new 
physics. The reported measurement of  $A_{LR}$ \cite{sld_Alr}
shows a deviation of about $2.8\sigma$ from the SM\footnote{With a
top quark mass  $m_t=175$ GeV and a Higgs mass $m_H=300$ GeV.}
prediction.  The effect of the SLC
measurement of $A_{LR}$ on possible new physics effects on the top
quark couplings depends on the way one incorporates $A_{LR}$
with LEP data.  If we include and average $A_{LR}$ with all LEP data,
the anomaly in $A_{LR}$ is almost washed away due to the large
number of LEP measurements consistent with the SM. One finds that
including the SLC measurement $A_{LR}$ with all LEP data yields a
slight decrease in the central value of  $\epsilon_1$ \cite{alta3} 
while keeping the fit on $\epsilon_b$ the same.  
As discussed in the previous section, the nonstandard coupling $\kln$ 
is mostly constrained by $\epsilon_b$. Therefore, no significant change 
in the allowed range of $\kln$ is expected. 
The effect of averaging the SLC and LEP data can be easily seen in the 
special model discussed previously ($\klc=\kln/2$).
In this case, the length of the 
allowed area is not affected since it is 
controlled by $\epsilon_b$. Since the uncertainty in 
$\epsilon_1^{\rm{exp.}}$
remains almost the same after including the $A_{LR}$ measurement, the
width of the allowed area is also 
hardly modified. The only effect will be
to shift the allowed area slightly 
downward (towards $2\kappa_R < \kappa_L$).
This conclusion is simply due to the preference for a more negative 
new physics contribution to accommodate the smaller 
value of $\epsilon_1^{\rm{exp.}}$.

We have seen that the precision LEP/SLC data can constrain the couplings
$\kln$, $\krn$ and $\klc$, without forcing them to be zero. 
For $\krc$ (the right--handed charged current) there is no constraint,
because its contribution to the relevant 
radiative corrections at LEP/SLC is
proportional to the bottom quark's mass. However, the nonstandard 
coupling $\krc$ can be studied using the $b\ra s\gamma$ measurement 
\cite{fuj} (c.f. Eq.~\ref{krcbound}). 

The important lesson from the above 
analysis is that the precision low energy
data do not exclude the possibility 
of having anomalous top quark interactions
with the gauge bosons.  Also, different models 
for the electroweak symmetry
breaking sector can induce different relations among the $\kappa$'s.
These relations can in turn be used  to discriminate between models
by comparing their predictions with experimental data.
In the next section, we examine how to improve our knowledge of
these non-standard couplings by 
direct measurements at current and future
colliders.

\section{ Direct measurement of dimension four anomalous couplings
at colliders}
\indent

In this section, we shall discuss how to measure the
dimension four anomalous couplings 
$\kappa_{L}^{\rm {NC}}$, $\kappa_{R}^{\rm {NC}}$,
$\kappa_{L}^{\rm {CC}}$, and $\kappa_{R}^{\rm {CC}}$
at hadron colliders and future electron collider.

Run I at the Fermilab Tevatron 
(a $\pbarp$ collider with $\sqrt{S}=1.8$\,TeV) is now complete,
and each experiment (CDF and~\D0 groups)
has accumulated an integrated luminosity of about 
110 $pb^{-1}$. Run II (the upgraded Tevatron with the Main Injector)
will begin in 1999, with a machine energy of 2 TeV 
and an integrated luminosity of about 2\,$\ifb$ per year.
The CERN Large Hadron Collider (LHC) is a $\pp$ collider with 
$\sqrt{S}=14$\,TeV and an integrated luminosity of about $10\sim 100$
\,$\ifb$ per year.
A future electron Linear Collider (LC) is also proposed to run 
at the top quark pair threshold (via $e^-e^+ \rightarrow t \bar t$
process) to study the detailed properties
of the top quark.

\subsection{At the Tevatron and the LHC} 
\indent\indent

At the Tevatron and the LHC, heavy top quarks are 
predominantly produced in pairs from the QCD process
$gg, q \bar q \ra t \bar t$. 
In addition, there are single-top quark events
in which only a single $t$ or $\bar t$ is produced.
A single-top quark signal can be produced from either
the $W$--gluon fusion process $qg (Wg) \ra t \bar{b}$ (or 
$ q' b \ra q t$~\cite{sally,wgtb}, 
the Drell-Yan-type $W^*$ process $q \bar q \rightarrow t \bar{b}$  
\cite{cortese,thesis,wstar}
and $W t$ production via $g b \ra W^- t$ \cite{galwt}.
The corresponding Feynman diagrams for these single-top
processes are shown in Figure~\ref{newdiag}.
The approximate cross sections (in pb) for single-top quark 
production (including both single-$t$ and 
single-$\bar t$ events) at the 
upgraded Tevatron (and the LHC) from the above four production
processes are 6.5(700), 2.0(200), 0.88(10)
and 0.2(70), respectively.

\begin{figure}
\centerline{\hbox{
\psfig{figure=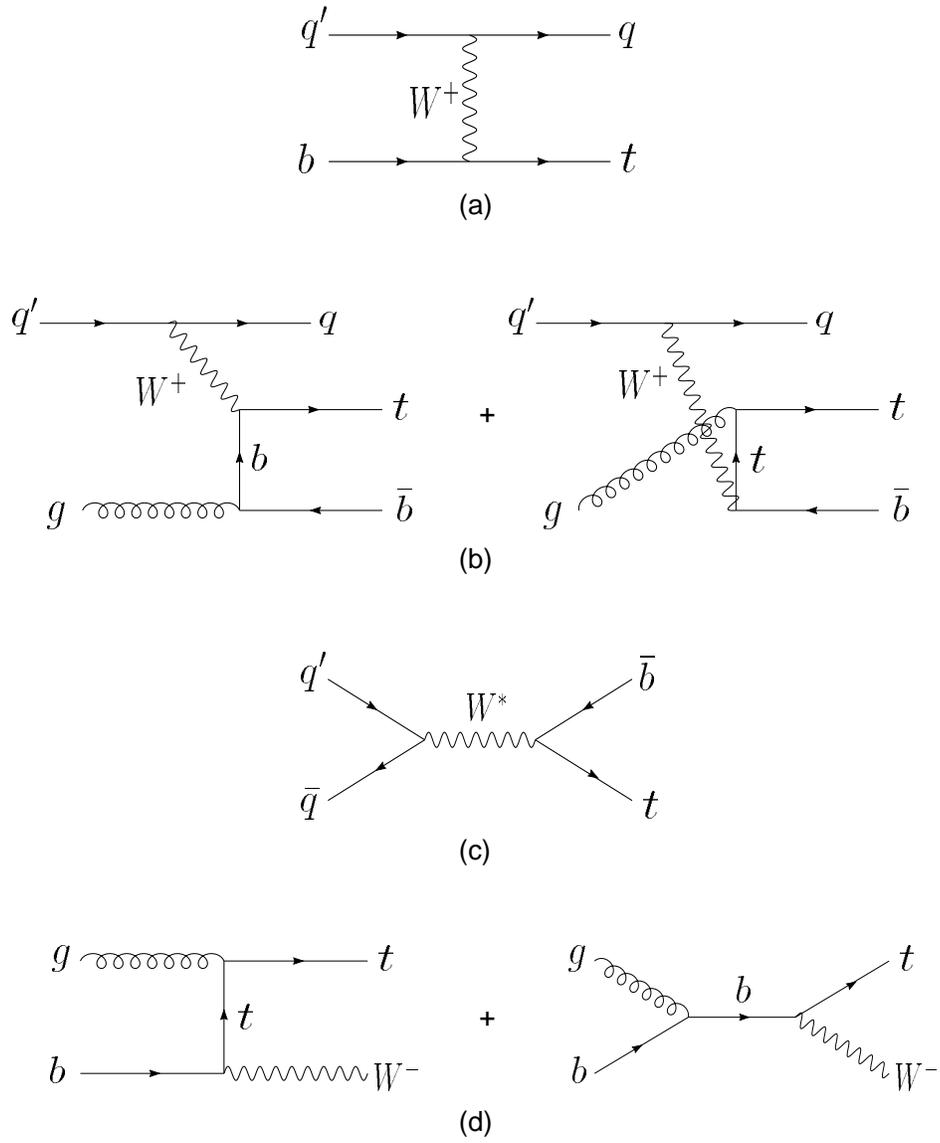,height=6in}}}
\caption{ Diagrams for various single-top quark processes.}
\label{newdiag}
\end{figure}

The relative magnitudes between the
dimension four anomalous couplings $\klc$ and $\krc$
can be measured from the decay of the 
top quark (produced from either of the above processes)
to a bottom quark and a $W$ boson.
These nonstandard couplings can be furthered
measured from counting the production 
rates of signal events with a single $t$ or $\bar t$. 
More details can be found in Refs.~\cite{wstar},~\cite{anlrev}
and ~\cite{tev2000}.

\subsubsection{From the decay of top quarks}
\indent\indent

In $t \bar t$ events, the final state with most kinematic
information is $W+4j$, where the $W$ 
is detected via its leptonic decay.  
These events are fully reconstructable. To reduce backgrounds, 
it is best to demand at least one $b$ tag.  The number of such 
events is about 500 per $\ifb$~\cite{tev2000}. 
Thus there will be on the order of 1000 tagged,
fully reconstructed top-quark events in
Run II, to be compared with 
the approximately 25 $W+4j$ single-tagged top
events in Run I.
To probe $\klc$ and $\krc$ from the decay of the 
top quark to a bottom quark and a $W$ boson, one needs to measure the 
polarization of the $W$ boson which can be determined by the 
angular distribution of the lepton 
(say, $e^+$ in the rest frame of $W^+$) in
the decay mode $t \ra b W^+ (\ra e^+ \nu)$. 
However, reconstructing the rest frame 
of the $W$-boson (in order to measure 
its polarization) could be a non-trivial matter due to the missing 
longitudinal momentum ($P_{\SST Z}$) (with a two-fold ambiguity)
of the neutrino ($\nu$) from $W$ decay.
Fortunately, as shown in Eq.~(\ref{mbe1}), 
one can determine the polarization
of the $W$-boson without reconstructing its rest frame by using the
Lorentz-invariant observable $m_{be}$, the invariant mass of 
$b$ and $e$ from $t$ decay.

The polar angle $\theta^*_{e^+}$
distribution of the $e^+$ in the rest frame
of the $W^+$ boson whose z-axis is defined to be the moving direction of
the $W^+$ boson in the rest frame 
of the top quark can be written in terms of
$m_{be}$ through the following derivation:
\begin{eqnarray}
\cos \theta^*_{e^+} &=& {{ E_e E_b - p_e \cdot p_b }\over
{|\vec{\bf p}_e| |\vec{\bf p}_b| }}       \nonumber \\
&\simeq & 1-{p_e \cdot p_b \over E_e E_b}
= 1-{2 m_{be}^2 \over m_t^2 - M_W^2}.  
\label{mbe1}
\end{eqnarray}
The energies $E_e$ and $E_b$ are evaluated in the rest frame of
the $W^+$ boson from the top quark decay and are given by
\begin{eqnarray}
E_e &=& {M_W^2+m_e^2-m_\nu^2 \over 2 M_W}, \qquad |\vec{\bf p}_e|=
\sqrt{E_e^2-m_e^2}, \nonumber \\ 
E_b &=& {m_t^2-M_W^2-m_b^2 \over 2 M_W}, \qquad |\vec{\bf p}_b|=
\sqrt{E_b^2-m_b^2}. 
\label{mbe2}
\end{eqnarray}
where we have not ignored the negligible masses $m_e$ and
$m_\nu$, of $e^+$ and $\nu_e$, for the sake of bookkeeping.

In Eq.~(\ref{mbe1}), the first line 
comes from exact definition, whereas 
the second line comes 
from applying Eq.~(\ref{mbe2}) in the limit $m_b=0$.
However, in practice two problems arise due to experimental limitations.
First, the measured momenta of the bottom quark and the charged lepton
will be smeared by detector effects, and second; it is difficult to do 
the identification of the right $b$ to reconstruct $t$.
There are three possible 
strategies to improve the efficiency of identifying
the correct $b$.  One is to 
demand a large invariant mass of the $t \bar t$
system so that $t$ is boosted and its decay products are collimated.
Namely, the right $b$ will be moving 
closer to the lepton from $t$ decay.
This can be easily enforced by demanding leptons with 
a larger transverse 
momentum.
Another strategy is to identify the 
soft (non-isolated) lepton from the $\bar b$
decay (with a branching 
ratio ${\rm Br}(\bar b \ra \mu^{+} X) \sim 10\%$).  
The third one is to statistically determine the electric charge of the 
$b$-jet (or $\bar b$-jet) to be $1/3$ (or $-1/3$) \footnote{
This is the kind of analysis performed at LEP to separate
a quark jet from a gluon jet.}.
How precisely can the invariant mass $m_{be}$ be measured is a question
yet to be answered.

For a massless $b$ (which is a good approximation for $m_b \ll m_t$),
the $W$ boson from top
quark decay can only be either 
longitudinally or left-handed polarized for 
a purely left-handed charged current 
($\krc=0$). For a purely right-handed 
charged current ($\klc=-1$) the $W$ boson 
can only be either longitudinally 
or right-handed polarized. 
(Note that the handedness of the $W$ boson is reversed for a massless
$\bar b$ from $\bar t$ decays.)
\begin{figure}
\centerline{\hbox{\psfig{figure=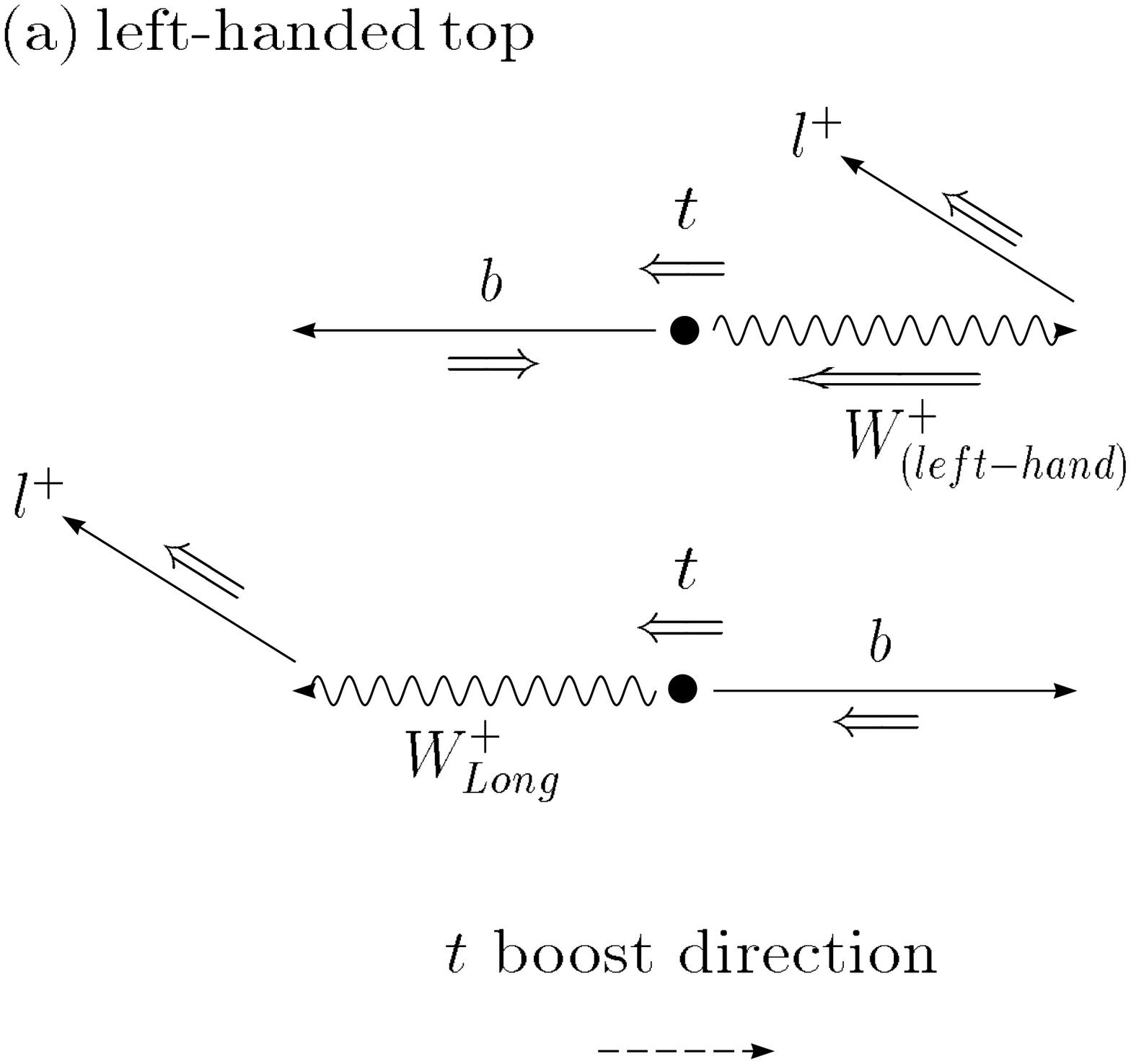,height=2.5in}
	\hspace{1.in}\psfig{figure=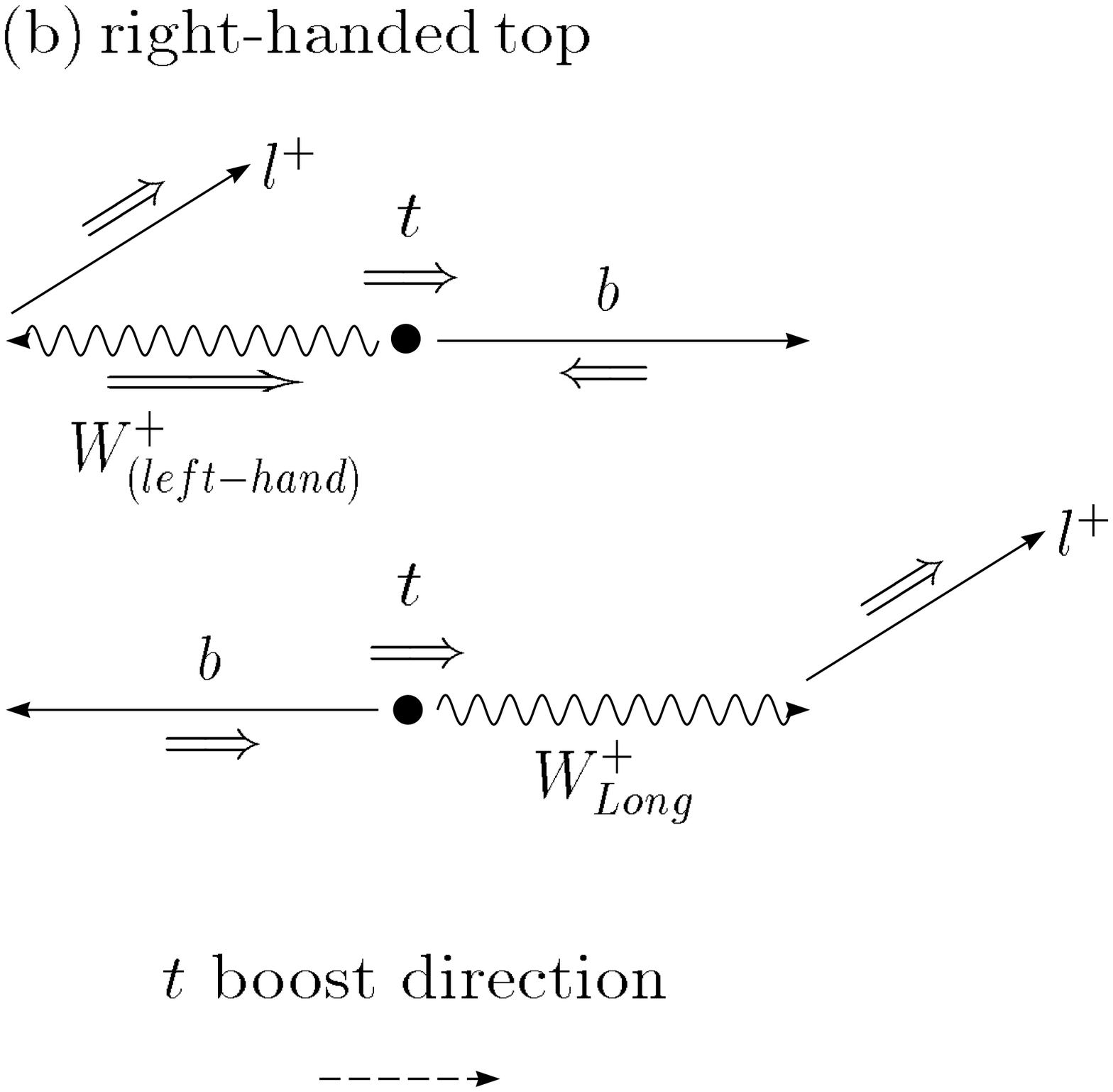,height=2.5in}}}
\caption{ For a left-handed $\tbw$ vertex. }
\label{left}
\end{figure}
\begin{figure}
\centerline{\hbox{\psfig{figure=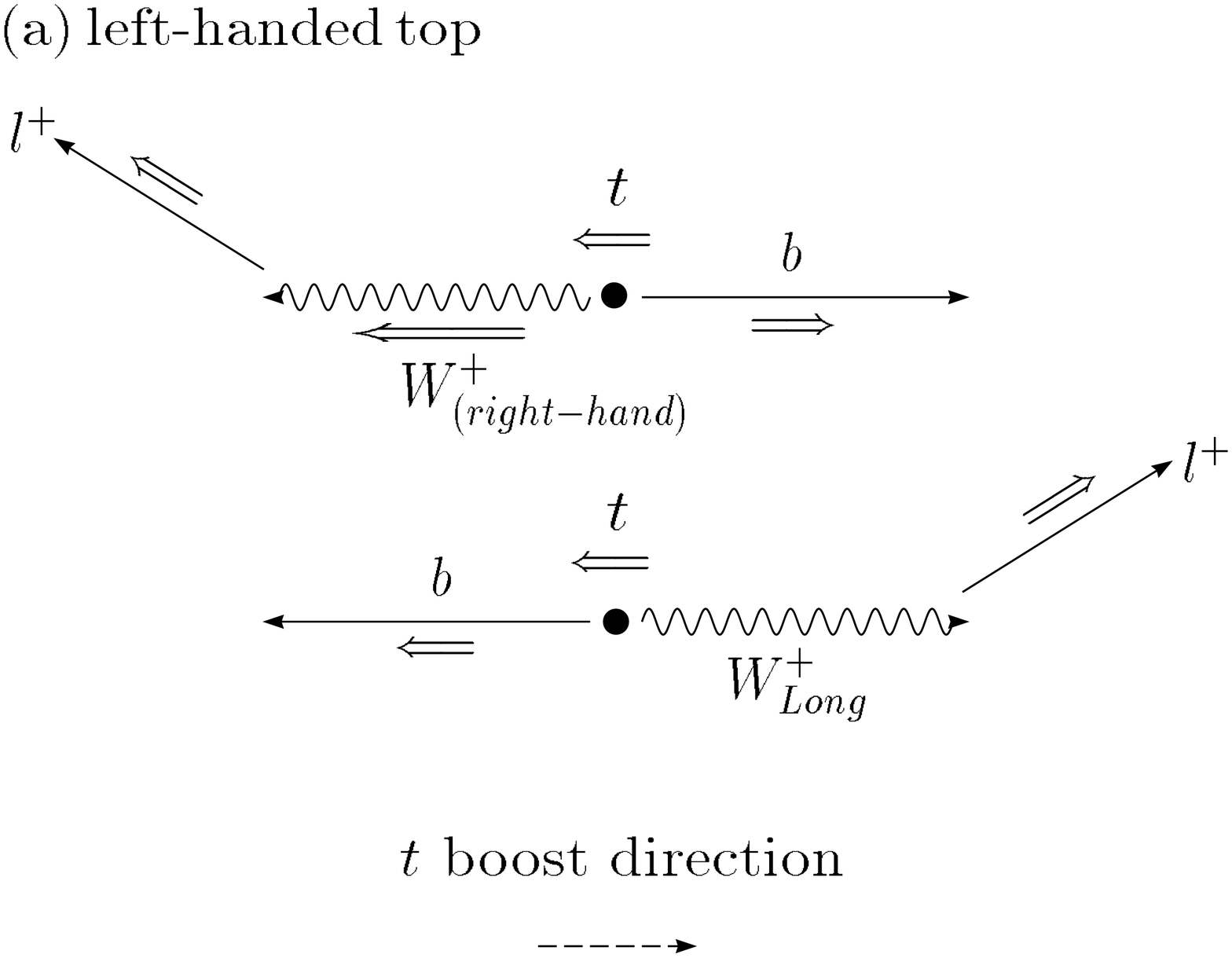,height=2.5in}
		  \psfig{figure=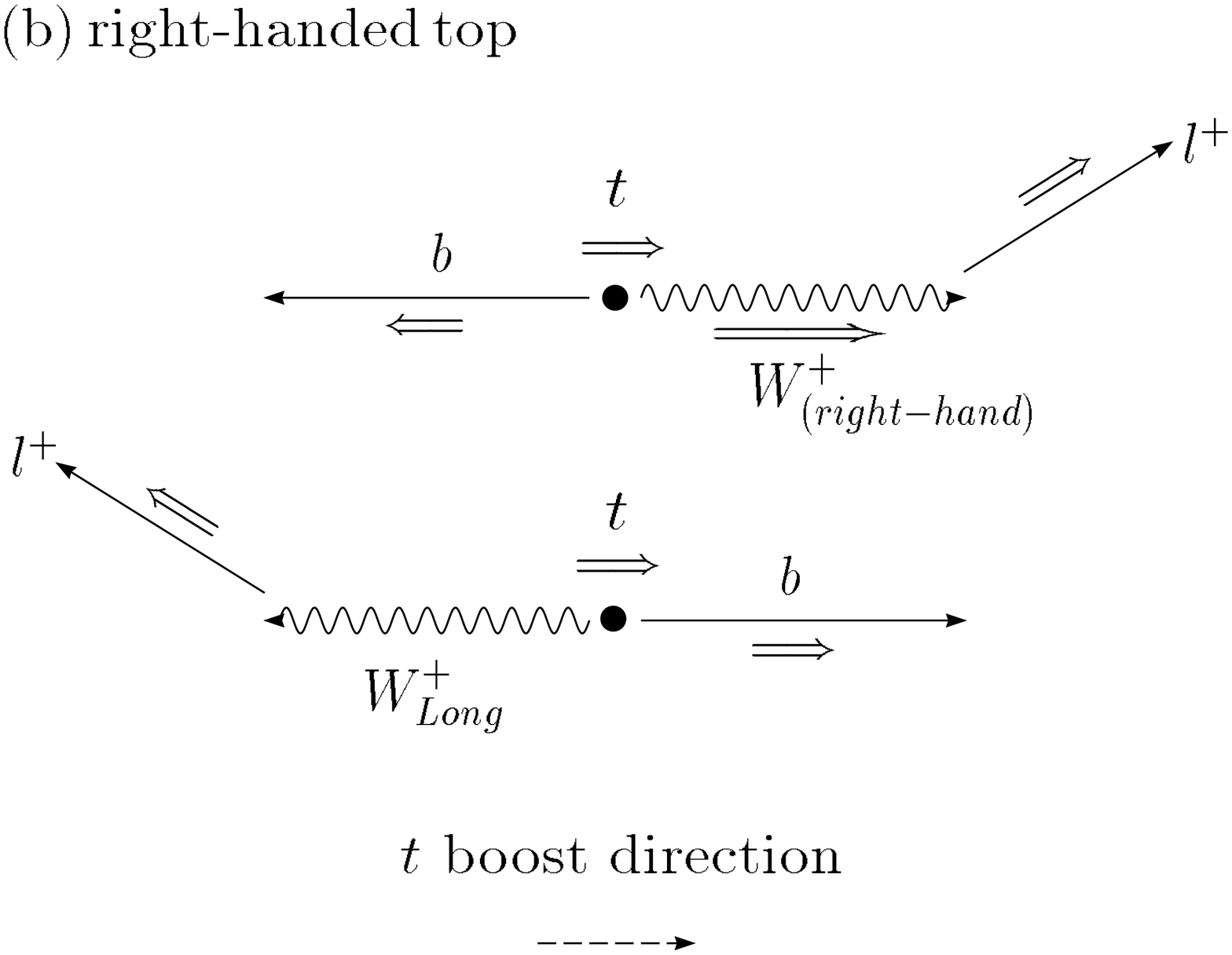,height=2.5in}}}
\caption{ For a right-handed $\tbw$ vertex. }
\label{right}
\end{figure}
This is the consequence of helicity conservation, as diagrammatically
shown in Figures~\ref{left} and ~\ref{right} for a polarized 
top quark. In these figures we show the preferred 
moving direction of the lepton coming from a polarized $W$-boson decay
 in the rest frame of a polarized top quark, 
for both cases of a left-handed
and a right-handed $\tbw$ vertex. 
As indicated in these figures, the invariant mass
$m_{b \ell}$ depends on the polarization of the $W$-boson from the decay
of a polarized top quark. 
\begin{figure}
\centerline{\hbox{\psfig{figure=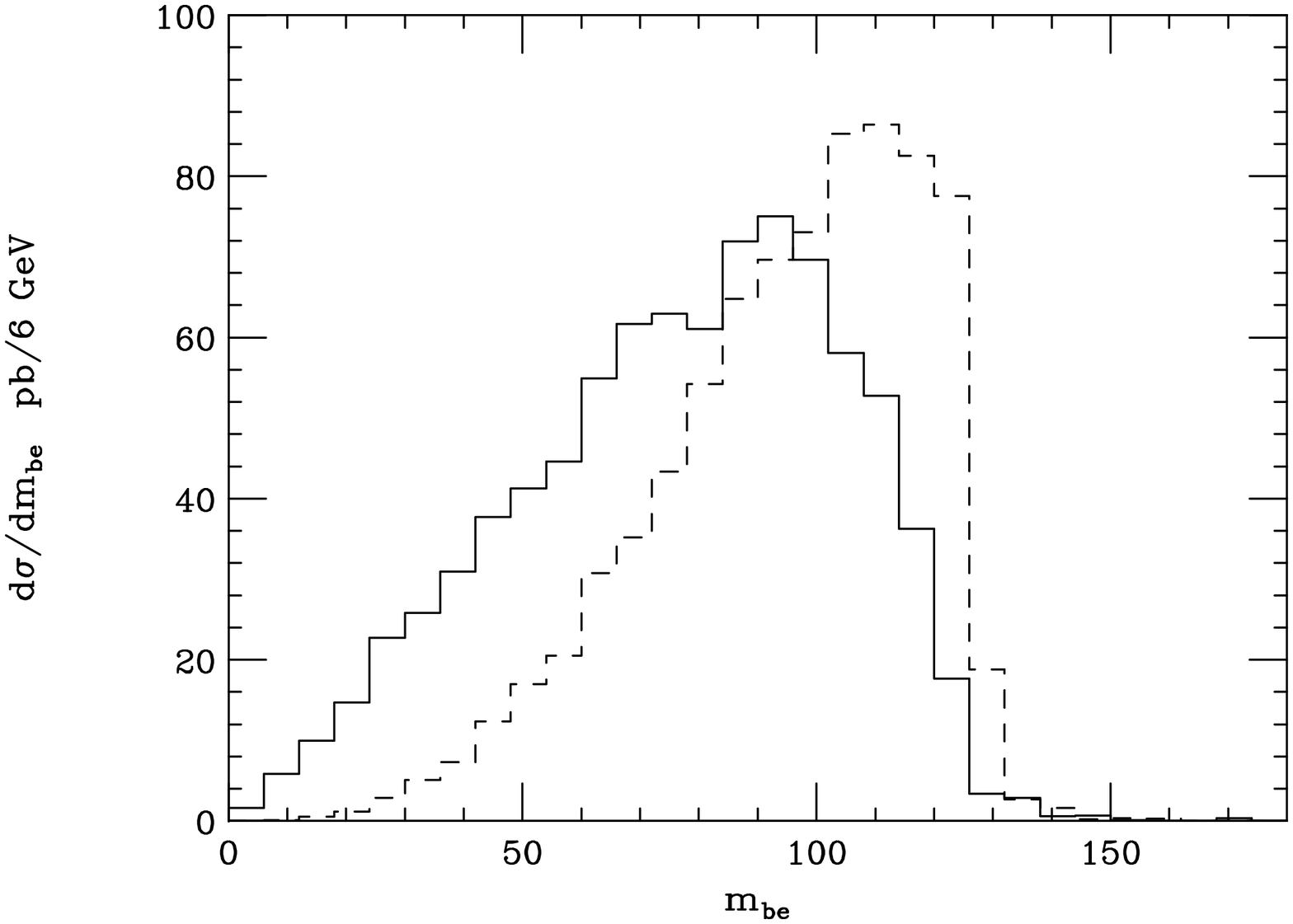,height=3.5in}}}
\caption{ $m_{b{\ell}}$ distribution for SM top quark 
(solid) and for a purely right-handed~$\tbW$ coupling (dash).}
\label{mbe}
\end{figure}
\begin{figure}
\centerline{\hbox{\psfig{figure=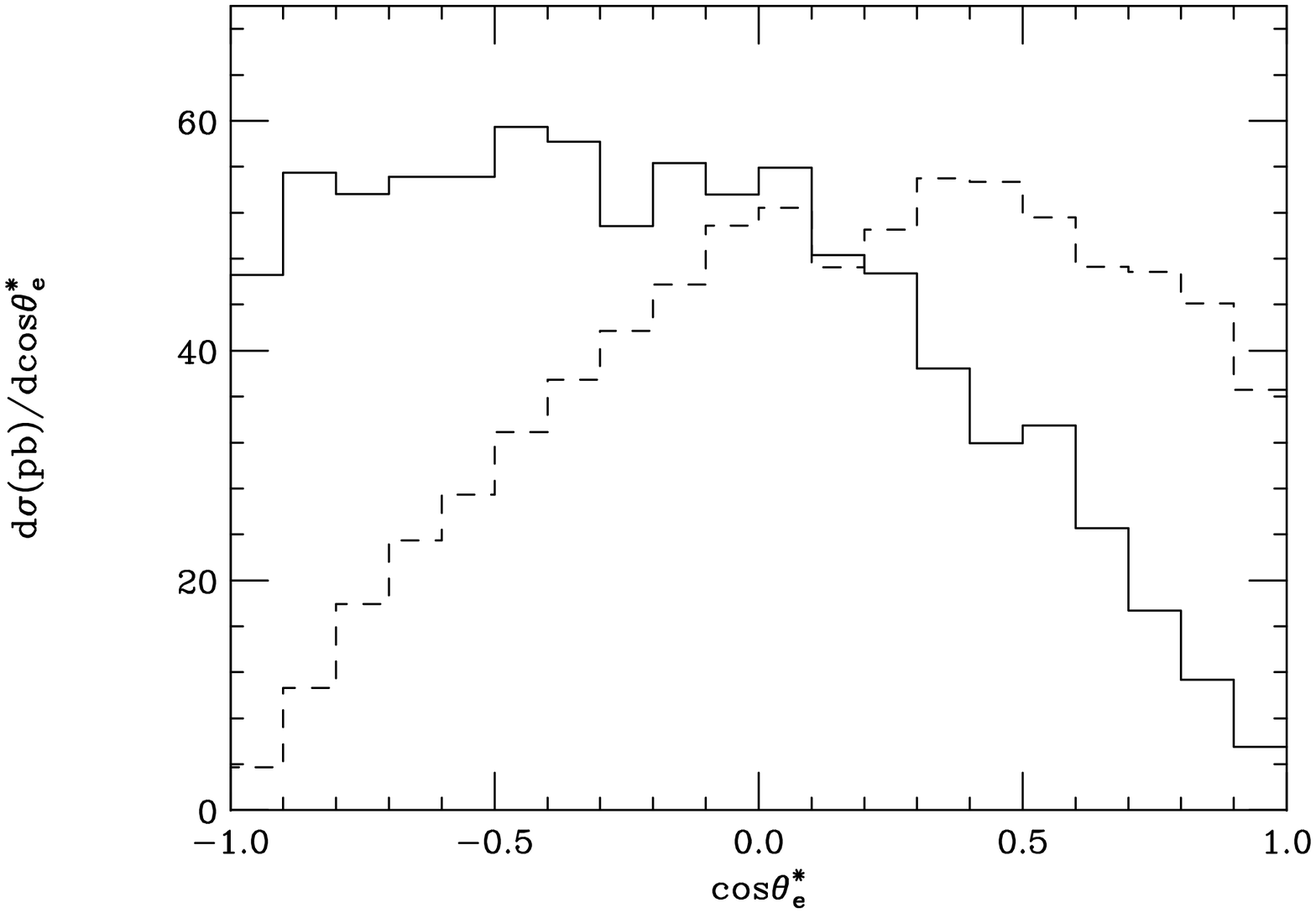,height=3.5in}}}
\caption{ $\cos \theta^*_{\ell}$ distribution for SM top quark
(solid) and for a purely right-handed~$\tbW$ coupling (dash).}
\label{thesta}
\end{figure}
Also, $m_{b \ell}$ is preferentially larger for a purely 
right-handed $\tbw$ vertex than for a purely left-handed one.
This is clearly shown in Figure~\ref{mbe}, in which 
the peak of the $m_{b{\ell}}$ 
distribution is shifted to the right and the distribution falls 
off sharply at the upper mass limit 
for a purely right-handed $\tbw$ vertex.
Their difference is shown, in terms of  $\cos \theta^*_{\ell}$, in
Figure~\ref{thesta}.
However, in both cases the fraction ($\flong$) of longitudinal $W$'s 
from top quark decay is enhanced by ${m_t}^2/{2{M_W}^2}$ as compared 
to the fraction of transversely polarized $W$'s \cite{toppol}, namely,
\beq
\flong = { {m_t^2 \over 2 M_W^2 } 
\over { 1 + {m_t^2 \over 2 M_W^2 } } } \,.
\enq
 Therefore, for a heavier 
top quark, it is more difficult to untangle the $\klc$ and $\krc$ 
contributions. 
\footnote{On the other hand, because of the very same reason,
the mass of a heavy top quark can be accurately measured from
$\flong$ irrespective of the nature of the $\tbw$ couplings
(either left-handed or right-handed).}
 
As noted above, studying the decay of the top quark can tell us
something about the relative size of the couplings $1+\klc$ and $\krc$.
To determine the values of $\klc$ and $\krc$, one has to provide 
additional information such as the decay width of $t \ra b W^+$
(which is about the total width of the top quark in the SM).\footnote{
The information 
(c.f. Eq.~\ref{krcbound}) on $\krc$ derived from the rare-decay process
$b \ra s \gamma$ could also be useful.}
If we {\it assume} the decay width of $t \ra b W^+$
is the same as the SM prediction (i.e., about 1.5\,GeV for a 175\,GeV
top quark), then the value of
$(\klc)^2 +(\krc)^2$ is fixed. Thus, combining with the 
information obtained from the previous analysis one can
decisively determine $\klc$ and $\krc$.
The important question  to ask then is how to measure
the decay width of $t \ra b W^+$, denoted as $\width$.

\subsubsection{Measuring the decay width  of $t \ra b W^+$}
\indent\indent

As shown in Ref.~\cite{steve}, the intrinsic width of the top quark 
cannot be measured at hadron colliders from
reconstructing the invariant mass of the jets from the decay of the 
top quark produced from the usual QCD processes ($\ggtt$) because
of the poor resolution of the jet energy as measured by the detector.
For a 175\,GeV SM top quark, its intrinsic width is about 1.5\,GeV, 
however the measured width from the invariant mass distribution of the 
top quark is unlikely to be much better than 10\,GeV \cite{tev2000}.
Is there a way to measure the top quark width $\width$ to within
a factor of 2 or better, at hadron colliders?
The answer is yes. It can be measured from single-top events.

The width $\width$ can be measured 
by counting the production rate of top
quark from the $W$-$b$ fusion process which is {\it equivalent} 
to the $W$-gluon fusion process by  properly treating the bottom 
quark and the $W$ boson as partons inside the hadron.
In the following we shall discuss how to correctly 
treat the $b$-quark as a parton inside the proton to properly 
resum all the large logs to all orders in $\alpha_s$. 
First, let us illustrate how to treat the $W$-boson as a parton
inside the proton. Consider the $q' b \ra q t$ process.
It can be viewed as the production of an on-shell $W$-boson
(i.e., effective-$W$ approximation)
which then rescatters with the $b$-quark to produce
the top quark. This factorization is similar to that in the
deep-inelastic scattering processes. The analytic expression
for the flux ($f_\lambda(x)$)
of the incoming $W_\lambda$-boson ($\lambda=0,+,-$ for longitudinal,
right-handed, or left-handed polarization) to rescatter 
with the $b$-quark
can be found in Ref.~\cite{wkteffw}.
The constituent cross section of $u b \ra d t$ is given by
\bea
\hat{\sigma}(u b \ra d t) & = &
\sum_{\lambda=0,+,-} \! f_\lambda\left(x={m^2_t \over \hat{s}}\right)
\left[ {16 \pi^2 m^3_t \over \hat{s} {(m^2_t-M^2_W)}^2 } \right]
\Gamma(t \ra b W_\lambda^+) \, ,
\nonumber 
\ena
where $M_W$ is the mass of $W^+$-boson and $\sqrt{\hat{s}}$ is 
the invariant mass of the hard part process.
Note that in order to derive the above result one has to assume that 
the dynamics of the hard part scattering, i.e., $b W^+(k_\mu) \ra t$,
does not change dramatically from an off-shell ($k^2 < 0$) to 
an on-shell ($k^2=M_W^2$) $W$-boson. 
Hence, the above equality is 
only valid under the effective-$W$ approximation even though 
the kinematic factors are correctly included.
Since the scattering rate of $W b \ra t$ is proportional to the
decay rate of $t \ra W b$, the production rate of single-top event from
the $W$-gluon fusion process measures the partial decay width
of the top quark $\width$. 
Furthermore, the branching ratio of $t \ra W b$ can be
measured\footnote{CDF group has 
reported a measurement of this branching ratio in \cite{incandela}.} 
from the ratio of the numbers of
double-$b$-tagged versus single-$b$-tagged $t \bar t$ events 
and the ratio of $(2\ell+\,jets)$ and $(1\ell+\,jets)$
rates in $t \bar t$ events for $t \ra b W^+(\ra \ell^+\nu)$ 
\cite{tev2000}.
Combining this model-independent 
measurement of the branching ratio Br($t \ra b W$)  
with the measurement of the partial decay width
$\width$ from the single-top production rate,
one can determine the total decay width 
$\Gamma_t=\Gamma(t \ra bW)/ {\rm Br}(t \ra b W)$
of the top quark, or equivalently,  the lifetime ($1/\Gamma_t$)
of the top quark.
At the Run-II of the Tevatron  we expect that the lifetime of
the top quark will be known to about 20\% $\sim$ 30\%.
Here, we have taken the values that 
the branching ratio Br($t \ra b W^+$)
can be measured to about 10\%~\cite{tev2000} and the cross section 
for $W$-gluon fusion process is known to about 15\% $\sim$ 20\%
(discussed in the next section).

Before closing this section, we comment on the importance of 
measuring the single-top production rate from the $W$-gluon fusion 
process. 
In the SM, the only nonvanishing coupling
 at the tree level is $\klc= 1$. 
These $\kappa$'s would have different values if new
physics exists. Nevertheless, 
the conclusion that the production rate of the 
$W$-gluon fusion event is proportional to the decay width of $t \ra W b$
holds irrespective of the specific forms of the anomalous couplings
(even including higher order operators).
Hence, measuring the single-top event rate from the $W$-gluon fusion
process is an {\it inclusive} method for 
detecting effects of new physics
which might produce large
modifications to the interactions of the top quark.
Strictly speaking, from the production rate of single-top events,
one measures the sum (weighted by parton densities)
of all the possible partial decay widths, such as
$\Gamma(t \ra b W^+) +\Gamma(t \ra s W^+) 
+\Gamma(t \ra d W^+) + \cdots$,
therefore, this measurement is actually measuring the width of
$\Gamma(t \ra XW^+)$ where $X$ can be more than one particle state 
as long as it originates from the partons 
inside the proton (or anti-proton).
If new physics strongly enhances the flavor-changing-neutral-current
$t$-$c$-$Z$, then the single-top production rate would also be enhanced
from the $Z$-$c$ fusion process $q c \ra q t$.

\subsubsection{The total production rate of $W$-gluon process}
\indent\indent

The calculation on the production rate of $W$-gluon fusion process 
involves a very important but not 
yet well-developed technique for handling the kinematics of 
a {\it heavy} $b$ parton inside a hadron. 
Thus, the kinematics of the top quark produced from this process can
not be accurately calculated.
However, the total event rate of the 
single-top quark production via this process
can be estimated using the method proposed in Ref.~\cite{wuki}.
The total rate for $W$-gluon fusion process
involves the ${\cal O}(\alpha^2)$
($2 \ra 2$) process $\ubdt$
plus the ${\cal O}(\alpha^2 \alpha_s)$ ($2 \ra 3$)
process $q' g (W^+ g) \ra q t \bar b $ 
(where the gluon splits to $b \bar b$)
minus the {\it splitting} piece $g \ra b \bar b \,\otimes\, \ubdt$
in which $b \bar b$ are nearly collinear.
\begin{figure}
\centerline{\hbox{\psfig{figure=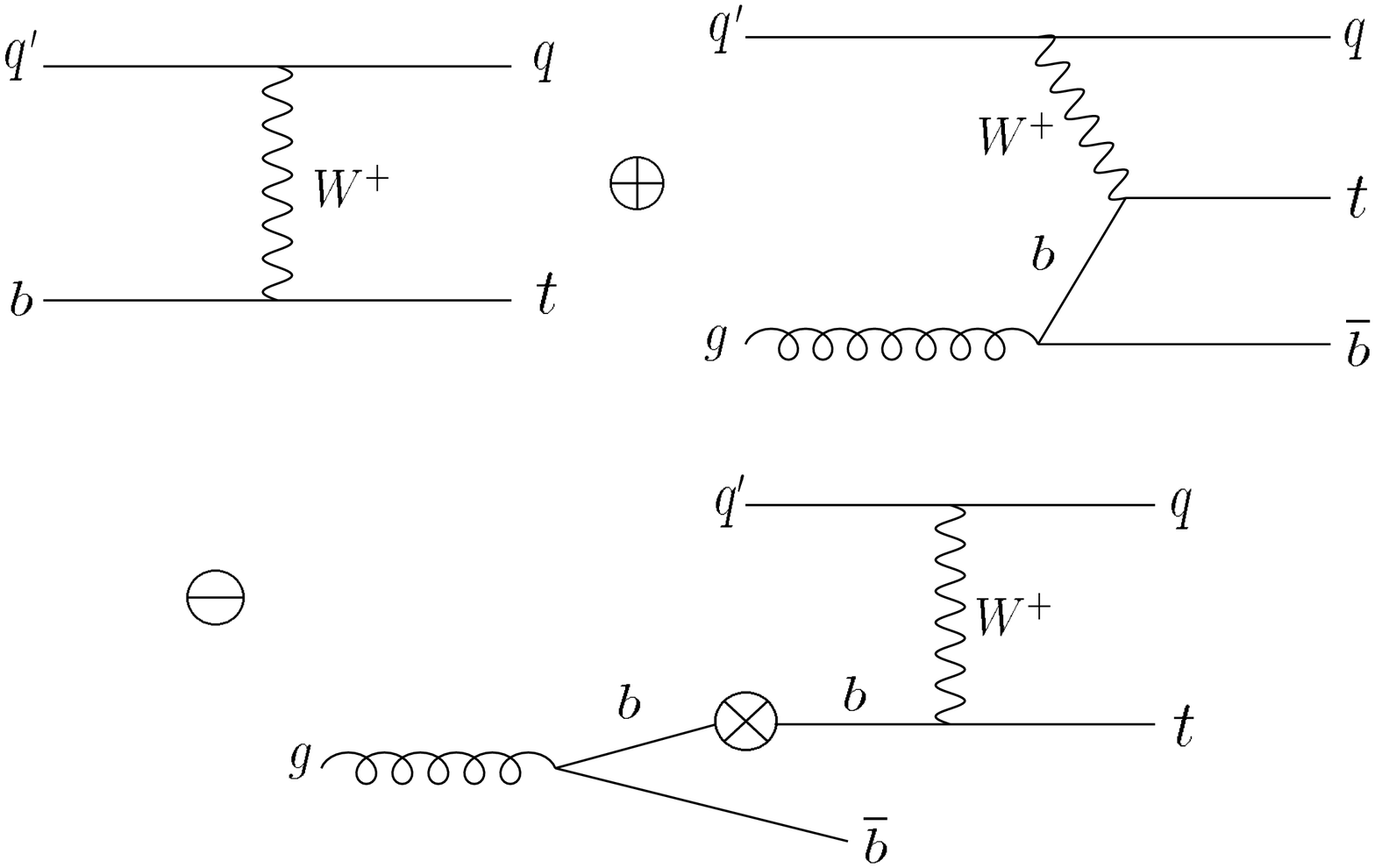,height=2.5in}}}
\caption{ Feynman diagrams illustrating the subtraction procedure
for calculating the total rate for $W$-gluon fusion:
$\ubdt\,\oplus\, q' g (W^+ g) \ra q t \bar b \,
\ominus\, (g \ra b \bar b \,\otimes\, \ubdt)$. }
\label{feynm}
\end{figure}
These processes are shown diagrammatically in Figure~\ref{feynm}.

The splitting piece is subtracted to avoid double counting the regime in
which the $b$ propagator in the ($2 \ra 3$) process closes 
to on-shell. This procedure is to resum the large logarithm 
$\alpha_s \ln (m_t^2/m_b^2)$ in the $W$-gluon fusion process
to all orders in $\alpha_s$ and include part of the higher order
${\cal O}(\alpha^2 \alpha_s)$ corrections to its production rate.
($m_b$ is the mass of the bottom quark.)
We note that to obtain the complete ${\cal O}(\alpha^2 \alpha_s)$ 
corrections beyond just the leading log contributions one should
also include virtual corrections to the ($2 \ra 2$) process, but
we shall ignore these non-leading contributions in this work\footnote{
In Ref.~\cite{nlost} it is shown that indeed these non-leading logs are
not important.}.
Using the prescription described above  
we find that when using the $\overline{{\rm MS}}$
parton distribution function (PDF) CTEQ2L \cite{pdf} the total rate
of the $W$-gluon fusion process is about  $25\%$ smaller than the
($2 \ra 2$) event rate either at the Tevatron or at the LHC.

\begin{figure}
\centerline{\hbox{
\psfig{figure=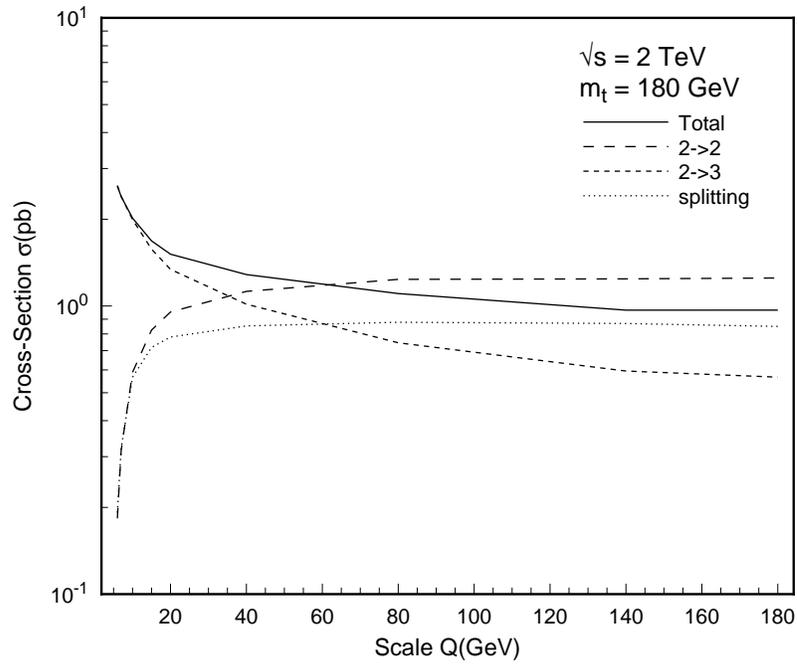,height=3.5in}}}
\caption{ Rate of $W$-gluon fusion process versus scale $Q$ 
for $m_t = 180\,$GeV and $\protect\rts=2$\,TeV. }
\label{scale}
\end{figure}

To estimate the uncertainty in the production rate 
due to the choice of the scale $Q$ in evaluating 
the strong coupling constant $\alpha_s$ and the 
parton distributions, we show
in Figure~\ref{scale} the scale dependence
of the $W$-gluon fusion rate for a SM top quark.  
As shown in the figure, although the individual rate from 
either ($2 \ra 2$), ($2 \ra 3$), or the splitting piece
is relatively sensitive to the choice of the scale,
the total rate as defined by $(2 \ra 2)\, + \, (2 \ra 3) \, - \, 
({\rm splitting\,piece})$ only varies by about 30\% 
for $M_W/2 < Q < 2 m_t$ at the Tevatron. (At the LHC, 
it varies by about 10\%).
This uncertainty reduces to about 10\% (at the Tevatron) for
$m_t/2 < Q < 2 m_t$.\footnote{
This conclusion is in good agreement with 
a complete next-to-leading-order calculation 
(different from the above resummation procedure) performed
in Ref.~\cite{nlowg} in which the theoretical error on the
total cross section at the Tevatron was estimated to be 
about 10\% for $Q$ ranging from $m_t/2$ to $2 m_t$.}
Based upon the results shown in Figure~\ref{scale}, we argue 
that $Q < M_W/2$ is probably not a good choice as the relevant 
scale for the production of the top quark from the $W$-gluon fusion
process because the total rate rapidly increases
by about a factor of 2 in the low $Q$ regime. 
In view of the prescription adopted in calculating the total rate,
the only relevant scales are the top quark mass $m_t$ and the 
virtuality of the $W$-line in the scattering amplitudes.
Since the typical transverse momentum
of the quark ($q$) which comes from the initial quark ($q'$) 
after emitting the $W$-line is about half of the $W$-boson mass,
the typical virtuality of the $W$-line is about 
$M_W/2 \sim 40$\,GeV. The scale $m_b \sim 5$\,GeV is thus
not an appropriate one to be used in calculating the $W$-gluon fusion
rate when using our prescription.
We note that in the ($2 \ra 2$) process the
$b$ quark distribution effectively contains sums to order 
$[\alpha_s \ln(Q/{\mb})]^n$ from $n$-fold collinear
gluon emission, whereas the subtraction term (namely, 
the splitting piece)
contains only first order in $\alpha_s 
\ln({\it Q}/{\mb})$.  Therefore, as 
${\it Q} \ra {\mb}$ the ($2 \ra 2$) 
contribution is almost cancelled by the
splitting terms. Consequently, as shown in Figure~\ref{scale}, the total
rate is about the same as the ($2 \ra 3$) rate for $Q \ra m_b$.
It is  easy to see also that based upon 
the factorization of the QCD theory
\cite{wuki} the total rates calculated via this prescription will not be
sensitive to the choice of $\overline{{\rm MS}}$ PDF although each
individual piece can have different results from different PDF's.

In conclusion, assuming $\krc=0$,
then $\klc$ can be constrained to within
 $-0.08 < \klc < 0.03$ 
assuming a 20\% uncertainty on the production rate
of single-top quark from the $W$-gluon fusion process at the 
Tevatron \cite{ehab}. This means that if 
we interpret {\mbox {($1+\klc$)}} as the 
CKM matrix element $|V_{tb}|$, 
then $|V_{tb}|$ can be bounded as $|V_{tb}| > 0.9$.\footnote{
This method is different from the one used in the recent CDF
measurement of $|V_{tb}|$ by measuring 
Br($t \ra b W^+$) and assuming 3 generations of quarks plus
unitarity~\cite{cdfvtb}. Our method does not require such assumption. 
}

\subsubsection{Other single-top production rates}
\indent\indent

Another single-top quark production mechanism 
is the Drell-Yan type process 
$q' \bar q \ra W^* \ra t \bar b$ whose production rate
can also provide information on $\klc$ and $\krc$.
Notice that the polarization of the
top quark produced from this process is different from the one in 
$W$--gluon fusion events~\cite{thesis}.  For instance, for a $175$ GeV
SM top quark produced at the Tevatron, $W$--gluon fusion produces
almost $100\%$ left handed top quarks, but the $W^{*}$ process
produces  $\sim 50\%$ polarized top quarks (i.e., $\frac{1}{4}$ of top
quarks are right handed and the rest are left handed).
Hence, these production rates depend on $\klc$ and $\krc$ 
differently. Furthermore, since the kinematics of the top quark 
produced from these two processes are different~\cite{thesis},
these two kinds of events can be separated at the Tevatron. 
In Ref.~\cite{wstar}, a careful study was carried out of how to 
measure $|V_{tb}|$ from the production rate of $W^*$ events.
It was concluded that $|V_{tb}|$ can be measured to about 10\% 
at the Tevatron if $\krc = 0$.
It was shown in Ref.~\cite{qcdwstar} that the production rate of 
$W^*$ events up to the next-to-leading 
order QCD corrections is well under control (better than $10\%$).
Hence, this process should provide a good measurement of $\klc$
and $\krc$.\footnote{
We note that the production rate of the $W^*$ process is not
directly proportional to the decay width of $t \ra b W^+$,
but the production rate of the $W$-gluon process is. 
}

We note that because the production cross sections of the single-top
events from the $W$-gluon fusion and the $W^*$ processes depend
differently on $\klc$ and $\krc$, they all have to be measured
and combined with the measurement of the decay kinematics of
the top quark to definitely constrain the anomalous couplings
$\klc$ and $\krc$.
At the LHC, the single-top production rate from 
$b g \ra Wt$ process is about 7 times the $W^*$ rate 
and should also be measured to
probe the interaction of the top quark with the $W$-boson.

\subsection{At the LC}
\indent\indent

 The best place to probe the couplings $\kln$ and $\krn$ 
associated with the
\ttz coupling is at the LC 
through $e^- e^+ \ra  \gamma, Z \ra t \bar{t}$
process because at hadron 
colliders the $t\ov {t}$ production rate is 
dominated by QCD interactions 
( $q\ov {q}, gg\;\ra\; t\ov {t}$ ).
A detailed Monte Carlo study on the measurement of these couplings
at the LC including detector effects and initial state radiation
can be found in Ref.~\cite{gal}.
The bounds were obtained by studying the
angular distribution and the polarization of the top quark
produced in $e^- e^+$ collisions.
Assuming a 50 $\rm{fb}^{-1}$
luminosity at $\sqrt{s}=500$\,GeV, we concluded that
within a 90\% confidence level, it should be possible to measure
$\kln$ to within about 8\%, while $\krn$ can be known
to within about 18\%.
A $1\,$TeV machine can do better than a $500\,$GeV machine in
determining $\kln$ and $\krn$ because the relative sizes of the
$t_R {(\overline{t})}_R$  and $t_L {(\overline{t})}_L$
production rates become small and the 
polarization of the $t \bar t$ pair
is purer. Namely, it is more likely to produce either
a $t_L {(\overline{t})}_R$ or a $t_R {(\overline{t})}_L$ pair.
A purer polarization of the $t \bar t$ pair makes $\kln$ and $\krn$
better determined. (The degree of top quark polarization
can be further improved by polarizing the electron beam
\cite{carl}.)
Furthermore, the top quark is
boosted more in a $1\,$TeV machine, thereby allowing a better
determination of its polar angle in the $t \bar t$ system
(because it is easy to find the right $b$ associated with the lepton
to reconstruct the top quark moving direction).

Finally, we remark that at the LC $\klc$ and $\krc$ can be studied
either from the decay of the top quark pair or from the single--top
quark production process, $W$--photon fusion 
process $e^{-}e^{+}(W\gamma) \ra t X $, 
or $ e^{-}\gamma (W\gamma) \ra {\bar t} X$, which is similar to the
$W$--gluon fusion process in hadron collisions.
 
\section{Dimension five anomalous couplings}
\indent\indent

So far we have discussed how to 
probe new physics effects that are expected to
give some information about the symmetry breaking mechanism, as
they can give rise to anomalous terms in the dimension 4 standard
gauge couplings of the top quark with the electroweak bosons.  Of 
course, this is not the only way in which these effects can become
apparent in future experiments.  A complete analysis should include
possible anomalous effective interactions of higher dimension.
In this section we will construct the complete set of independent 
operators of the first higher order operators with
dimension 5, such that the complete effective Lagrangian 
relevant to this study will be:
\begin{equation}
{\cal L}_{eff}\; = \;{\cal L}^{B}\;+\; 
{\cal L}^{(4)}\;+\;{\cal L}^{(5)}\; ,
\end{equation}
where ${\cal L}^{(5)}$ denotes the dimension 5 operators.

Our next task is to find all the possible dimension five hermitian 
interactions that involve the top quark and the fields 
${\cal W}_{\mu}^{\pm}$,  ${\cal Z}_{\mu}$ and ${\cal A}_{\mu}$.  
Notice that the gauge transformations associated with these and the 
composite fermion fields ( Eq. ~(\ref{eq1}) ) 
are dictated simply by the 
$\rm{U(1)}_{em}$ group.    We will follow a 
procedure similar to the one in 
Ref. \cite{buchmuller}, which consists of constructing all possible 
interactions that satisfy the required gauge invariance 
($\rm{U(1)}_{em}$ in this work), and that 
are not equivalent to each other. 
The criterion for equivalence is based on the equations of motion 
and on partial integration.  As for the five dimensions 
in these operators, three will come from the fermion fields, and the 
other two will involve the gauge bosons.  
To make a clear and systematic 
characterization,  let us recognize the only three possibilities for 
these two dimensions:

\noindent
(1) Operators with two boson fields.\\
(2) Operators with one boson field and one derivative.\\
(3) Operators with two derivatives.\\

\noindent
(1)  {\bf Two boson fields.}
\indent 
First of all, notice that the ${\cal A}_{\mu}$ 
field gauge transformation 
( Eq. ~(\ref{atransf}) ) 
will restrict the use of this field to covariant derivatives only.  
Therefore, except for the field 
strength term $\cal A_{\mu\nu}$ only the 
$\cal Z$ and $\cal W$ fields can appear multiplying the fermions in any 
type of operators.   
Also, the only possible Lorentz structures are given in terms of the 
$\sigma_{\mu \nu}$ and $g_{\mu \nu}$ tensors.  
We do not need to consider the tensor product of $\gamma_{\mu}$'s since
\begin{equation} 
\aslash \bslash =  g_{\mu \nu} a_{\mu} b_{\nu} - i\sigma_{\mu \nu} 
a_{\mu} b_{\nu}\; .\label{idesig}
\end{equation}  
Finally,  we are left with only three possible combinations: 
(1.1) two ${\cal Z}_{\mu}$'s, (1.2) two ${\cal W}_{\mu}$'s,  and 
(1.3) one of each.\\
\indent
(1.1)  Since $\sigma_{\mu \nu}$ is antisymmetric, only the 
$g_{\mu \nu}$ part is non-zero\footnote{ 
In the next section we will write explicitly the hermitian conjugate 
($h.c.$) parts.}:
\begin{eqnarray}
O_{g{\cal Z}{\cal Z}} = \bar t_L  
t_R {\cal Z}_{\mu}{\cal Z}^{\mu} + h.c.
\label{ozz}
\end{eqnarray} 
\indent
(1.2) Here, the antisymmetric part is non-zero too:
\begin{eqnarray}
O_{g{\cal W}{\cal W}} &=& \bar t_L  t_R {\cal W}_{\mu}^{+} 
{\cal W}^{- \mu } + h.c. \label{opgww} \\
O_{\sigma {\cal W}{\cal W}} &=& 
\bar t_L  {\sigma}^{\mu\nu} t_R {\cal W}_{\mu}^{+} 
{\cal W}^{-}_{\nu } + h.c. \label{opsww}
\end{eqnarray} 
\indent
(1.3) In this case we have two different quark fields, therefore we can 
distinguish two different combinations of chiralities:
\begin{eqnarray}
O_{g{\cal W}{\cal Z} L (R)} &=& 
\bar t_{L(R)}  b_{R(L)} {\cal W}_{\mu}^{+} {\cal Z}^{\mu } + h.c.
\label{ogwz} \\
O_{\sigma {\cal W}{\cal Z} L (R)} &=& 
\bar t_{L(R)} {\sigma}^{\mu\nu} b_{R(L)} {\cal W}^{+}_{\mu}
{\cal Z}_{\nu} + h.c. \label{oswz}
\end{eqnarray} 
\\
(2) {\bf One boson field and one derivative.}
\indent The obvious distinction arises:  
(2.1) the derivative acting on a fermion 
field, and (2.2) the derivative acting on the boson.

\indent
(2.1) The covariant derivative for the fermions is given by
\footnote{To simplify notation we will 
use the same symbol $D_{\mu}$ for 
all covariant derivatives.  Identifying which derivative we 
are referring to 
should be straightforward, e.g. $D_{\mu}$ in 
Eq. (\ref{derf}) is different 
from $D_{\mu}$ in Eq. (\ref{covderw}).}
 (see Eqs.~(\ref{atransf}) and~(\ref{eq1}))
\begin{eqnarray}
D_{\mu} f = (\partial_{\mu} + i Q_f s_w^2 
{\cal A}_{\mu}) f \; , \nonumber \\
{\overline {D_{\mu} f}} = 
{\bar f}(\stackrel{\leftarrow}{\partial}_{\mu} 
- i Q_f s_w^2 {\cal A}_{\mu}).\label{derf}
\end{eqnarray}
Notice that the covariant derivative depends 
on the fermion charge $Q_f$, 
hence the covariant derivative for the 
top quark is not the same as for the bottom quark; partial integration 
could not relate two operators involving derivatives 
on different quarks. 
Furthermore, by looking at the equations 
of motion we can immediately see 
that operators of the form, for example, 
$\bar f {\zslash} {\Dslash} f $ or
${\bar f}^{(up)} {\wslash}^{+} {\Dslash} {f}^{(down)}$, are 
equivalent to operators with two bosons, which have all been considered 
already.  Following the latter statement and bearing in 
mind the identity 
of Eq.~(\ref{idesig}) 
we can see that only one Lorentz structure needs to be considered here, 
either one with $\sigma_{\mu \nu}$ or one with $g_{\mu\nu}$.  
Let us choose the latter.  
\begin{eqnarray}
O_{{\cal W}D b L (R)} &=&{\cal W}^{+ \mu} \bar t_{L(R)} 
D_{\mu} b_{R(L)}  + h.c.  \label{odbw} \\
O_{{\cal W}D t R (L)} &=&{\cal W}^{- \mu} \bar b_{L(R)} 
D_{\mu} t_{R(L)}  + h.c. \label{odtw} \\
O_{{\cal Z}D f } &=& {\cal Z}^{\mu} \bar t_{L} D_{\mu} t_{R} + h.c.
\label{odfz}
\end{eqnarray}
Of course, the $\cal A$ field did not appear.  
Remember that its gauge transformation prevents 
us from using it on anything that is not a covariant derivative or a 
field strength ${\cal A}_{\mu\nu}$. \\  
 
\indent
(2.2) Since $\cal W$  transforms as a field with electric charge one, 
the covariant derivative is simply given by (see Eq. ~(\ref{wtransf}) ):
\begin{eqnarray}
D_{\mu} {\cal W}_{\nu}^{+} &=& 
(\partial_{\mu} + i s_w^2 {\cal A}_{\mu}) 
{\cal W}_{\nu}^{+} \nonumber \\
D^{\dagger}_{\mu} {\cal W}_{\nu}^{-} &=& 
(\partial_{\mu} - i s_w^2 {\cal A}_{\mu}) {\cal W}_{\nu}^{-}\label{Dw}
\end{eqnarray}

Obviously, since the neutral 
$\cal Z$ field is invariant under the $G$ group 
transformations ( see Eq. ~(\ref{ztransf}) ), 
we could always add it to our 
covariant derivative:
\begin{eqnarray}
D^{(\cal Z)}_{\mu} {\cal W}_{\nu}^{+} &=& 
(\partial_{\mu} + i s_w^2 {\cal A}_{\mu} + i a {\cal Z}_{\mu}) 
{\cal W}_{\nu}^{+} \,\label{Dw2} \nonumber 
\end{eqnarray}
where $a$ stands for any complex constant.  
Actually, considering this second  
derivative would insure the generality 
of our analysis, since for example 
by setting $a = c_w^2$ and 
comparing with Eqs. ~(\ref{b1}) and ~(\ref{b2})
 we would automatically include the field strength term
\footnote{From Eqs. ~(\ref{wdef}) and ~(\ref{wstrength}), we  
write ${\cal W}^{\pm}_{\mu\nu} = 
\frac{1}{\sqrt{2}}({\cal W}^1_{\mu\nu} \mp i {\cal W}^2_{\mu\nu})$.}
\begin{eqnarray}
{\cal W}_{\mu \nu}^{\pm} = 
{\partial}_{\mu} {\cal W}^{\pm}_{\nu} - 
{\partial}_{\nu} {\cal W}^{\pm}_{\mu} 
\pm i ( {\cal W}^{\pm}_{\mu} {\cal W}^{3}_{\nu} - 
{\cal W}^{3}_{\nu} {\cal W}^{\pm}_{\mu} ) = 
D^{(\cal Z)}_{\mu} {\cal W}_{\nu}^{\pm} - 
D^{(\cal Z)}_{\nu} {\cal W}_{\mu}^{\pm} .
\end{eqnarray} 
However, this extra term in the 
covariant derivative would only be redundant.  
We can always decompose any given operator written in terms of  
$D_{\mu}^{(\cal Z)}$ into the sum of the same operator in terms of the 
original $D_{\mu}$ plus another operator of the form  
$O_{g{\cal W}{\cal Z} L (R)} $ or $O_{\sigma {\cal W}{\cal Z} L (R)}$ 
( c.f. Eqs. ~(\ref{ogwz}) and ~(\ref{oswz}) ).  
Therefore, we only need to consider 
the covariant derivative ~(\ref{Dw}) 
for the charged boson and still maintain the generality of our 
characterization.
For the neutral ${\cal Z}$ boson we have the simplest situation, 
the covariant derivative is just the ordinary one, 
\begin{eqnarray}
D_{\mu} {\cal Z}_{\nu} &=& \partial_{\mu} {\cal Z}_{\nu} .
\end{eqnarray}

The case for the ${\cal A}$ boson is nevertheless different. 
Being the field that makes possible 
the $\rm{U(1)}_{em}$ covariance in the 
first place, it can not be given any covariant derivative itself.  
For ${\cal A}$, we have the field strength: 
\begin{eqnarray}
{\cal A}_{\mu \nu} &=& 
{\partial}_{\mu}{\cal A}_{\nu} - {\partial}_{\nu}{\cal A}_{\mu}\; , 
\nonumber
\end{eqnarray}

Finally, we can now write the operators with the covariant 
derivative-on-boson terms.  
Unfortunately, no equations of motion can help us reduce the number of 
independent operators in this case, and we have to bring up both the 
$\sigma_{\mu\nu}$ and the $g_{\mu\nu}$ Lorentz structures.
\begin{eqnarray}
O_{\sigma D {\cal Z}} &=& 
\bar t_L \sigma^{\mu \nu} t_R {\partial}_{\mu} {\cal Z}_{\nu} + h.c. \\
O_{g D {\cal Z}} &=& \bar t_L  t_R {\partial}_{\mu} 
{\cal Z}^{\mu} + h.c. \\
O_{\sigma D {\cal W} L(R)} &=& 
\bar t_{L(R)} \sigma^{\mu \nu} b_{R(L)} D_{\mu} 
{\cal W}^{+}_{\nu} + h.c. \\
O_{g D {\cal W}L(R)} &=& 
\bar t_{L(R)}  b_{R(L)} D_{\mu} {\cal W}^{+ \mu} +h.c. \\
O_{ {\cal A}} &=& \bar t_L \sigma^{\mu \nu} t_R  
{\cal A}_{\mu \nu} + h.c.
\label{oa1}
\end{eqnarray}
\\
(3) {\bf Operators with two derivatives.}\\

As it turns out, all operators of 
this kind are equivalent to the ones already 
given in the previous cases.  Here, we 
shall present the argument of why 
this is so.  First of all, we only have two possibilities, 
(3.1) one derivative acting on each fermion field,  
and (3.2) both derivatives acting on the same fermion field.\\

\indent
(3.1)  Just like in the  case (2.1) above, we first notice that an
operator of the form ${\bar f}\stackrel{\leftarrow}{\Dslash} {\Dslash} f$
can be decomposed into operators of the previous cases
(1.1), (1.2) and (1.3) by means of the equations of 
motion.  Therefore, we only have to consider 
one of two options, 
either ${\overline {D_{\mu} f}} \sigma^{\mu \nu} D_{\nu} f$, or   
${\overline {D_{\mu} f}} g^{\mu \nu} D_{\nu} f$.  
Let us choose the latter.   
By means of partial integration we can see that the term 
$(\partial_{\mu} \bar f) \partial^{\mu} f$ yields 
the same action as the term 
$-\bar f \partial^{\mu} \partial_{\mu} f$, and 
we only need to consider the 
case in which the covariant derivatives act on the same $f$,
which is just the type of operator to be considered next.\\

\indent
(3.2) By using the equations of motion twice we can relate the 
operator  ${\bar f} {\Dslash} {\Dslash} f$ to operators of the type 
(1.1), (1.2) or (1.3).   Either ${\bar f} 
\sigma^{\mu \nu} D_{\mu} D_{\nu} f$, 
or ${\bar f} D^{\mu} D_{\mu} f$ needs to be considered.  
This time we choose the former, which 
can be proved to be nothing but the 
operator $O_{ {\cal A}}$ itself ( Eq.~(\ref{oa}) ).

\subsection{Hermiticity and CP invariance}
\indent\indent

The list of operators above is complete in the sense that it includes 
all non-equivalent dimension five interactions that 
satisfy gauge invariance. 
It is convenient now to analyze their CP properties.  
In order to make our study 
more systematic and clear we will re-write this 
list again, but this time we will 
display the added hermitian conjugate part 
in detail.  By doing this the CP transformation 
characteristics will be most clearly presented too.

Let us divide the list 
of operators in two: those with only the top quark, 
and those involving both top and bottom quarks.

\subsubsection{\bf Interactions with top quarks only}
\indent\indent

Let's begin by considering the operator $O_{g {\cal Z} {\cal Z}}$.  
We will include an arbitrary constant coefficient, denoted as $a$, 
which in principle could be complex:
\begin{eqnarray}
O_{g{\cal Z}{\cal Z}} &=& 
a \bar t_L  t_R {\cal Z}_{\mu}{\cal Z}^{\mu} + 
a^{*} \bar t_R  t_L {\cal Z}_{\mu}{\cal Z}^{\mu} \nonumber \\
&=& Re(a) \bar t t  {\cal Z}_{\mu}{\cal Z}^{\mu}  + 
Im(a) i \bar t \gamma_5 t  {\cal Z}_{\mu}{\cal Z}^{\mu} \nonumber
\end{eqnarray}
Our hermitian operator has naturally split into two independent parts: 
one that preserves parity (scalar), and 
one that does not (pseudoscalar).  
Also, the first part is CP 
even whereas the second one is odd. The natural 
separation of these two parts happens to be a common feature of all 
operators with only one type of fermion field.  Nevertheless, 
not always will the parity conserving part  also be the CP even one, 
as we shall soon see.

Below, the complete list of all 
7 operators with only the top quark is given.  
In all cases the two independent terms 
are included; the first one is CP even, 
and the second one is CP odd.
\begin{eqnarray}
O_{g{\cal Z}{\cal Z}} &=& {\frac {1}{\Lambda}}
Re(a_{zz1}) \bar t t {\cal Z}_{\mu}{\cal Z}^{\mu} 
\; +\; {\frac {1}{\Lambda}}
Im(a_{zz1}) i \bar t \gamma_5 t  
{\cal Z}_{\mu}{\cal Z}^{\mu}\label{fir} \\
O_{g{\cal W}{\cal W}} &=& {\frac {1}{\Lambda}}
Re(a_{ww1}) \bar t  t {\cal W}_{\mu}^{+} {\cal W}^{- \mu } \; + \; 
{\frac {1}{\Lambda}} 
Im(a_{ww1}) i \bar t \gamma_5 t {\cal W}_{\mu}^{+} {\cal W}^{- \mu }
\label{ogww} \\
O_{\sigma {\cal W}{\cal W}} &=& {\frac {1}{\Lambda}}
Im(a_{ww2}) i \bar t  {\sigma}^{\mu\nu} t {\cal W}_{\mu}^{+} 
{\cal W}^{-}_{\nu }\; +\; {\frac {1}{\Lambda}} Re(a_{ww2}) \bar t  
{\sigma}^{\mu\nu} \gamma_5 t {\cal W}_{\mu}^{+} {\cal W}^{-}_{\nu } 
\label{osww} \\
O_{{\cal Z}D f } &=&{\frac {1}{\Lambda}} Im(a_{z3}) i \bar t D_{\mu} t 
{\cal Z}^{\mu} \; + \; {\frac {1}{\Lambda}} Re(a_{z3}) \bar t D_{\mu} 
\gamma_5 t {\cal Z}^{\mu}  \\
O_{g D {\cal Z}} &=& {\frac {1}{\Lambda}}
Im(a_{z4}) i \bar t \gamma_5 t {\partial}_{\mu} {\cal Z}^{\mu} \; +\; 
{\frac {1}{\Lambda}}Re(a_{z4}) \bar t  t 
{\partial}_{\mu} {\cal Z}^{\mu}  \\
O_{\sigma D {\cal Z}} &=& {\frac {1}{\Lambda}}
Re(a_{z2}) \bar t \sigma^{\mu \nu} t 
{\partial}_{\mu} {\cal Z}_{\nu} \; +\; 
{\frac {1}{\Lambda}} 
Im(a_{z2}) i \bar t \sigma^{\mu \nu} \gamma_5 t {\partial}_{\mu} 
{\cal Z}_{\nu}  \\
O_{{\cal A}} &=& {\frac {1}{\Lambda}}Re(a_{\cal A}) 
\bar t \sigma^{\mu \nu} t  
{\cal A}_{\mu \nu}\; +\; {\frac {1}{\Lambda}} 
Im(a_{\cal A}) i \bar t \sigma^{\mu \nu}\gamma_5 t 
{\cal A}_{\mu \nu} \; .\label{oa}
\end{eqnarray}
Notice that in the operator $O_{g D {\cal Z}}$ the parity 
violating part happens to be CP even.  This is because under a CP 
transformation a scalar term $\bar t t$ 
remains intact, i.e. it does not 
change sign,  whereas a pseudoscalar term 
$\bar t \gamma_5 t$ changes sign.  
The gauge bosons change sign too, and 
this is what makes the scalar part of 
the operator to change sign under CP.  Compare with the operator 
$O_{g{\cal Z}{\cal Z}}$, there we have two bosons; two changes of sign 
that counteract each other.  Therefore, it is 
the scalar part that is CP 
even in $O_{g{\cal Z}{\cal Z}}$.   Furthermore, based on the naive 
dimensional analysis (NDA) the coefficients of these operators are of 
order $1/ \Lambda$.  Therefore, the normalized coefficients (the $a$'s) 
are expected to be of order $1$.

\subsubsection{Interactions with both top and bottom quarks}
\indent\indent

Below, we show the next list of 12 operators with both top and bottom 
quarks. Again, we include an arbitrary complex coefficient\footnote{
${\overline {D_{\mu} f}}_{R(L)}$ stands for 
$({D_{\mu} f_{R(L)}})^{\dagger}\gamma_0$; $\bar f_{R(L)}$ stands for 
$(f_{R(L)})^{\dagger}\gamma_0$.}:
\begin{eqnarray}
O_{g{\cal W}{\cal Z} L (R)} &=& {\frac {1}{\Lambda}}
a_{wz1L(R)} \bar t_{L(R)}  b_{R(L)} 
{\cal W}_{\mu}^{+} {\cal Z}^{\mu } \; +\;
{\frac {1}{\Lambda}} 
a^{*}_{wz1L(R)} \bar b_{R(L)}  t_{L(R)} 
{\cal W}_{\mu}^{-} {\cal Z}^{\mu } \\
O_{\sigma {\cal W}{\cal Z} L (R)} &=& {\frac {1}{\Lambda}}a_{wz2L(R)} 
\bar t_{L(R)} {\sigma}^{\mu\nu} b_{R(L)} 
{\cal W}^{+}_{\mu}{\cal Z}_{\nu} + 
{\frac {1}{\Lambda}} a^{*}_{wz2L(R)} 
\bar b_{R(L)} {\sigma}^{\mu\nu} t_{L(R)} 
{\cal W}^{-}_{\mu}{\cal Z}_{\nu} \\
O_{{\cal W}D b L (R)} &=& {\frac {1}{\Lambda}}
a_{bw3L(R)} {\cal W}^{+ \mu} \bar t_{L(R)} D_{\mu} b_{R(L)} \; +\; 
{\frac {1}{\Lambda}} 
a^{*}_{bw3L(R)} {\cal W}^{- \mu} 
{\overline {D_{\mu} b}}_{R(L)}  t_{L(R)}  \\
O_{{\cal W}D t R (L)} &=& {\frac {1}{\Lambda}} a_{w3R(L)} 
{\cal W}^{- \mu} \bar b_{L(R)} D_{\mu} 
t_{R(L)} \; +\; {\frac {1}{\Lambda}}
a^{*}_{w3R(L)} {\cal W}^{+ \mu} 
{\overline {D_{\mu} t}}_{R(L)} b_{L(R)}   \\
O_{\sigma D {\cal W} L(R)} &=& {\frac {1}{\Lambda}}a_{w2L(R)} 
\bar t_{L(R)} \sigma^{\mu \nu} b_{R(L)} D_{\mu} {\cal W}^{+}_{\nu} \; + 
{\frac {1}{\Lambda}} a^{*}_{w2L(R)} 
\bar b_{R(L)} \sigma^{\mu \nu} t_{L(R)} 
D^{\dagger}_{\mu} {\cal W}^{-}_{\nu} \\
O_{g D {\cal W}L(R)} &=& {\frac {1}{\Lambda}} a_{w4L(R)} 
\bar t_{L(R)}  b_{R(L)} D_{\mu} {\cal W}^{+ \mu} \; 
+\;{\frac {1}{\Lambda}} 
a^{*}_{w4L(R)} \bar b_{R(L)}  
t_{L(R)} D^{\dagger}_{\mu} {\cal W}^{- \mu}
\label{las}  
\end{eqnarray}
In this case, if $a$ is real ($a = a^*$) 
then $O_{g{\cal W}{\cal Z} L (R)}$ 
and $O_{\sigma D {\cal W} L(R)}$ are both CP even, but 
$O_{\sigma {\cal W}{\cal Z} L (R)}$, $O_{{\cal W} D b L (R)}$, 
$O_{{\cal W} D t R (L)}$ and 
$O_{g D {\cal W}L(R)} $ are odd.  Just the other 
way around if $a$ is purely imaginary.   

The dimension five lagrangian ${\cal L}^{(5)}$ is simply the 
sum of all these 19 operators ( Eqs.~(\ref{fir}) to ~(\ref{las}) ): 
\begin{equation}
{\cal L}^{(5)} = \sum_{i=1,19} O_i \; .
\end{equation}
  
To study the possible effects on the 
production rates of top quarks in  high
energy collisions, only the CP conserving parts which give
imaginary vertices (like the SM) are relevant.   
The amplitude squared will 
depend linearly on the CP even 
terms, but only quadratically on the CP odd 
terms, because the {\it no-Higgs} SM  (${\cal L}^{(4)}$) 
interactions\footnote{Since in the unitary gauge ${\cal L}^{(4)}$
reproduces the SM without the physical 
Higgs boson, we will refer to it as 
the {\it no-Higgs} SM.} are CP even when ignoring the 
CP-violating phase in the Cabibbo-Kobayashi-Maskawa 
(CKM) mixing elements.  
     
However, this does not mean 
that it is not possible to probe the CP violating
phase in the operators. Later on in the next section
we will show one observable that depends linearly on the CP odd 
coefficients.   From now on, the appropriate CP even part  
(either real or imaginary) is assumed for each 
coefficient .   To simplify notation we 
will use the same label; $a_{zz1}$ 
will stand for $Re(a_{zz1})$, $a_{wz2L(R)}$ will stand for 
$Im( a_{wz2L(R)} )$, and so 
on, the only exception will be $a_{\cal A}$, 
whose real part is recognized as 
proportional to the magnetic moment of the 
top quark, and will be denoted by $a_m$.   
It is thus understood that all coefficients below are real numbers.

In conclusion, the dimension 5 Lagrangian 
consists of 19 independent operators 
which are listed from Eq. (\ref{fir}) to Eq. (\ref{las}).   
Since the top quark is heavy 
(its mass is of the order of the weak scale), 
it is likely to interact strongly with the Goldstone bosons which 
are equivalent to the longitudinal 
weak gauge bosons in the high energy regime.
(This is known as the Goldstone Equivalence Theorem \cite{et}.)  
Hence, we shall study in 
the rest of this paper how to probe these anomalous
couplings from the production of top 
quarks via the $V_L V_L$ fusion process, 
where $V_L$ stands for the longitudinally 
polarized $W^{\pm}$ or $Z$ bosons.

\section{Probing the dimension 5 anomalous couplings at the colliders}
\indent\indent

As it is suggested by the very form of these operators, we decide to 
probe their potential contribution to high 
energy scattering processes like
longitudinal vector boson ($V_L V_L$) 
fusions (see Figure~\ref{fusion}),
and study how they can
affect the production rates of top quarks in both the LHC and the LC.
For simplicity, in this study 
we shall take all the non-standard dimension
four couplings to be zero.  A 
general result including these operators are
given in Ref.~\cite{top5}.

\begin{figure}
\centerline{\hbox{
\psfig{figure=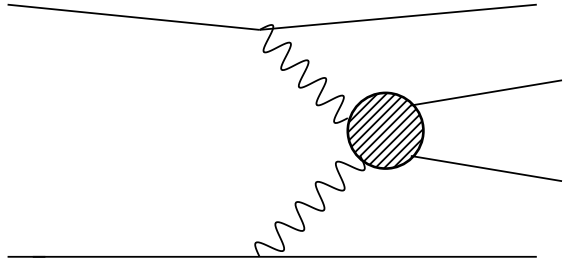,height=1.5in}}}
\caption{ Production of $t\bar{t}$  ($t\bar{b}$ or $b\bar{t}$) from 
$W^{+}_L W^{-}_L$ or $Z_L Z_L$ ($W^{+}_L Z_L$ or
$W^{-}_L Z_L$) fusion processes.} 
\label{fusion}
\end{figure}

Before doing any calculation at all, we can make an estimate of 
the expected sizes of these tree 
level amplitudes according to their high 
energy behavior.  A general power counting rule has been 
given that estimates the high energy behavior of a scattering amplitude
$T$ to be \cite{poc}
\begin{eqnarray}
T=&& c_T v^{D_T}\displaystyle 
\left(\frac{v}{\Lambda}\right)^{N_{\cal O}}
\left(\frac{E}{v}\right)^{D_{E0}}
\left(\frac{E}{4 \pi v}\right)^{D_{EL}}
\left(\frac{M_W}{E}\right)^{e_v} H\left(\ln( {E/\mu})\right) ~~, \\ 
 D_{E0}=&& 2+\sum_n {\cal V}_n
(d_n+\frac{1}{2}f_n-2)~, ~~~~ 
D_{EL}=2L~\nonumber,
\label{eqcounting}
\end{eqnarray}
where ${D_T}=4-e=0$ ($e$ is the number 
of external lines; 4 in our case),
${N_{\cal O}}= 0$  for all dimension 4 operators and  
${N_{\cal O}}= 1$  for all dimension 5 operators based upon the naive 
dimensional analysis (NDA)\footnote{NDA counts $\Sigma$ as $\Lambda^0$,
$D_\mu$ as $\frac{1}{\Lambda}$, and fermion
fields as $\frac{1}{v\sqrt{\Lambda}}$. Hence, ${\cal W}^{\pm}$,
${\cal Z}$ and ${\cal A}$ are also counted as
$\frac{1}{\Lambda}$. After this counting, one 
should multiply the result by
$v^2 \Lambda^2$. Notice that up to the 
order of intent, the kinetic term of 
the gauge boson fields and the mass term of the fermion fields are two 
exceptions to the NDA, and are of order $\Lambda^0$.}
 \cite{georgi,howard}, $L=0$ is the number of loops in the diagrams,  
$ H\left(\ln( {E/\mu})\right) = 1$ 
comes from the loop terms (none in our case), ${e_v}$ accounts 
for any external $v_\mu$-lines\footnote{$v_\mu$ is equal to 
$\epsilon^{(0)}_{\mu}-\frac{k_\mu}{M_V}$, where $k_\mu$ is 
the momentum of 
the gauge boson with mass $M_V$ 
and $\epsilon^{(0)}_{\mu}$ is its longitudinal
polarization vector.}(none in 
our case of $V_L V_L \ra t\ov t,\;t\ov b$), 
${\cal V}_n$ is the number of vertices of type $n$ that contain 
$d_n$ derivatives and $f_n$ fermionic lines.  The dimensionless 
coefficient $~c_T~$ contains 
possible powers of gauge couplings ($g,g^\prime$) and Yukawa 
couplings ($y_f$) from the vertices 
of the amplitude $~T~$, which can be directly counted.

At high energy the longitudinal components of the gauge bosons 
will dominate the production of top 
quarks coming from vector boson fusions, 
we will thus concentrate on their contribution from now on.  
According to the Goldstone boson Equivalence Theorem (ET) \cite{et}, 
in the high energy limit we 
can substitute the longitudinal external weak
boson lines with the corresponding Goldstone boson lines, 
and then perform a much easier calculation.        
In Figure~\ref{power} we show these 
diagrams for the $Z_L  Z_L  \ra t\bar {t}$
process (see also Figure~\ref{fzz}).  Let us 
start with the {\it no-Higgs} SM  
dimension 4 Lagrangian contribution to the 
process $Z_L Z_L  \ra t\bar {t}$.   It is convenient 
to use an alternative non-linear parameterization that is equivalent 
in the sense that it produces the 
exact same matrix elements \cite{cole}, 
but with the advantage that the couplings of the fermions with 
the Goldstone bosons do not contain derivatives, and we do not 
have to worry for high energy {\it gauge} cancellations.   The desired
form of the SM Lagrangian is given by
\begin{eqnarray}
{\cal L}^{(4)}_{SM} =&& {\ov {\Psi}_L} i 
\gamma^{\mu} D^L_{\mu} {{\Psi}_L} +
{\ov {\Psi}_R} i \gamma^{\mu} D^R_{\mu} {{\Psi}_R} - 
\left( {\ov {\Psi}_L} \Sigma M {{\Psi}_R} + h.c. \right) \nonumber \\
&&- {{1}\over {4}} W^a_{\mu\nu} W^{a\mu\nu} - {{1}\over {4}} B_{\mu\nu}
B^{\mu\nu} +  {{v^2}\over {4}} 
Tr\left( D_{\mu} \Sigma^{\dagger} D^{\mu} \Sigma 
\right)\;,\label{noder}\\ 
M =&&
\left(
\begin{array}{r}
m_t \;\;\;\; 0\\
 0\;\;\;\; m_b
\end{array}
\right)\; ,\nonumber \\
D^L_{\mu} =&& \partial_{\mu} - i g  {\tau^a \over 2} W^a_{\mu}- 
i g^{\hspace{.5mm}\prime} {Y\over 2} B_{\mu}\; , \nonumber \\
D^R_{\mu} =&& \partial_{\mu} -i g^{\hspace{.5mm}\prime}Q_f B_{\mu}\; .
\nonumber 
\end{eqnarray}
In the above equation, $Y=\frac{1}{3}$ is the hypercharge quantum
number for the quark doublet, $Q_f$ is the charge of the 
fermion, $\Psi_L$
is the linearly realized left handed quark doublet, and $\Psi_R$ is the 
right handed singlet for top or bottom quarks 
( see Eqs. (\ref{psi}) and (\ref{psr}) ).
As we have just said, the advantage of using this parameterization 
for  the {\it no-Higgs} SM case is that once we have made the 
ET substitution in the diagrams all the high energy cancellations 
due to gauge boson self-interactions 
(that have nothing to do with the symmetry breaking sector) will 
be already taken care of, and we will thus be evaluating directly 
the high energy behavior of a SM with no Higgs boson.          

\begin{figure}
\centerline{\hbox{
\psfig{figure=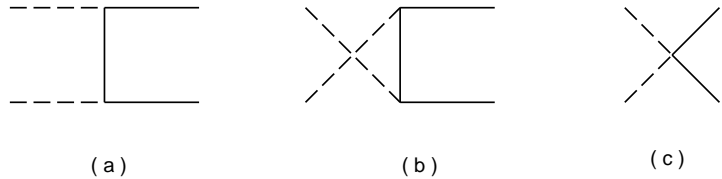,height=1in}}}
\caption{The corresponding Goldstone boson diagrams for 
$Z_L Z_L \rightarrow t\bar {t}$, i.e. $\phi^0\phi^0\ra t\ov t$.}
\label{power}
\end{figure}

When we expand the $\Sigma$ matrix field up to the second power 
( see Eq.~(\ref{sigfield}) ) in the mass term of Eq.~(\ref{noder}), 
we will notice two things: ({\it i}) 
the first power term gives the usual mass 
term and associates the coefficient $c_T\;=\; m_t / v$ to each vertex; 
({\it ii}) the second power term generates the four-point diagram
(Figure ~\ref{power}(c)) 
with a coefficient  $c_T\;=\; m_t^2 / v^2$ associated to its vertex. 
As it is well known, a mass term always involves 
a chirality flip, therefore we readily 
recognize that this diagram will  
only participate when the chiralities 
of the top and anti-top are different.   
 Since in the high energy regime the 
mass is much smaller than the energy 
of the fermions, different chiralities mean equal helicities
(for the particle-antiparticle pair).  Hence, for the case of opposite
helicities we only count the power dependance for diagrams  2(a) and
2(b), and take the highest one.  
For final state fermions of equal helicities
we consider all three diagrams. 

The results are the following:  for diagrams  2(a) and 
2(b) we have $D_{E0}=2+(-1)+(-1)=0$ ,  
thus the amplitude $T_{\pm \mp}$ is of order $ m_t^2/v^2$; which is 
the contribution given by the coefficients $c_T$ from both vertices.  
On the other hand, diagram  2(c) has $D_{E0}=2-1=1$;  
the equal helicities amplitude  $T_{\pm \pm}$ will be driven by this 
dominant diagram, therefore  $T_{\pm \pm} \; = \; m_t E/v^2$. 

For the other processes; $W^{+}_L W^{-}_L \rightarrow t \bar {t}$  and 
$W^{+}_L Z_L \rightarrow t \bar {b}$, the analysis is the same, except 
that there is an extra {\it s-channel} diagram 
(see Figures~\ref{fww} and
~\ref{fwz} ) that usually behaves just like the four point diagram  
~\ref{power}(c).

For the dimension 5 anomalous operators 
we do not expect {\it a priori} any 
{\it gauge} cancellations at high $E$. We therefore expect the 
parameterization used for our effective operators to reflect  
this high energy behavior.  
Actually, the chiral lagrangian parameterization 
given by Eq.~(\ref{eq2}), 
which organizes the new physics effects in the momentum expansion
\cite{georgi,howard}, is the only framework that allows the existence
of such dimension 5 gauge invariant operators.   On the other hand,
we know that as far as the 
 {\it no-Higgs} SM contribution to these {\it anomalous} 
amplitudes is concerned, it is better to use the equivalent 
parameterization of Eq.~(\ref{noder}), because the right high energy 
contribution from this model is made easily and consistently.  
We will therefore use the appropriate 
couplings from ${\cal L}^{(4)}_{SM}$ 
and ${\cal L}^{(5)}$ in our next power counting analysis.  

In principle,  we can evaluate the contribution to the diagrams 
2(a) and 2(b)  when both vertices are anomalous, 
but this would be suppressed by two powers of the cut-off scale; 
we will just ignore it.    We are only 
considering the contribution from one dimension 5 coupling at a time.

As an example, let us take the operator 
with derivative on fermion $O_{{\cal Z}D f }$ and apply the expansions 
of the composite fields:
\begin{eqnarray}
O_{{\cal Z}D f } =&& a_{z3} i 
\bar t D_{\mu} t {\cal Z}^{\mu} \nonumber \\
=&& -\frac{g}{c_w} a_{z3} i {\ov {\psi}_t} {\partial}_{\mu} 
{\psi}_t Z^{\mu} + \frac{2}{v} a_{z3} i {\ov {\psi}_t} {\partial}_{\mu} 
{\psi}_t \partial^{\mu} {\phi}^3 + \cdot \cdot \cdot \label{ogzexp}
\end{eqnarray}
Where $\psi_t$ denotes the usual linearly realized top quark field, and 
$\cdot \cdot \cdot$ includes the photon field of the covariant 
derivative, the higher (than zeroth) order terms from the spinor field 
expansion, and so on.  The important 
thing to notice is that the second  
term in the right hand side of Eq.~(\ref{ogzexp}) is the only one to 
consider, since all the 
others either have a lower energy dependance (less derivatives) or 
do not even represent the effective vertices relevant to the 
process of interest.  Without having to perform similar expansions 
for all the other operators we can infer that the important effective 
vertex will always contain two derivatives; the coefficients $c_T$ will 
be $ \frac{2}{v} a_{O}$ for the 
three-point operators with one derivative 
on the Goldstone boson field, and $\frac{4}{v^2} a_{O}$ for four-point 
operators (with two derivatives on the Goldstone boson fields).  
Nevertheless, two operators are an exception to this rule: 
$O_{\sigma D {\cal Z}}$  and   $O_{\sigma D {\cal W} L(R)}$.  
For these operators the {\it leading} term vanishes; let us expand the 
second one (for example):
\begin{eqnarray}
a_{w2L(R)}\; \bar t_{L(R)}\sigma^{\mu \nu} b_{R(L)}\; 
D_{\mu} {\cal W}^{+}_{\nu}
=&& -g \; a_{w2L(R)} \; {\ov 
{\psi}_{tL(R)}}\sigma^{\mu \nu} {\psi}_{bR(L)}\;  
{\partial}_{\mu} W^{+}_{\nu} \nonumber \\
+&& \frac{2}{v}\; a_{w2L(R)} \; {\ov {\psi}_{tL(R)}}\sigma^{\mu \nu} 
{\psi}_{bR(L)} {\partial}_{\mu} \partial_{\nu} {\phi}^{+}\; + 
\cdot \cdot \cdot \nonumber
\end{eqnarray}
As shown, we have the important term vanishing 
because of the contraction 
between the antisymmetric $\sigma^{\mu \nu}$ and the symmetric 
${\partial}_{\mu} \partial_{\nu} {\phi}^{+}$.    Hence, this 
operator will not 
contribute to the leading high energy behavior 
for top quark productions via 
$V_L V_L$ fusion at tree level.  It 
can only contribute to the part of the 
amplitudes that vanish as $g\ra 0$.
   
As in the previous case of the 
{\it no-Higgs} SM, we expect a distinction 
between the  $T_{\pm \mp}$ and  $T_{\pm \pm}$ amplitudes.   
Actually, the situation is the same except for 
the fact that the anomalous vertex generated by the anomalous 
operators will yield a  $(d_n+\frac{1}{2}f_n-2) = 1$ 
factor, whereas the  
dimension 4 SM operators yield a ($-1$) factor.  
Therefore, $D_{E0}=2+1+(-1)=2$ for the first 
two diagrams 2(a) and 2(b) and 
thus $T_{\pm \mp}$ is of expected to be of order
\begin{eqnarray}
T_{\pm \mp} \sim {2 a_O}{{m_t}\over {v}}  
{{v}\over {\Lambda}} {\left( {E}\over {v}\right)}^2\; .\nonumber 
\end{eqnarray} 
On the other hand, diagram 2(c) may be generated by either 
the direct contribution of four-point operators 
or the contribution from 
three-point operators when considering higher order 
terms in their expansion. 
In each case the anomalous operators contribute with a 
$(d_n+\frac{1}{2}f_n-2)=1$ factor which yields $D_{E0}=2+1=3$,  
and the predicted value for $T_{\pm \pm}$ is
\begin{eqnarray}
T_{\pm \pm} \sim {4 a_O}  
{{v}\over {\Lambda}} {\left( {E}\over {v}\right)}^3\; .\nonumber 
\end{eqnarray} 
 
Naturally, this power counting formula can not predict the fact 
that sometimes an amplitude can be zero due to the different 
helicities of spinors.  For instance, 
by performing the calculation of the amplitudes with 
external gauge bosons in the CM frame we 
can easily verify that the product of the spinors 
${\ov u_t}[\lambda_t =\pm1] v_{\ov t} [\lambda_{\ov t} =\mp1]$ 
vanishes\footnote{${u_t}[\lambda_t =+1]$ denotes the spinor of a top 
quark with right handed helicity.}.   
This means that contributions from operators 
of the {\it scalar}-type, like 
$O_{g{\cal Z}{\cal Z}}$,  $O_{g{\cal W}{\cal W}}$,  
$O_{{\cal Z}D f }$,  $O_{g{\cal W}{\cal Z} L (R)}$,  and  
$O_{{\cal W}D t R (L)}$ will vanish for $T_{\pm\mp}$ amplitudes. 
Also, we have the relation ${\epsilon}_{\mu} p^{\mu} = 0$ applicable 
to all three polarizations of external on-shell boson lines, which  
will make the contribution of operators with derivative on boson 
and scalar Lorentz contraction like  $O_{g D {\cal Z}} $  and  
$O_{g D {\cal W}L(R)}$ to vanish in {\it t-channel} and  
{\it u-channel} diagrams (Figure~\ref{fzz}).  In principle, one would
think that the exception could be the {\it s-channel} type diagram.  
Actually, this is the case for the operator $O_{g D {\cal W}L(R)}$
which is able to contribute significantly on the single top production 
process $W^{+}_L Z_L \rightarrow t {\bar {b}}$  
(see Table~\ref{opsderb}).  However, for the 
 $O_{g D {\cal Z}} $ operator even this diagram vanishes; as can 
be easily verified by performing the calculation in the CM frame. 
 One will see that  the result of 
making the Lorentz contraction between the boson propagator 
$-g_{\mu\nu}+k_{\mu} k_{\nu} / M_Z^2$ and the tri-boson coupling 
is identically zero in the process 
$W^{+}_L W^{-}_L \rightarrow t {\bar {t}}$.   
Therefore, for the $O_{g D {\cal Z}} $ operator all the 
possible Feynman diagrams vanish, so it does not contribute 
to the $t\ov t$ production rate.

In Tables \ref{ops4}, \ref{opsderf}, and \ref{opsderb} we 
show the leading contributions (in powers of the CM energy $E$) 
of all the operators for each different process; those cells 
with a dash mean that no anomalous vertex generated by 
that operator intervenes in the given process, and those cells 
with a zero mean that the anomalous vertex intervenes in the 
process but the amplitude vanishes for any of the reasons 
explained above.   

\begin{table}[htbp]
\begin{center}
\vskip -0.06in
\begin{tabular}{|l||c||c||c||c||c||c|} \hline \hline
Process & ${\cal L}^{(4)}$ & $O_{g{\cal Z}{\cal Z}}$
&$O_{g{\cal W}{\cal W}}$ &
$O_{\sigma {\cal W}{\cal W}}$ & $O_{g{\cal W}{\cal Z} L (R)} $ &
 $O_{\sigma {\cal W}{\cal Z} L (R)}$  \\ 
$\;$ & $\;$  & $a_{zz1}\times$ & $a_{ww1}\times$ &  $a_{ww2}\times$ &  
$a_{wz1L(R)}\times$ &  $a_{wz2L(R)}\times$ \\
 \hline\hline
$Z_L Z_L \rightarrow t \bar {t}$ & $m_t {E / {v^2}}$ & 
${E^3 / {v^2}{\Lambda}}$ & $-$& 
$-$ & $-$& $-$ \\  \hline
$W^{+}_L W^{-}_L \rightarrow t \bar {t}$  & $m_t {E / {v^2}}$ &$-$ & 
${E^3 / {v^2}{\Lambda}}$ & 
${E^3 / {v^2}{\Lambda}}$ & $-$ & $-$ \\ \hline
$W^{+}_L Z_L \rightarrow t \ov b$ & 
${m_t^2  / {v^2}}$ &$-$ & $-$ & $-$ & 
${E^3 / {v^2}{\Lambda}}$ & ${E^3 / {v^2}{\Lambda}}$ \\ \hline \hline
\end{tabular}
\end{center}
\vskip 0.08in
\caption{The leading high energy terms for the 4-point operators.}
\label{ops4}
\end{table}

\begin{table}[htbp]
\begin{center}
\vskip -0.06in
\begin{tabular}{|l||c||c||c||c|}\hline \hline
Process &   ${\cal L}^{(4)}$ &
$O_{{\cal Z}D f } $ & $O_{{\cal W}D t R }$ &  $O_{{\cal W}D t L}$ \\ 
$\;$  & $\;$  &  $a_{z3}\times$ & $a_{w3R}\times$ & $a_{w3L}\times$ 
\\ \hline\hline
$Z_L Z_L \rightarrow t \bar {t}$ & $m_t {E / {v^2}}$ & 
${E^3 / {v^2}{\Lambda}}\;$& $-$ & $-$ \\ \hline
$W^{+}_L W^{-}_L \rightarrow t \bar {t}$ & $m_t {E / {v^2}}$ & 
${E^3 / {v^2}{\Lambda}}$ & ${E^3 / {v^2}{\Lambda}}\;$ & 
${{m_b E^2} / {v^2}{\Lambda}}\rightarrow 0$  \\ \hline 
$W^{+}_L Z_L \rightarrow t \ov b$ & ${m_t^2 /{v^2}}$ & 
${E^3 /{v^2}{\Lambda}}\;$ & 
${E^3 / {v^2}{\Lambda}}$ & ${E^3 / {v^2}{\Lambda}}$  \\ \hline \hline
\end{tabular}
\end{center}
\vskip 0.08in
\caption{The leading high energy terms for the operators 
with derivative-on-fermion.}
\label{opsderf}
\end{table}

\begin{table}[htbp]
\begin{center}
\vskip -0.06in
\begin{tabular}{|l||c||c||c||c||c||c|} \hline \hline
Process & ${\cal L}^{(4)}$ & $O_{g D {\cal Z}}$&$O_{g D {\cal W}L(R)}$&
$O_{\sigma D {\cal Z}}$&
$O_{\sigma D {\cal W} L(R)} $& $O_{ {\cal A}}$ \\
$\;$ & $\;$  & $a_{z4}\times$ & 
$a_{w4}\times$ &  $a_{z2}\times$ &  
$a_{w2}\times$ &  $a_{m}\times$ \\ \hline\hline
$Z_L Z_L\rightarrow t \bar {t}$ & $m_t {E / {v^2}}$ & 
$0$ & $-$& 
$g^2 E/{\Lambda}$ & $-$& $-$ \\  \hline
$W^{+}_L W^{-}_L \rightarrow t \bar {t}$  & $m_t {E / {v^2}}$ &$0$ & 
$0$ & ${E^3 / {v^2}{\Lambda}}$ & 
$g^2 E/{\Lambda}$ & ${E^3 / {v^2}{\Lambda}}$ \\ \hline
$W^{+}_L Z_L \rightarrow t \ov b$ & ${m_t^2  / {v^2}}$ &$0$ & 
$E^3/v^2{\Lambda}$ & $g^2 E/{\Lambda}$ & 
${E^3 / {v^2}{\Lambda}}$ & $-$ \\ \hline \hline
\end{tabular}
\end{center}
\vskip 0.08in
\caption{The leading high energy terms for the operators 
with derivative-on-boson.}
\label{opsderb}
\end{table}

In conclusion, based on the NDA \cite{georgi,howard}
and the power counting rule \cite{poc}, 
we have found that the leading high 
energy behavior in the $V_L V_L 
\ra t\ov t\;or\;t\ov b$ scattering amplitudes 
from the {\it no-Higgs} SM operators (${\cal L}^{(4)}_{SM}$)
can only grow as $\frac{m_t E}{v^2}$ 
(for $T_{++}$ or $T_{--}$, $E$ is 
the CM energy of the top quark system), 
whereas the contribution from the dimension 
5 operators (${\cal L}^{(5)}$) 
can grow as $\frac{E^3}{v^2\Lambda}$ in the 
high energy regime.   Let us 
compare the above results with those 
of the $V_LV_L \ra V_LV_L$ scattering 
processes.  For these $V_LV_L \ra V_LV_L$ 
amplitudes the leading behavior 
at the lowest order gives $\frac{E^2}{v^2}$, and 
the contribution from the 
next-to-leading order (NLO) bosonic operators gives 
$\frac{E^2}{\Lambda^2}\frac{E^2}{v^2}$ \cite{poc}.   This indicates 
that 
the NLO contribution is down by a factor of $\frac{E^2}{\Lambda^2}$ in 
$V_LV_L \ra V_LV_L$.   On the 
other hand, the NLO fermionic contribution 
in $V_L V_L \ra t\ov t\;or\;t\ov b$ is only down by a factor 
$\frac{E^2}{m_t \Lambda}$ which compared to 
$\frac{E^2}{\Lambda^2}$ turns 
out to be bigger by 
a factor of $\frac{\Lambda}{m_t}\sim 4\sqrt{2}\pi$ for 
$\Lambda \sim 4 \pi v$.   
Hence, we expect that the NLO contributions in 
the $V_L V_L \ra t\ov t\;or\;t\ov b$ processes can be better measured 
(by about a factor of $10$) than 
the $V_LV_L \ra V_LV_L$ counterparts for 
some class of electroweak symmetry breaking 
models in which the NDA gives 
reasonable estimates of the coefficients.  

As to be shown later, 
the coefficients of the NLO fermionic operators in 
${\cal L}^{(5)}$ can be determined via top quark production to order   
$10^{-2}$ or $10^{-1}$.   In contrast, the coefficients of the NLO 
bosonic operators  are usually determined to 
about an order of $10^{-1}$  
or $1$ \cite{et,sss} via $V_LV_L \ra V_LV_L$ processes.  
Therefore, we conclude that the top 
quark production via the longitudinal 
gauge boson fusions $V_L V_L \ra t\ov t\;or\;t\ov b$ 
at high energy may be more sensitive for probing some 
symmetry breaking mechanism than the scattering of longitudinal 
gauge bosons alone ($V_LV_L \ra V_LV_L$).

Our next step is to study the production rates of $t\ov t$ 
pairs and single-$t$ or single-$\ov t$ 
events at future colliders like LHC 
and LC.   We will also estimate how 
accurate these NLO fermionic operators 
can be measured via the $V_L V_L \ra t\ov t\;or\;t\ov b$ processes.

\subsection{Underlying custodial symmetry}
\indent\indent

To reduce the number of 
independent parameters in this study, we shall make the same
assumption of an underlying custodial symmetric theory that gets broken
in such a way that only the 
couplings that involve the top quark get modified;
as was done for the case of ${\cal{L}}^{(4')} $ 
( see Eq.~(\ref{custol4}) and the
discussion there ).  The analysis for the operators 
with derivatives is exactly
the same.  The custodial symmetric dimension 5 Lagrangian has the same
$SU(2)$ structure\footnote{Notice that the composite 
left and right handed
doublets $F_{L,R}$ transform in the same way under global
$SU(2)_R \times SU(2)_L$, $F_{L,R} \ra F_{L,R}^{'} = R F_{L,R}$ with
R in $SU(2)_R$.} as ${\cal L}^{(custodial)}$ in Eq.~(\ref{custi}):
\beq
{\cal L}^{(5deriv)}\;=\; \kappa^{(5)}_{1g} \overline{F_L}  g_{\mu\nu}
D_\mu {\cal W}_{\nu}^{a} \tau^a F_R \, + \, \kappa^{(5)}_{1\sigma}
\overline{F_L} \sigma^{\mu\nu} D_\mu {\cal W}_{\nu}^{a}
\tau^a F_R \; +h.c.\, ,
\enq
and the symmetry breaking term 
will also be similar to ${\cal L}^{(EWSB)}$
in Eq.~(\ref{4ewsb}).  Therefore the conclusion is 
the same, that in order to
keep the couplings $\bbz$ unaltered we have to conform to the matrix
structure of ${\cal L}^{(4')}$ in Eq.~(\ref{custol4}), and this 
will impose the
condition
\beq
a_{z(2,3,4)} = \sqrt{2}  a_{w(2,3,4)L(R)}\,  
\enq
to all the operators with derivatives.

For the case of 4-point operators 
the situation is somewhat different.  The
custodial Lagrangian in this case is of the form:
\begin{eqnarray}
{\cal L}^{(5custod)}=&&\kappa^{4pt.}_{1g} \overline{F_L} g^{\mu\nu}
{\cal W}_{\mu}^{a} \tau^a {\cal W}_{\nu}^{b} \tau^b F_R \; + \; 
\kappa^{4pt.}_{1\sigma} \overline{F_L}
\sigma^{\mu\nu} {\cal W}_{\mu}^{a} \tau^a {\cal W}_{\nu}^{b} \tau^b F_R
\nonumber \\
=&& \kappa^{4pt.}_{1g} \overline{F_L} g^{\mu\nu}
\left(
\begin{array}{r}
{\cal W}_{\mu}^{3} {\cal W}_{\nu}^{3} +
2 {\cal W}_{\mu}^{+} {\cal W}_{\nu}^{-} \;\;\;\;\;\;\;\;\;\;\;\; 0\\
 0 \;\;\;\;\;\;\;\;\;\;\;\; {\cal W}_{\mu}^{3} {\cal W}_{\nu}^{3} +
2 {\cal W}_{\mu}^{+} {\cal W}_{\nu}^{-}
\end{array}
\right) 
F_R  \nonumber \\
&& \qquad\qquad 
  + \kappa^{4pt.}_{1\sigma} \overline{F_L} \sigma^{\mu\nu}
\left(
\begin{array}{r}
2  {\cal W}_{\mu}^{+} {\cal W}_{\nu}^{-} \;\;\;\;\;\;\;\;\;\;\; 0\\
 0 \;\;\;\;\;\;\;\;\;\;\;2 {\cal W}_{\mu}^{+} {\cal W}_{\nu}^{-}
\end{array}
\right) 
F_R \, ,
 \label{custod5}
\end{eqnarray}
and for the symmetry breaking Lagrangian we can consider two terms:
\begin{eqnarray}
{\cal L}^{(5EWSB)}=&& \sum_{c=g,\sigma}
c^{\mu\nu} \,\left( \;\; \kappa^{4pt.}_{2c} \overline{F_R} \tau^3
{\cal W}_{\mu}^{a} \tau^a {\cal W}_{\nu}^{b} \tau^b F_L\,+\,
\kappa^{4pt.\dagger}_{2c} \overline{F_L} {\cal W}_{\mu}^{a} \tau^a
{\cal W}_{\nu}^{b} \tau^b \tau^3 F_R \right. \nonumber \\
&& \qquad\qquad 
+ \left. \kappa^{4pt.}_{3c} \overline{F} {\cal W}_{\mu}^{a} 
\tau^a \tau^3
{\cal W}_{\nu}^{b} \tau^b F \; \; \right) 
\end{eqnarray}
where $\kappa^{4pt.}_{3c}$ is real and $\kappa^{4pt.}_{2c}$ is complex.
As it turns out, in order to set the anomalous couplings of the bottom
quark equal to zero, we have to choose $\kappa^{4pt.}_{3c}=0$, and
$\kappa^{4pt.}_{2c}$ real and half the size of $\kappa^{4pt.}_{1c}$
($\kappa^{4pt.}_{1c}=2 \kappa^{4pt.}_{2c}$ for $c=g,\sigma$). The
non-standard 4-point dimension 5 interactions will then have the
structure
\beq
\left(
\begin{array}{r}
c^{\mu\nu} {\cal W}_{\mu}^{3} {\cal W}_{\nu}^{3} +
2 c^{\mu\nu} {\cal W}_{\mu}^{+} {\cal W}_{\nu}^{-} 
\;\;\;\;\; 0\\
 0 \;\;\;\;\;\;\;\;\;\;\;\;\;\;\;\;\;\;\;\;\;\;\;\; 0
\end{array}
\right) 
\enq
where $c^{\mu\nu}$ is either $g^{\mu\nu}$ or $\sigma^{\mu\nu}$.

In conclusion, by assuming that our 
dimension 5 interactions are the result
of an underlying custodial symmetric theory that is broken in such a way
that only the couplings of the top 
quark get modified from the SM values,
we can impose the following conditions to the effective coefficients:
\begin{eqnarray}
a_{z(2,3,4)} =&& \sqrt{2}  a_{w(2,3,4)L(R)}\; , \nonumber \\
2 a_{zz1}  =&& \, a_{ww1}\; , \\
a_{wz1L(R)} =&& a_{wz2L(R)} = 0\; . \nonumber
\end{eqnarray}

\subsection{ Production rates for 
$Z_L Z_L$, $W_L W_L$, and $W_L Z_L$ fusion processes}
\indent\indent

Below, we present the helicity amplitudes for each process.  
We have simplified our analysis as much as possible by considering only 
the leading terms in powers of $E$, and by assuming an approximate 
$\rm{SU(2)}$ custodial symmetry.  
 As a rule,  the next to leading contribution in the $E$ expansion is 
always two powers down as compared to the leading contribution.  
Also, the special case of custodial symmetry is assumed in this study;
the amplitudes for the most general case are presented elsewhere
\cite{top5}.   

\subsubsection{$Z_L Z_L \rightarrow t\bar{t}$}
\indent\indent

Comparing with the results for $W_L W_L$ and  $W_L Z_L$ fusions, 
this is the amplitude that takes the simplest form 
with no angular dependance. 
This means that any new physics effects coming through 
this process only modify the S-partial 
wave amplitude.   The notation for 
the amplitudes indicates the helicity of 
the outgoing fermions: the first 
(second)  symbol ($+$ or $-$) refers 
to the fermion on top (bottom) part of 
the diagram.  A right handed fermion is 
labelled by '$+$', and a left handed 
fermion by '$-$'\footnote{For example, the anomalous amplitude
$azz_{++}$ stands for the anomalous contribution to the amplitude for
the production of right handed $t$ and $\ov t$ via $Z_L Z_L$ fusion.}.  
Figure~\ref{fzz} shows the diagrams that contribute to this process.  
We take only one anomalous vertex at a time. 

\begin{figure}
\centerline{\hbox{
\psfig{figure=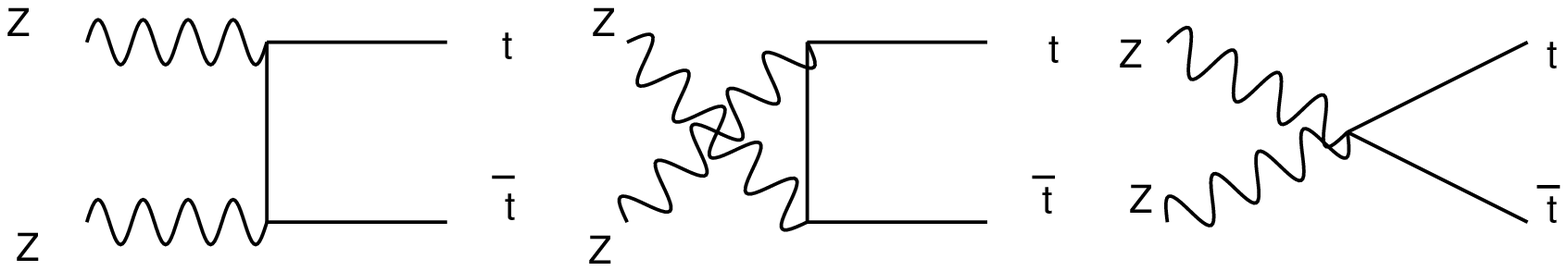,height=1in}}}
\caption{Diagrams for the $Z Z \rightarrow t\bar{t}$ process.}
\label{fzz}
\end{figure}

The leading contributions to the various helicity amplitudes from the 
dimension 5 operators are 
($E\,=\,\sqrt{s}$ is the CM energy of the $V_L V_L$
system):

\begin{eqnarray}
azz_{++} =&& -azz_{--} \;\;\;=\;\; 
 -{{{{\it E}^3}}\over {{v^2}}} {X\over {\Lambda}}\; , \nonumber \\
azz_{+-} =&& azz_{-+} \;\;\;=\;\;0 \; ,
\label{azz}
\end{eqnarray}
where
\begin{equation}
X = {\it a_{zz1}}+\left({1\over 2}-{4\over 3} 
s^2_w\right){\it a_{z3}}\; .
\label{X}
\end{equation}

Notice how at this 
stage it is impossible to distinguish the effect of the 
coefficient ${\it a_{zz1}}$ from the effect 
of the coefficient ${\it a_{z3}}$.  
However, in the next section we will show how we can still combine this 
information with the results of the other processes, 
and obtain bounds for each coefficient.

\subsubsection{ $W^{+}_L W^{-}_L \rightarrow t\bar{t}$}
\indent\indent

The amplitudes of this 
process are similar to the ones of the previous process 
except for  the presence of two $s-channel$ diagrams  
(see Figure~\ref{fww}), whose off-shell $\gamma$ and $Z$  
propagators allow 
for the contribution from the magnetic moment of the top quark and the 
operator with derivative on boson 
$O_{\sigma D {\cal Z}}$ (${\it a_{z2}}$),   
respectively.   Also, since these two operators 
are not of the $scalar$-type, 
we have a non-zero contribution to the $T_{\pm\mp}$ amplitudes, and an 
angular dependance that will help in distinguishing the effect of their 
coefficient $X_m$ from the effect of the coefficient for the $scalar$ 
operators  $X^{'}$.   Throughout this study, the angle of scattering 
$\theta$ in all processes is 
defined to be the one subtended between the center of mass momentum of 
the incoming gauge boson that appears 
on the top-left part of the Feynman 
diagram ($W^{+}$ in this case) and 
the momentum of the outgoing fermion  
appearing on the top-right part of the same diagram ($t$ in this case). 

\begin{figure}
\centerline{\hbox{
\psfig{figure=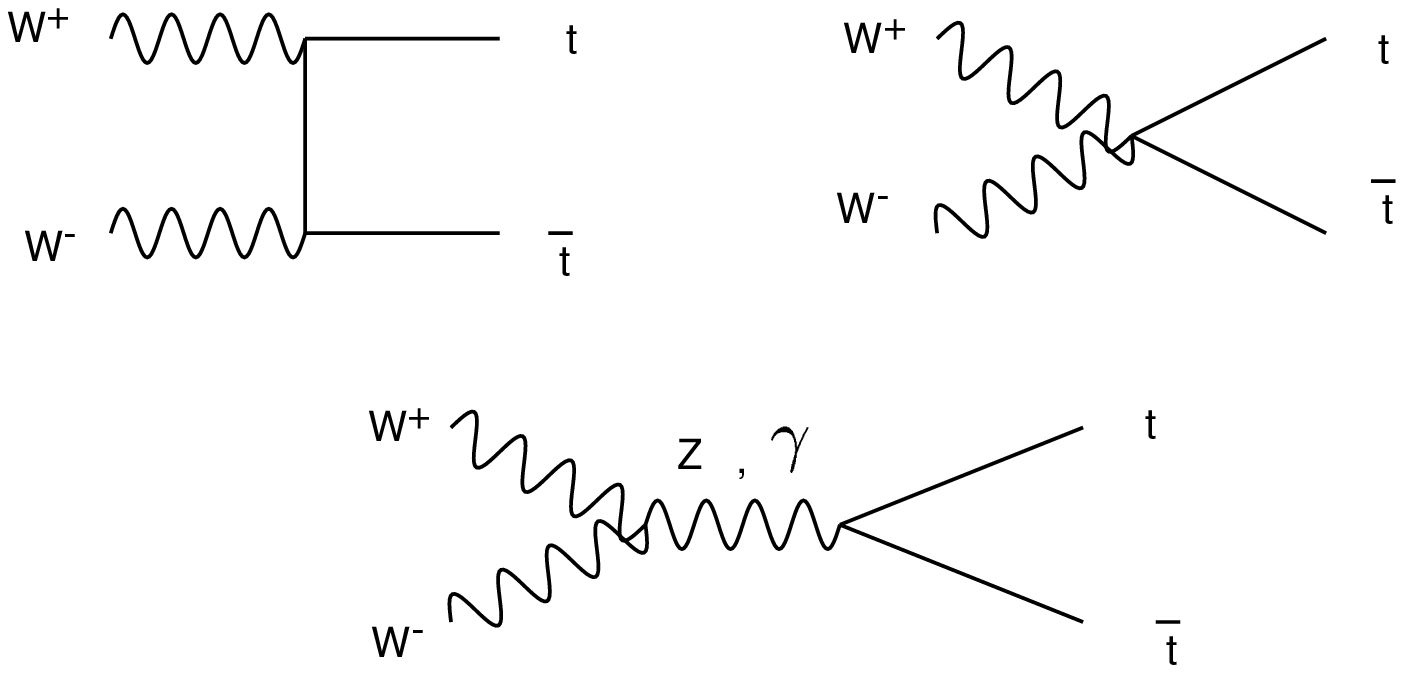,height=2in}}}
\caption{Diagrams for the $W W \rightarrow t\bar{t}$ process.}
\label{fww}
\end{figure}

The leading contributions to the various helicity amplitudes for this 
process from the dimension 5 operators are:

\begin{eqnarray}
aww_{++} =&& -aww_{--} \;\;\;=\;\; {{-2 \,{{\it E}^3}}\over {v^2}} 
{{\left(X^{'} +2 X_m c_{\theta}\right)}\over {\Lambda}}\; , \nonumber \\
aww_{-+} =&& {{8{\it E}^2} \over {v^2}} m_t s_{\theta} 
{{X_m-\frac{1}{8}{\it a_{z3}}}\over {\Lambda}} \; ,\nonumber\\
aww_{+-} =&& {{{8{\it E}^2} \over {v^2}} m_t s_{\theta}\,
{{\left( X_m - \frac{1}{4}{\it a_{z3}} \right)}\over{\Lambda}}}\; ,
\end{eqnarray}
where
\begin{eqnarray}
X^{'}=&& {\it 2 a_{zz1}}+ \frac{1}{4}{\it a_{z3}} \; ,\nonumber\\
X_m=&& {\it a_{m}} - \frac{1}{2}{\it a_{z2}}
+ \frac{1}{4}{\it a_{z3}}+ \frac{1}{2}{\it a_{ww2}}\; .\label{Xprime}
\end{eqnarray}

Notice that the 
angular distribution of the leading contributions in the 
$T_{\pm\pm}$ amplitudes consists of the flat component (S-wave) and the 
$d^1_{0,0}=\cos {\theta}$ component (P-wave).   The $T_{\pm\mp}$  
helicity amplitudes only contain the 
$d^1_{0,\pm 1}=-\frac{\sin {\theta}}{\sqrt{2}}$ component.  
This is so because the initial 
state consists of longitudinal gauge bosons 
and has zero helicity. 
The final state is a fermion pair so 
that the helicity of this state can be 
$-1$,  $0$, or $+1$.   Therefore, in 
high energy scatterings the anomalous 
dimension 5 operators only modify the leading 
contributions on the S-type and 
P-type partial waves of the scattering amplitudes.   
We also note that, as 
expected, $T_{\pm\pm}$ has an $E^3$ leading behavior, 
whereas $T_{\pm\mp}$ 
only has an $E^2$ contribution.

\subsubsection{$W^{+}_L Z_L \rightarrow t\bar{b}$}
\indent\indent

Finally, we have the amplitudes for 
the single-top quark production process
$W^{+}Z \rightarrow t\bar{b}$ (which are just the 
same as for the conjugate
process $W^{-}Z \rightarrow b\bar{t}$).  Figure~\ref{fwz} shows 
the diagrams that
participate in this process.

\begin{figure}
\centerline{\hbox{
\psfig{figure=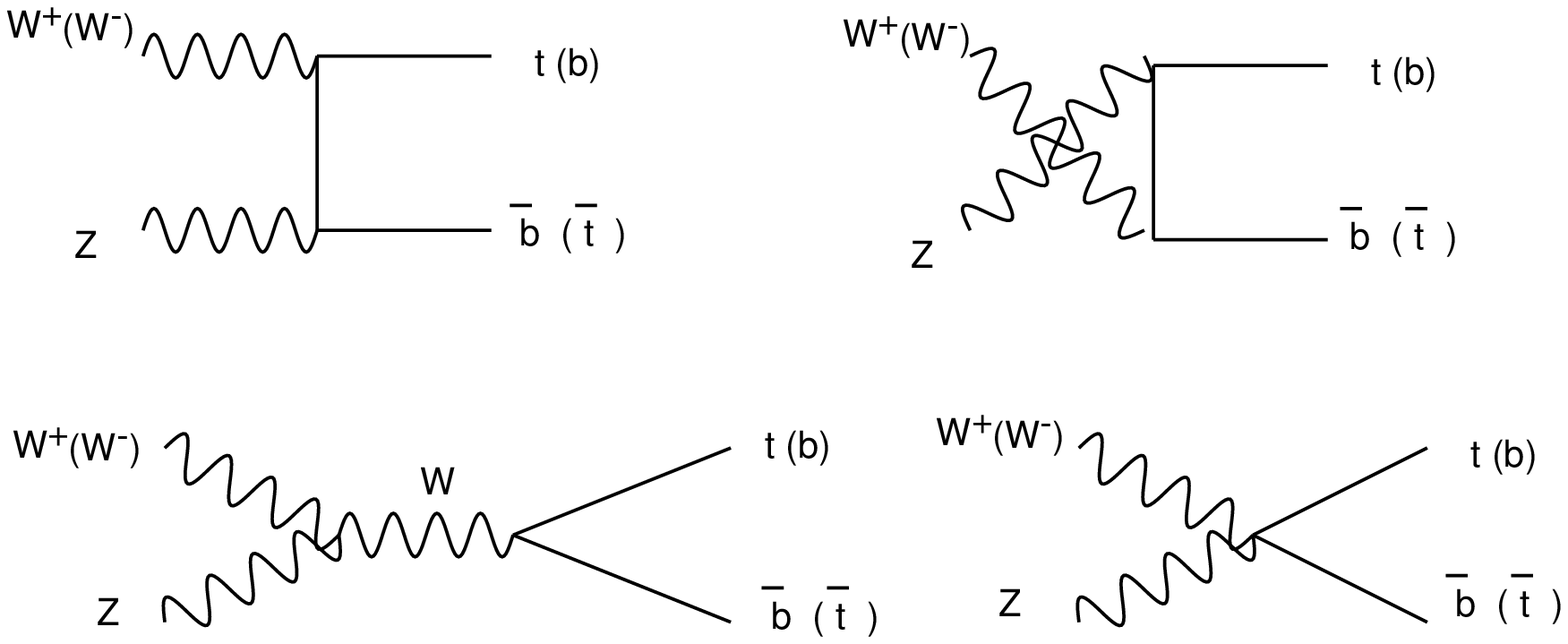,height=2in}}}
\caption{Diagrams for the $W Z \rightarrow t\bar{b}$ process.}
\label{fwz}
\end{figure}

The leading contributions to the various helicity amplitudes for this 
process from the dimension 5 operators are\footnote{For the approximate
custodial symmetry case, the 4-point 
vertex diagram does not contribute.}:

\begin{eqnarray}
awzt_{++} =&& {2 \sqrt{2} {\it E}^3 \over {v^2}} 
{{\left( X_1 - X_2 c_{\theta} - 
(\frac{2 s^2_w}{3} -1-c_\theta) X_3 \right)}
\over {\Lambda}}\; ,\nonumber\\
awzt_{--} =&& {-2 \sqrt{2} {\it E}^3 \over {v^2}} 
{{\left( X_1 - X_2 c_{\theta}-(\frac{2 s^2_w}{3}\;+c_{\theta}-1) 
X_3 \right)}\over {\Lambda}}\; ,\nonumber\\
awzt_{+-} =&& {{2\sqrt{2} {\it E}^2}\over {v^2}} m_t s_{\theta}\,
{{\left( X_3+  X_2 \right)}\over {\Lambda}} \; ,\nonumber\\
awzt_{-+} =&& {{2\sqrt{2} {\it E}^2}\over {v^2}} m_t s_{\theta}\,
{{\left( X_2 -3 X_3 \right)}\over {\Lambda}}\; ,\label{awz}
\end{eqnarray}
where
\begin{eqnarray}
X_1 =&& {1 \over 2}  s^2_w a_{z4}  \; ,\nonumber\\
X_2 =&& {1 \over 2}  c^2_w  a_{z2} \; , \nonumber\\
X_3 =&& {1 \over 8}  a_{z3} \, .
\label{X123}  
\end{eqnarray}

 Here, as in the previous case, we 
can distinguish the effect from 
the three coefficients $X_1$, $X_2$, and $X_3$ 
by looking at their different angular 
contribution; $X_1$'s is flat, $X_2$'s and 
$X_3$ is in part through $s_{\theta}$ 
and in part through $c_{\theta}$.  

In order to simplify 
our numerical analysis we have made the approximation 
$aww_{+-}\simeq aww_{-+}$.  We have 
verified numerically that the error thus  
introduced, is of about 
$5 \%$ or less; depending on the difference between 
$X_m$ and ${\it a_{z3}}$ or 
the value chosen for the angle of scattering 
$\theta$.

\subsection{Top quark production rates from $V_L V_L$ fusions}
\indent\indent

As discussed above, the top 
quark productions from $V_L V_L$ fusion processes 
can be more sensitive to 
the electroweak symmetry breaking sector than the 
longitudinal gauge boson productions 
from $V_L V_L$ fusions.   In this section 
we shall examine the possible increase 
(or decrease) of the top quark events 
at the future hadron collider LHC 
(a ${\rm pp}$ collider with $\sqrt{s}=14$ TeV and 
$100 \; {\rm {fb}}^{-1}$ of integrated luminosity) and the 
future electron linear collider LC 
(an $e^{-} e^{+}$ collider with $\sqrt{s}=1.5$ TeV and $200 \; 
{\rm {fb}}^{-1}$ of integrated luminosity)\footnote{This is 
another energy phase of the proposed LC.}.  

To simplify 
our discussion we shall assume an approximate custodial symmetry 
and use the 
helicity amplitudes given in the previous section to compute the 
production rates for 
$t\ov t$ pairs and for single-$t$ or $\ov t$ quarks.  We 
shall adopt 
the effective-W approximation method \cite{effw} and use the 
CTEQ3L parton 
distribution function with the factorization scale chosen to 
be the mass 
of the $W$-boson \cite{cteq3}.    For this study we do not intend 
to do a 
detailed Monte Carlo simulation for the detection of the top quark; 
therefore, we shall only impose 
a minimal set of cuts on the produced $t$ or 
$b$.   The rapidity 
of $t$ or $b$ produced from the $V_L V_L$ fusion process 
is required 
to be within $2$ (i.e. $|y^{t,b}|\leq 2$) and the transverse 
momentum of $t$ 
or $b$ is required to be at least $20$ GeV.  To validate the 
effective-$W$ approximation, 
we also require the invariant mass $M_{VV}$ to 
be larger than $500$ GeV.

Since we are working in the high energy regime $E\gg v$, the leading 
contributions (proportional to $E^3$) to the 
$V_L V_L \ra t\ov t\;or\;t\ov b$ scattering amplitudes that come from 
the dimension 5 operators in ${\cal {L}}^{(5)}$ and the leading 
contributions (proportional to $E^1$) from the {\it no-Higgs} SM 
Lagrangian ${\cal {L}}^{(4)}_{SM}$ become a very good approximation.

It is apparent 
from the helicity amplitudes listed in the previous section 
that the top quark production rates from $V_L V_L$ fusion depend on a 
few independent dimension 5 
operators.  For instance, for $W^{\pm}_L Z_L$ 
fusion the production rates depend on the combined coefficients 
($X1, \;X2,\;X3$),  for $W^{+}_L W^{-}_L$ 
fusion depends on ($X^{'},\;X_m$), 
 and for $Z_L Z_L$ fusion only depends on ($X$). 

As noted before, 
in all the $T_{\pm\pm}$ amplitudes, the dimension 5 operators 
will only modify the constant term (S-wave) and the $\cos{\theta}$ 
(P-wave: $d^1_{0,0}$) dependence in 
the angular distributions for the leading 
$E^3$ contributions, whereas all the $T_{\pm\mp}$ amplitudes have a 
$\sin{\theta}$ (P-wave: $d^1_{0,\pm 1}$) dependence 
in their leading $E^2$ 
contributions.   In general, the contributions 
to these partial waves do not 
cancel.  Hence, let us examine 
the dependence of the top quark production 
rates as a function of the 
coefficients of the operators that only contribute 
the S-partial wave of 
the scattering amplitudes.  Namely, they are $X^{'}$, 
$X$ and 
$X_1$ for the $W^{+}_L W^{-}_L$, $Z_L Z_L$ and $W^{\pm}_L Z_L$ fusion 
processes respectively\footnote{In $W^{+}_L Z_L\ra t\ov b$, $X_3$ 
contributes to both, 
the S- and the P-partial waves.}.   The predicted top 
quark event rates as 
a function of these coefficients are given in Figures~\ref{flhc} 
and~\ref{flc} for the 
LHC and the LC, respectively.   In these plots, neither the 
branching ratio nor the detection efficiency have been included.

The {\it no-Higgs} SM event rates are given in Figures~\ref{flhc}
and~\ref{flc} for $X=0$. 
At the LHC, there are in total 
about 1500 $t\ov t$ pair and single-$t$ or 
$\ov t$ events predicted by the {\it no-Higgs} SM.   
The $W^{+}_L W^{-}_L$ fusion rate is 
about a factor of $2$ larger 
than the $Z_L Z_L$ fusion rate, and about an 
order of magnitude larger than the $W^{+}_L Z_L$ fusion rate.  
The $W^{-}_L Z_L$ rate, which is not 
shown here, is about a factor of $3$ 
smaller than the $W^{+}_L Z_L$ rate due 
to smaller parton luminosities at a 
${\rm pp}$ collider. 
The large slopes of the $W^{+}_L W^{-}_L \ra t{\ov t}$ and 
$Z_L Z_L \ra t\ov t$ curves, as a function of $X$ 
indicate that the scattering processes are sensitive 
enough to probe the 
anomalous couplings  $X^{'}$ and $X$ respectively.

\begin{figure}
\centerline{\hbox{
\psfig{figure=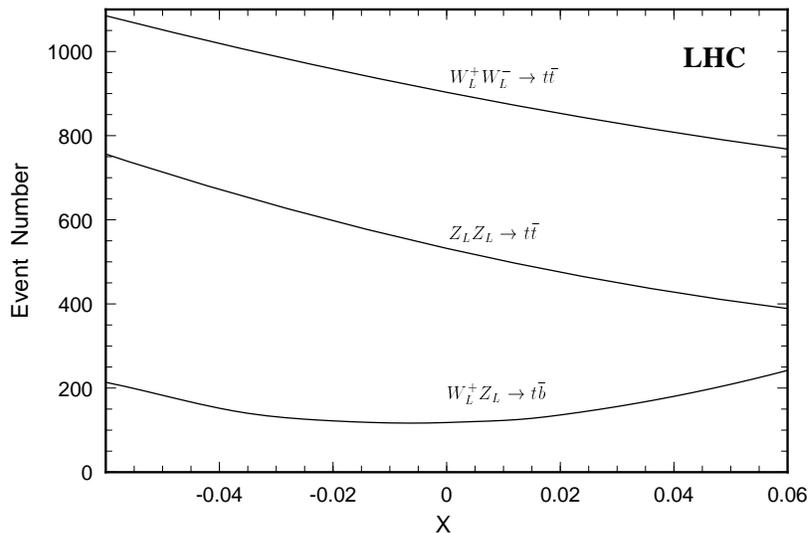,height=3in}}}
\caption{ Number of events at the LHC for  $W^{+}_L W^{-}_L, Z_L Z_L$
and $W^{+}_L Z_L$ fusion. The variable $X$ stands for the effective
coefficients $X$, $X^{'}$ and $X_1$
(Eqs.~(\protect\ref{X}),~(\protect\ref{Xprime}) and 
(\protect\ref{X123}) 
respectively).}
\label{flhc}
\end{figure}

\begin{figure}
\centerline{\hbox{
\psfig{figure=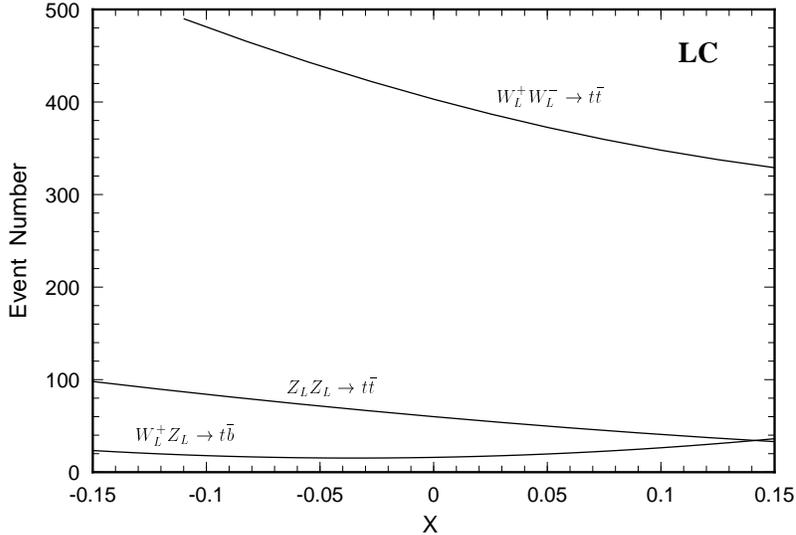,height=3in}}}
\caption{Number of events at the LC for $W_L W_L, Z_L Z_L$ and $W_L Z_L$
fusion. The variable $X$ stands for the effective coefficients
$X$, $X^{'}$ and $X_1$
(Eqs.~(\protect\ref{X}),~(\protect\ref{Xprime}) and 
(\protect\ref{X123}) respectively).}
\label{flc}
\end{figure}

For the LC, because of the 
small coupling of $Z$-$e$-$e$, the event rate for 
$Z_L Z_L \ra t\ov t$ is small.   For 
the {\it no-Higgs} SM, the top quark 
event rate at LC is about half of that at the LHC and yields 
about 550 $t\ov t$ pair and single-$t$ or $\ov t$ events.  
Again, we see that the 
$W^{+}_L W^{-}_L \ra t{\ov t}$  rate is sensitive 
to the dimension 5 operators that correspond to $X^{'}$,  
but the $Z_L Z_L \ra t\ov t$ 
rate is less sensitive\footnote{Needless to say, 
the $W^{-}_L Z_L$ rate is the same as the $W^{+}_L Z_L$ rate at an 
$e^{+}e^{-}$ LC}.   Since the 
detection of the top quark at the LC would be easier than at the LHC 
(i.e. the detection efficiency would be 
larger), we conclude that the LC 
and the LHC can have the same sensitivity for probing the NLO fermionic 
operators via the $W^{+}_L W^{-}_L \ra t{\ov t}$ process. 

The production rates shown in Figure~\ref{flc} are for an unpolarized
$e^{-}$ beam at the LC.  Assuming a longitudinally polarized $e^{-}$
beam at the LC,  the $W^{+}_L W^{-}_L \ra t{\ov t}$ rate will be doubled
because the coupling of the $W$ boson to the electron is 
purely left handed so the 
parton luminosity of the $W$ from the electron 
beam will be doubled if 
this beam is polarized.  However, this is not true 
for the parton luminosity of $Z$ because in this case the $Z$-$e$-$e$ 
coupling is nearly purely axial-vector ($1-4s^2_w \approx 0$) and the 
production rate of 
$Z_L Z_L \ra t\ov t$ does not strongly depend on whether 
the electron beam is polarized or not.  
As shown in Figure~\ref{flc} , if the 
coefficient of the anomalous dimension
5 operators is as large as ${{0.1}\over {\Lambda}}$ in magnitude 
then their effect can in principle\footnote{Of course, a complete study
including signal versus background, detector efficiency, etc., would be
necessary in order to confirm this.}
 be identified in the measurement of 
$W^{+}_L W^{-}_L$ 
fusion rate at the LHC and the LC.  A similar conclusion 
also holds for the $Z_L Z_L$ and $W^{\pm}_L Z_L$ fusion processes 
with somewhat less sensitivity.  
It is useful to ask for the bounds on the coefficients of the anomalous 
dimension 5 operators if the 
measured production rate at the LHC and the LC 
is found to be in agreement with the {\it no-Higgs} SM predictions  
(i.e. with $X=0$).   
At the $95\%$ C.L. we summarize the bounds on the $X$'s in 
Table \ref{bounds}. 
Here, only the statistical error 
is included.  In practice, after including 
the branching ratios of the relevant decay modes and the detection 
efficiency of the events, 
these bounds will become somewhat weaker, but we 
do not expect an order of 
magnitude difference.   Also, these bounds shall 
be improved by carefully analysing angular correlations when data is 
available. 

\begin{table}[htbp]
\begin{center}
\vskip -0.06in
\begin{tabular}{|l||c||c|} \hline \hline
Process &  LHC ( $pp$ ) &
 LC ( $e^{+}e^{-}$ )   \\ \hline\hline
$W^{+(-)}_L Z_L \rightarrow t \bar {b}\; (b \bar {t})$  & 
$-.035<X_1<.025$ & $-.13<X_1<.07$   \\ \hline
$W^{+(-)}_L Z_L \rightarrow t \bar {b}\; (b \bar {t})$ & 
$-.045<X_2<.10$ & $-.12<X_2<.35$  \\ \hline
$W^{+(-)}_L Z_L \rightarrow t \bar {b} \;(b \bar {t})$ & 
$-.19<X_3<.12$ & $-.65<X_3<.35$   \\ \hline
$W^{+}_L W^{-}_L \rightarrow t \bar {t}$ & $-.022<X^{'}<.017$ &  
$-.06<X^{'}<.07$  \\ \hline
$W^{+}_L W^{-}_L \rightarrow t \bar {t}$ & $-.11<X_m<.06$ &  
$-.28<X_m<.13$  \\ \hline
$Z_L Z_L \rightarrow t \bar {t}$ & $-.015<X<.017$ & 
$-.07<X<.08$  \\ \hline \hline
\end{tabular}
\end{center}
\vskip 0.08in
\caption{The range of parameters for which the total number of events 
deviates by less than $2\sigma$ from the {\it no-Higgs} SM prediction.}
\label{bounds}
\end{table}

As shown in Table \ref{bounds}, these coefficients can be probed to 
about an order of $10^{-2}$ to $10^{-1}$.  For this Table, we have 
only consider an unpolarized $e^{-}$  beam for the LC.  To obtain the 
bounds we have set all the anomalous coefficients to be zero except the 
one of interest.  (The definitions of the combined coefficients $X$, 
$X^{'}$, $X_1$, $X_2$ and $X_3$ are given in the previous section.)

If the LC 
is operated at the $e^{-} e^{-}$ mode with the same CM energy of 
the collider, then it can not be used to probe the effects for 
$W^{+}_L W^{-}_L \rightarrow t \bar {t}$, 
but it can improve the bounds on the combined coefficients $X_1$, $X_2$ 
and $X_3$ because the event rate will increase by a factor of $2$ for 
$W^{-}_L Z_L \ra b{\ov t}$ production.  

By combining the limits of these ranges of parameters we can find the 
corresponding limits on the range of the effective coefficients 
$a_{zz1}, \, a_{z2}, \, a_{z3}, \, a_{z4}$, and   $a_m$.  For example, 
if we consider the limits for the LC, we will see that the limits 
for $a_{z2}$ and $a_{z3}$ are directly given by $X_2$ and $X_3$,  
respectively.  Then, we 
can combine this information with the limits given 
for $X^{'}$ and $X_m$, to find the limits for $a_{zz1}$ 
($-.38< a_{zz1}< .24$) and for $a_m$ ($-.34< a_m< .3$).  If we consider
one anomalous coupling at a time, then these bounds can be largely
improved, for instance $-0.03<a_{zz1}<0.04$ from $X^{'}$ (for
$W^{+}_L W^{-}_L \rightarrow t \bar {t}$ scattering).

The above results are for the LC with a $1.5$ TeV CM energy.  
To study the possible new effects in the production rates of  
$W^{+}_L W^{-}_L \rightarrow t \bar {t}$ at the LC with different CM 
energies $E=\sqrt{s}$ 
we plot the production rates for various values of $X^{'}$ 
in Figure~\ref{fwwpro} (Again, $X^{'}=0$ stands 
for the {\it no-Higgs} SM).     
If $X^{'}$ can be as large as $-1.0$, then a $1$ TeV  LC will already 
observe the anomalous rate via $W^{+}_L W^{-}_L$ fusion\footnote{If 
$X^{'}$ is too big , partial wave unitarity is violated at this order.}.
For $X^{'}\;=0.5$ the event rate at $1.5$ TeV is down by about 
a factor of $2$ from the SM event rate\footnote{For positive values of 
$X^{'}$ the rate tends to diminish below the SM rate, but then at some 
value near $0.5$ the rate begins to grow back up towards the SM rate.}.

\begin{figure}
\centerline{\hbox{
\psfig{figure=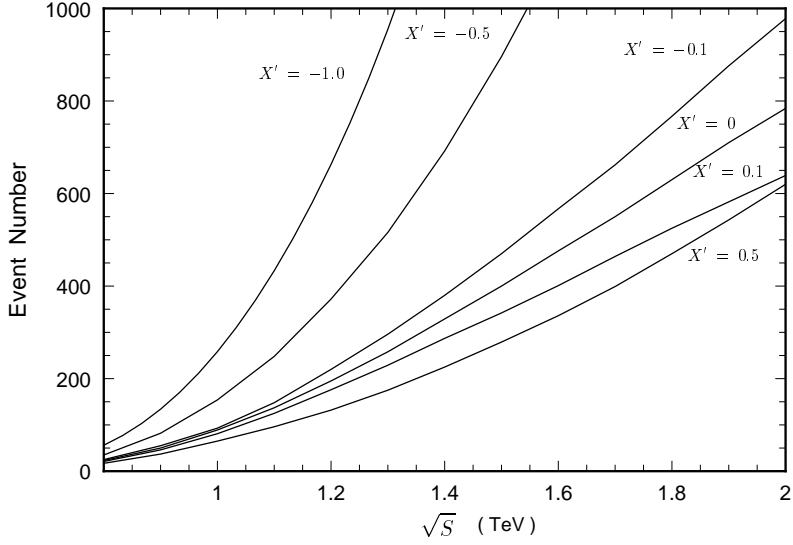,height=3in}}}
\caption{Number of $t\ov t$ events at the LC from $W^{+}_L W^{-}_L$
fusion for different values of the effective coefficient $X^{'}$ as a
function of the CM energy.}
\label{fwwpro}
\end{figure}

\subsection{CP violating effects due to dimension 5 interactions}
\indent\indent

The complete set of anomalous dimension 5 operators listed in 
${\cal {L}}^{(5)}$ consists of CP-invariant and 
CP-violating operators.  
In our study of the top quark production rates due to these anomalous 
operators up to order $\frac{1}{\Lambda}$, we have only considered the 
CP-even part of the operators.  Their contribution, like
the one from the {\it no-Higss} SM at tree level, is real.  
If the coefficient of the CP-violating part is not zero, then it will 
contribute to the imaginary part 
of the helicity amplitude, and one has  
to examine CP-odd observables to probe these operators. 

To illustrate this point, let us 
consider the CP-odd part of the four-point 
scalar type operator $O_{g{\cal W}{\cal W}}$ 
and the electric dipole moment 
term of $O_{{\cal A}}$ ( Eqs.~(\ref{oa}) and (\ref{ogww}) ).   After 
including the contributions from the 
{\it no-Higss} SM and the above two 
CP-odd operators, 
the helicity amplitudes for 
the $W^{+}_L W^{-}_L \rightarrow t \bar {t}$ 
process in the $W^{+}_L W^{-}_L$ CM frame are:
\begin{eqnarray}
a_{\pm \pm} =&& \pm {{m_t E}\over {v^2}} + 
i 2 {E^3\over {v^2}} {{\left( {\tilde a}_{ww1} + 
2 a_d c_{\theta} \right)}\over {\Lambda}}\; , \nonumber \\
a_{+-} =&& {{2\,{{{\it m_t}}^2}\,{\it s_{\theta}}}\over 
    {\left( {{{{{\it m_b}}^2}}\over {2\,{E^2}}} + 
        \left( 1 - {\it c_{\theta}} \right) \,
\left( 1 - 
{{{{{\it m_t}}^2}}\over {2\,{E^2}}} \right)\right) {v^2}}}\; , \\ 
a_{-+} =&& 0\; . \nonumber
\end{eqnarray}
Where by $a_d$ and 
${\tilde a}_{ww1}$ we refer to the imaginary parts of 
the coefficients of $O_{\cal A}$ 
and  $O_{g{\cal W}{\cal W}}$, respectively. 

One of the CP-odd observables that 
can measure $a_d$ and ${\tilde a}_{ww1}$ 
is the transverse polarization $P_{\perp}$ of 
the top quark which is the 
degree of polarization of the top quark in the direction  
perpendicular to the plane of the 
$W^{+}_L W^{-}_L \rightarrow t \bar {t}$ scattering process.  

It was shown in Ref. \cite{kanetop} that 
\begin{equation}
P_{\perp}\; =\; \frac{2 Im \left( {a_{++}^{*}a_{-+}+a_{+-}^{*} a_{--}}
\right)} {|ww_{++}|^2+|ww_{+-}|^2+|ww_{-+}|^2+|ww_{--}|^2}\;.\nonumber 
\end{equation}
which gives, up to the order $\frac{1}{\Lambda}$,
\begin{equation}
P_{\perp} \;{\cong} \;
{{4 s_{\theta} E}\over {{\left( {{{{{\it m_b}}^2}}\over {2\,{E^2}}} + 
   \left( 1 - {\it c_{\theta}} \right) \,
   \left( 1 - {{{{{\it m_t}}^2}}\over {2\,{E^2}}} \right)  \right)}} } 
{{\left( {\tilde a}_{ww1}+2 a_d c_{\theta}\right)}\over 
{\Lambda}}.\nonumber 
\end{equation}
Where $E\,=\,\sqrt{s}$ is the CM energy 
of the $W^{+} W^{-}$ system, and $P_{\perp}$ is 
defined to be a value between $-1$ and $1$.  
For instance, for $E=1.5$ TeV, $\Lambda = 3$ TeV, and 
$\theta = \frac{\pi}{2}$ (or $\frac{\pi}{3}$), we obtain 
$P_{\perp}\;=\; 4{\tilde a}_{ww1}$  
(or ${4 {\sqrt{3}}} ({\tilde a}_{ww1}+a_d$)).   Since $P_{\perp}$ is 
bounded to be $1$ by definition, this requires 
$\mid {\tilde a}_{ww1}\mid  < \frac{1}{4}$ and  
$\mid {\tilde a}_{ww1}+a_d\mid  < \frac{1}{4\sqrt{3}}$.   

If we consider a $1.5$ TeV $e^{+} e^{-}$ collider 
with a number of $t\ov t$ 
events from $W^{+}_L W^{-}_L$ fusion of approximately 100
( which is approximately the {\it no-Higgs} SM rate for a $W^{+} W^{-}$ 
invariant mass between  800 GeV 
and 1100 GeV ) and assume that $P_{\perp}$ can be measured to about  
$\frac{1}{\sqrt{100}}=10\%$, then an agreement between data and the 
{\it no-Higgs} SM prediction ($P_{\perp}=0$ at tree level) implies that 
$\mid {\tilde a}_{ww1}\mid \,\leq \, 0.04$ for the case that $a_d=0$.

\section{Conclusions}
\indent\indent

Because top quark is heavy ($m_t \sim v/\sqrt{2}$), it is likely that
the interaction of the top 
quark can deviate largely from the SM predictions
if the electroweak symmetry breaking and the generation
of fermion masses are closely related.
In this study, we have 
applied the electroweak chiral Lagrangian to probe 
new physics beyond the SM by studying the couplings of the top quark 
to gauge bosons.  We have restricted ourselves to only consider the
interactions of the top and bottom quarks and not the flavor changing
neutral current vertices like $t$-$c$-$Z$.  Furthermore, seeing the
heaviness of the top quark as a possible indication that any new
physics effects associated to the symmetry breaking (and mass
generating) sector will manifest themselves preferably on this
particle, we have considered only the couplings that involve the
top quark as showing possible deviations from the standard values.
(The vertex $\bbz$ is considered unmodified.)   We introduced 4
effective coefficients: two that represent the non-standard couplings
associated to the left and right handed charged currents $\klc$ and
$\krc$, and two more for the anomalous left and right handed neutral
currents $\kln$ and $\krn$.
Then, we used the precision LEP data to set bounds on the couplings
$\kln$, $\krn$, and $\klc$, and we also discussed how the SLC
measurement of $A_{LR}$ can modify these constraints.  The right
handed charged current coupling $\krc$ has to be constrained by means
of the CLEO measurement on $b\ra s \gamma$.  Last,  we showed
how to improve our knowledge about the top quark nonstandard 
couplings at current and future colliders such as at the Tevatron, 
the LHC, and the LC.

Because of the non--renormalizability 
of the electroweak chiral Lagrangian 
one can only estimate the size of 
these nonstandard couplings by studying the 
contributions to LEP/SLC observables at the order of
${m_t^2}\ln{\Lambda}^{2}$, where $\Lambda = 4 \pi v \sim 3$ TeV 
is the cutoff scale of the 
effective Lagrangian.  Nevertheless, this does not
mean we can not extract useful information.  For instance,
by assuming that the \bbz vertex 
is not modified, we found that $\kln$ is already 
constrained to be $-0.05 < \kln < 0.17$ ($0.0 < \kln < 0.15$) 
by LEP/SLC data at the 95\% C.L. for a 160 (180) GeV top quark.
Although $\krn$ and $\klc$ are allowed to be in the full
range of $\pm 1$, the precision LEP/SLC data do impose some correlations
among $\kln$, $\krn$, and $\klc$. 
($\krc$ does not contribute to the LEP/SLC
observables of interest in the limit of $m_b=0$.)

Inspired by the experimental fact $\rho \approx 1$, reflecting 
the existence of an approximate custodial symmetry, we 
related $\kln$ and $\klc$. Then, the remaining two free parameters
$\kappa_L=\kln$ and $\kappa_R=\krn$ get to be strongly 
correlated as well ($\kappa_L \sim 2 \kappa_R$).

We noted that the relations among the $\kappa$'s
can be used to test different models of 
electroweak symmetry-breaking. For instance, a heavy SM Higgs boson 
($m_H \gg m_t$) will
modify the couplings \ttz and \tbw of
a heavy top quark at the scale $m_t$ such that
$\kln = 2\klc$, $\kln =-\krn$, and $\krc=0$.
Another example is the effective model discussed in Ref.~\cite{pczh}
where, $\krc=\klc=0$, in which the low energy precision data 
impose the relation $\kappa_L \sim\kappa_R$. 
On the other hand, the simple commuting extended technicolor model 
presented in Ref.~\cite{sekhar3} predicts that the nonstandard 
top quark couplings are of the same order as the nonstandard 
bottom quark couplings, and are thus small.

Undoubtedly, direct detection of the top quark at the Tevatron,
the LHC, and the LC is crucial to measuring the couplings of 
\tbw and \ttz. At hadron 
colliders, $\klc$ and $\krc$ can be measured by 
studying the polarization of the $W$ boson from top quark decay
in $t \bar t$ events, and 
from the production rate of the single top quark event
via $W$-gluon fusion, $W^*$ or $W t$ processes.
The LC is the best machine to measure $\kln$ and $\krn$ which can be 
measured from studying the angular distribution and the polarization
of the top quark produced in $e^- e^+$ collision.


If a strong dynamics of the electroweak symmetry breaking mechanism can
largely modify the dimension 4 anomalous couplings, it is natural to ask
whether the same dynamics can also give large dimension 5 anomalous
couplings.  In the framework 
of the electroweak chiral Lagrangian, we have
found that there are 19 independent dimension five operators 
associated with the top quark and the bottom quark system.  
The high energy behavior, two powers in $E$ above the 
{\it no-Higgs} SM, for the 
$V_L V_L \ra t\ov t,\;t\ov b,\;({\rm or}\;b\ov t)$ processes,  
gives them a good possibility to manifest themselves 
through the production of $t {\bar t}$ pairs or single-$t$ or $\ov t$ 
events at the LHC and LC in high energy collisions.    
Since in the high energy 
regime a longitudinal gauge boson is equivalent to 
the corresponding would-be Goldstone boson (cf. Goldstone Equivalence
Theorem \cite{et}), the production of top quarks 
via $V_L V_L$ fusions can
probe the part of the electroweak symmetry breaking sector 
which modifies
the top quark interactions.
To simplify our discussion on the accuracy for the measurement of these 
anomalous couplings at future colliders, we have taken the dimension 4
anomalous couplings to be zero for this part of the study.  Also we have
considered a special class of new physics effects in which an underlying
custodial $SU(2)$ symmetry is assumed that gets broken in such a way as
to keep the vertices of the bottom quark unaltered (as was done for the
dimension 4 case).  This approximate custodial symmetry then
relates some of the coefficients of the anomalous operators.
Then we study the contributions of these couplings to the production
rates of the top quark. 
We find that for the leading contributions at high
energies, only the 
S- and P-partial wave amplitudes are modified by these
anomalous couplings if 
the magnitudes of the coefficients of the anomalous
dimension 5 operators are allowed to be as large as $1$ 
(as suggested by the naive dimensional analysis \cite{georgi,howard}),
then we 
will be able to make an unmistakable identification of their effects to
the production 
rates of top quarks via the longitudinal weak boson fusions.
However, if the measurement of the top quark 
production rate 
is found to agree with the SM prediction, then one can bound 
these coefficients to be at most of order $10^{-2}$ or $10^{-1}$.   
This is about a factor 
$\frac{\Lambda}{m_t} \simeq \frac{3 
{\rm {TeV}}}{175 {\rm {GeV}}}\sim O(10)$ 
more stringent than in the case of the study of NLO 
bosonic operators via the $V_L V_L \ra V_L V_L$ scattering processes 
\cite{et,poc,sss}.    Hence, for those models of electroweak symmetry 
breaking for which the 
naive dimensional analysis gives the correct size for
the coefficients of 
dimension 5 effective operators, the top quark production
via $V_L V_L$ fusions can be a more sensitive probe to EWSB
than the longitudinal
gauge boson pair production via $V_L V_L$ 
fusions which is commonly studied.  
For completeness, we also briefly 
discuss how to study the CP-odd operators
by measuring the CP-odd observables.  
In this paper we study their effects on the transverse 
(relative to the 
plane of $W^{+}_L W^{-}_L \rightarrow t\bar t$ scattering) 
polarization of the top quark. 

In conclusion, the production 
of top quarks via $V_L V_L$ fusions at the 
LHC and the LC 
should be carefully studied when data is available because 
it can be sensitive to the electroweak symmetry breaking mechanism, even
more than the commonly studied $V_L V_L \ra V_L V_L$ processes in
some models of strong dynamics.

\medskip
\section*{Acknowledgments }
C.P.Y. would like to 
thank Marek Jezabek and all those involved in organizing
the Cracow school for their warm 
hospitality.  We thank Douglas Carlson and
Hong-Jian He for helpful discussions.
F. Larios  was supported in part by the Organization of American 
States, and by the  Sistema Nacional de Investigadores.  C.P.Y.  was 
supported in part by the NSF grants  PHY-9309902 and PHY-9507683.

\end{document}